\documentclass[useAMS,usenatbib,twocolumn]{mnras}
\usepackage{graphics}
\usepackage{graphicx}
\usepackage{epstopdf}
\usepackage{amsmath}
\usepackage{amssymb}
\usepackage{xspace}
\usepackage{commath}
\usepackage{times}
\usepackage{soul}
\usepackage{bm}
\usepackage{mathtools}

\usepackage{hyperref}
\hypersetup{dvips, colorlinks=true, linkcolor=black, citecolor=black, filecolor=blue, urlcolor=black}

\usepackage{acronym}
\newacro{PDF}{probability distribution function}
\newacroplural{PDFs}{probability distribution functions}
\newcommand{\PDF}{\ac{PDF}}
\newcommand{\PDFs}{\acp{PDF}}
\newacro{DF}{distribution function}
\newacroplural{DFs}{distribution functions}
\newcommand{\DF}{\ac{DF}}

\newacro{BH}{black hole}
\newacroplural{BHs}{black holes}
\newcommand{\BH}{\ac{BH}}

\newacro{IMBH}{intermediary mass black hole}
\newacroplural{IMBHs}{intermediary mass black holes}
\newcommand{\IMBH}{\ac{IMBH}}
\newcommand{\IMBHs}{\acp{IMBH}}
\newacro{VRR}{vector resonant relaxation}
\newcommand{\VRR}{\ac{VRR}}
\newacro{SRR}{scalar resonant relaxation}
\newcommand{\SRR}{\ac{SRR}}
\newacro{NR}{non-resonant relaxation}
\newcommand{\NR}{\ac{NR}}

\newcommand{\rd}{\mathrm{d}}
\newcommand{\re}{\mathrm{e}}

\newcommand{\rt}{\mathrm{t}}
\newcommand{\MBH}{M_{\bullet}}
\newcommand{\hbL}{\widehat{\mathbf{L}}}
\newcommand{\bK}{\mathbf{K}}
\newcommand{\mJ}{\mathcal{J}}
\newcommand{\bx}{\mathbf{x}}
\newcommand{\tp}{t^{\prime}}
\newcommand{\fb}{f^{\mathrm{b}}}
\newcommand{\deltaD}{\delta_{\mathrm{D}}}
\newcommand{\bKp}{\mathbf{K}^{\prime}}
\newcommand{\Tc}{T_{\mathrm{c}}}
\newcommand{\Cb}{C^{\mathrm{b}}}
\newcommand{\bQ}{\mathbf{Q}}
\newcommand{\bO}{\mathbf{\Omega}}
\newcommand{\taup}{\tau^{\prime}}
\newcommand{\bA}{\mathbf{A}}
\newcommand{\bB}{\mathbf{B}}
\newcommand{\bD}{\mathbf{D}}
\newcommand{\bI}{\mathbf{I}}
\newcommand{\erf}{\mathrm{erf}}
\newcommand{\Msun}{M_{\odot}}
\newcommand{\mstar}{m_{\star}}
\newcommand{\Min}{\mathrm{Min}}
\newcommand{\Max}{\mathrm{Max}}
\newcommand{\ain}{a_{\mathrm{in}}}
\newcommand{\aout}{a_{\mathrm{out}}}
\newcommand{\ein}{e_{\mathrm{in}}}
\newcommand{\eout}{e_{\mathrm{out}}}
\newcommand{\phip}{\phi^{\prime}}
\newcommand{\bM}{\mathbf{M}}
\newcommand{\rh}{r_{\mathrm{h}}}
\newcommand{\mmin}{m_{\mathrm{min}}}
\newcommand{\amin}{a_{\mathrm{min}}}
\newcommand{\amax}{a_{\mathrm{max}}}
\newcommand{\emin}{e_{\mathrm{min}}}
\newcommand{\emax}{e_{\mathrm{max}}}
\newcommand{\ellmax}{\ell_{\mathrm{max}}}
\newcommand{\mO}{\mathcal{O}}
\newcommand{\bKt}{\mathbf{K}_{\mathrm{t}}}
\newcommand{\mt}{{m_{\mathrm{t}}}}
\newcommand{\at}{a_{\mathrm{t}}}
\newcommand{\et}{e_{\mathrm{t}}}

\newcommand{\be}{\mathbf{e}}
\newcommand{\mbh}{m_{\bullet}}
\newcommand{\Tdiff}{T_{\mathrm{diff}}}
\newcommand{\phidisc}{\phi_{\mathrm{disc}}}

\begin{document}

\title[VRR and neighbour separation]{
Young stellar cluster dilution near supermassive black holes:\\
the impact of Vector Resonant Relaxation on neighbour separation
}

\author[J. Giral Mart\'inez, J.-B. Fouvry, C. Pichon]{Juan Giral Mart\'inez$^{1,2}$, Jean-Baptiste Fouvry$^{1}$ and Christophe Pichon$^{1,3,4}$
\vspace*{6pt}\\
\noindent
$^{1}$ CNRS and Sorbonne Universit\'e, UMR 7095, Institut d'Astrophysique de Paris, 98 bis Boulevard Arago, F-75014 Paris, France\\
\noindent$^{2}$ Ecole polytechnique, Route de Saclay, 91120 Palaiseau, France \\
\noindent$^3$ IPHT, DRF-INP, UMR 3680, CEA, Orme des Merisiers Bat 774, 91191 Gif-sur-Yvette, France\\
\noindent$^{4}$ Korea Institute of Advanced Studies (KIAS) 85 Hoegiro, Dongdaemun-gu, Seoul, 02455, Republic of Korea
}

\maketitle

\begin{abstract}
We investigate the rate of orbital orientation dilution of young stellar clusters
in the vicinity of supermassive black holes.
Within the framework of vector resonant relaxation,
we predict the time evolution of the two-point correlation function
of the stellar orbital plane orientations
as a function of their initial angular separation
and diversity in orbital parameters (semi-major axis, eccentricity).
As expected, the larger the spread in initial orientations
and orbital parameters,
the more  efficient the dilution of a given set of co-eval stars,
with a characteristic timescale set up
by the coherence time of the background potential fluctuations.
A  Markovian prescription which matches numerical simulations
allows us to efficiently probe the underlying kinematic properties of the 
unresolved nucleus when requesting consistency
with a given dilution efficiency, 
imposed  by the observed stellar disc
within the one arcsecond of Sgr A*.
As a proof of concept, we compute maps of constant
dilution times as a function 
of the semi major axis cusp index
and fraction of intermediate mass black holes
in the old background stellar cluster.
This computation suggests that vector resonant relaxation
should prove useful in this context since it impacts
orientations on timescales comparable to the stars' age.
\end{abstract}


\section{Introduction}
\label{sec:intro}

Most nearby galaxies harbour a supermassive \BH\
in their centre, surrounded by a dense stellar cluster~\citep{Genzel2010,KormendyHo2013}.
In the last few years,  outstanding instrumental  developments have led  
to observational breakthrough in this context:
 the strings of gravitational wave detections following 
the coalescence of black holes~\citep{Ligo2016},
the first shadow image of the horizon 
via radio interferometry of M87~\citep{EHT2019},
as well as the first measurement of the relativistic precession of stars
in our own Galaxy~\citep{Gravity2020}.
These past successes should undoubtedly be followed by others,
accompanying the upcoming thirty meter-class optical instruments~\citep{Do2019}
and space interferometry~\citep{Lisa2006}. These datasets 
will allow us to put more stringent constraints
on the vicinity of massive black holes embedded in  galactic centres,
e.g.\@, to identify dark relics
such as intermediate mass black
holes~\citep{PortegiesMcMillan2002}.

Owing to the infinite potential well generated by the central \BH\@,
galactic nuclei, even in isolation,  are  stellar systems
that involve a wide range of dynamical timescales and
processes~\citep{RauchTremaine1996,HopmanAlexander2006,Alexander2017}.
(i) Since the \BH\ dominates the gravitational potential,
stars follow nearly Keplerian ellipses. This is the dynamical time.
(ii) The deviations from a Keplerian potential due to the stellar mass
and relativistic corrections lead to the in-plane precessions
of the Keplerian ellipses. This is the precession time.
(iii) Subsequently, because of the non-spherical components of the potential
fluctuations, the orbital orientations of the stars get reshuffled.
The Keplerian ellipses' angular momentum vectors
change in orientations,
without changing in magnitude (i.e.\ eccentricity) nor in energy (i.e.\ semi-major axis).
This is the timescale for \VRR\@~\citep[][and references therein]{KocsisTremaine2015,FouvryBarOr2019}.
(iv) On longer timescales, resonant torques between the precessing stars
lead to a diffusion of the stars' angular momentum magnitude.
This is the timescale for \SRR\@~\citep[][and references therein]{BarOrFouvry2018}.
(v) Finally, on the longest timescale, the slow build-up of close encounters
between stars allow for the relaxation of their Keplerian energy.
This is the timescale for \NR\@~\citep{BahcallWolf1976,CohnKulsrud1978,ShapiroMarchant1978,BarOrAlexander2016,Vasiliev2017}.

In the present paper, we are interested in the process of \VRR\@ as an 
astrophysical tool to probe the 
structure of galactic centres.
This dynamical process, through which stars see their orbital orientations vary,
plays a crucial role in numerous dynamical phenomena in galactic nuclei.
Indeed, \VRR\ allows for example for the warping of stellar discs~\citep{KocsisTremaine2011},
or for the enhancement of binary mergers rates~\citep{Hamers2018}.
Moreover, because it is the only mechanism that can efficiently shuffle orbital orientations,
a detailed study of \VRR\ is also a mandatory step to understand
the formation and survival of anisotropic, i.e.\ non-spherical, structures
in galactic nuclei.
This is in particular the case of the `clockwise' disc observed
within SgrA*~\citep{Bartko2009,Yelda2014,Gillessen2017},
whose presence constrains the efficiency with which \VRR\
can dissolve anisotropic stellar structures.

Following its first description in~\cite{RauchTremaine1996},
\VRR\ has been extensively studied in various ways.
It was tackled using full numerical simulations~\citep{Eilon2009},
orbit-averaged simulations~\citep{KocsisTremaine2015},
as well as kinetic predictions~\citep{FouvryBarOr2019,FouvryBarOrAxi2019}.
Because \VRR\ is a rather fast dynamical process
(${ \sim 10^{6} \,\mathrm{yr} }$ for the S-cluster of SgrA*),
it also prove important to characterise the thermodynamical equilibria
of that process~\citep{Roupas2017,TakacsKocsis2018,Tremaine2020}.
On that front,~\cite{SzolgyenKocsis2018} recently showed
how \VRR\ equilibria can exhibit strong anisotropic mass segregation
leading to the formation of black hole discs in galactic nuclei.

In the present paper, we set out to characterise the efficiency with which \VRR\
can lead to the dissolution of anisotropic stellar structures in galactic nuclei,
 building upon~\cite{FouvryBarOr2019}.
As such, we are interested in determining how efficiently
stars with similar orbital orientations,
e.g.\@, stars orbiting within the same stellar disc,
can diffuse away from each other as a result of the \VRR\ process.
We call this process `neighbour separation'.
The paper is organised as follows.
In section~\ref{sec:VRR},
we recall the key equations of \VRR\@,
as well as the statistical properties of the potential fluctuations
present in the system.
In section~\ref{sec:NeighCalc},
we compute the average rate of separation of two nearby test stars.
Section~\ref{sec:Piecewise} relies on a Markovian
assumption to improve upon this prediction
to capture the (very) slow dilution of (very) similar orientations,
while Appendix~\ref{app:Virtual} shows how one can design
an effective Markov process to generate virtual separations
that statistically match that dynamics.
In section~\ref{sec:Applications},
we subsequently illustrate how this formalism can be used to put constraints on  the unresolved cluster orbiting a super massive BH
in a setting inspired by the  SgrA*'s stellar disc.
Finally, we conclude in section~\ref{sec:Conclusion}.

Details of the involved calculations or numerical
validations are spread through appendices.
In particular, we present a toy model (Appendix~\ref{sec:ToyModel})
which allows us to both qualitatively capture
key aspects of \VRR\@,
and produce virtual dilutions of neighbour test stars.

\section{VRR dynamics}
\label{sec:VRR}

\begin{figure}
    \centering
   \includegraphics[trim = 0cm 2.7cm 0cm 1.3cm , clip , width=0.35 \textwidth]{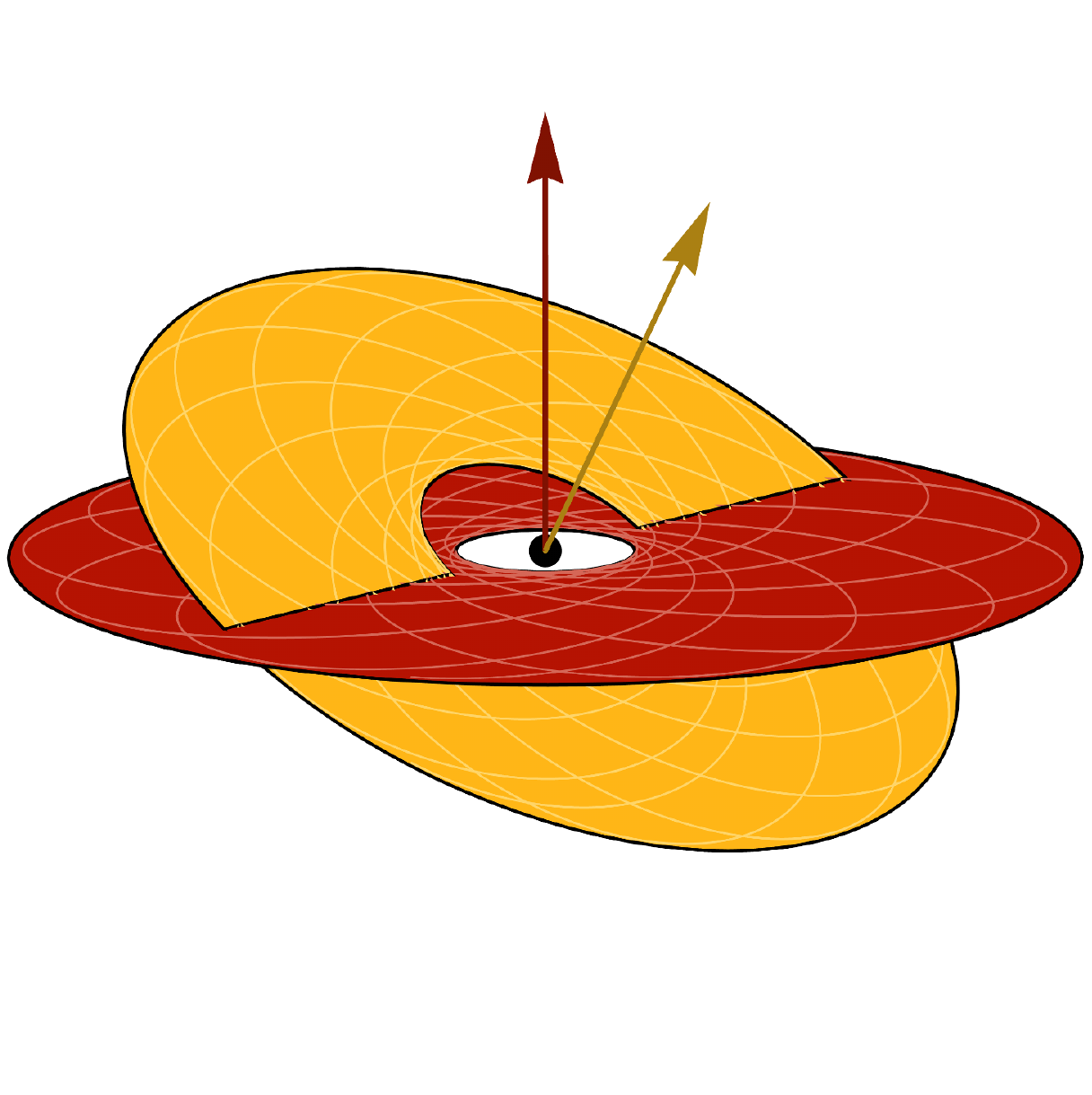}
   \caption{Illustration of the interaction between two annuli
   with different orbital parameters ${ \bK = (m , a, e) }$.
   The torque between the two annuli only depends
   on the relative angle, ${ \cos (\phi) }$, between the two normals.
  VRR predicts the process through which the orientation
  of these orbital annuli diffuse in time.
   }
   \label{fig:Annuli}
\end{figure}

We consider a set of ${ N \gg 1 }$  stars orbiting a supermassive \BH\
of mass $\MBH$,
that we call the background bath  of particles. They represent the unresolved  population (low mass stars, intermediate mass black holes, etc) which contribute to the clumpiness of the  gravitational potential. 
Provided that one considers the dynamics of stars on timescales
longer than the in-plane precession,
but shorter than the relaxation time for eccentricity (by \SRR\@)
and energy (by \NR\@),
one can smear out stars along their respective mean anomalies and pericentre phase.
After this double orbit-average, stars are replaced by annuli,
extending between the pericentre and apocentre of every stellar orbit.
as illustrated in Fig.~\ref{fig:Annuli}.
To every annuli is then associated a set of conserved quantities ${ \bK = (m , a , e) }$,
with $m$ the individual mass, $a$ the semi-major axis, and $e$ the eccentricity.
In that limit, the only remaining dynamical quantity is the instantaneous orbital orientation
given by the unit vector $\hbL$.
 \VRR\  then describes the dynamics
of a set of long-range coupled unit vectors ${ \hbL_{i} }$.

\subsection{Hamiltonian of VRR}
\label{sec:H_VRR}

Following the notations from~\cite{FouvryBarOr2019},
the effective single-particle Hamiltonian of \VRR\ reads
\begin{equation}
H = - \sum_{j = 1}^{N} \bigg\langle \frac{G m m_{j}}{|\bx (t) - \bx_{j} (\tp)|} \bigg\rangle_{t , \tp} ,
\label{H_full}
\end{equation}
where the sum over $j$ runs over all the background particles,
and the average ${ \langle \, \rangle_{t , \tp} }$ operates over
the fast Keplerian motions and the in-plane precessions
of both the test particle, and the background particles,
hereby replacing stars with annuli.

Following Eq.~{(1)} of~\cite{FouvryBarOr2019}, this Hamiltonian
can be rewritten under the shorter form
\begin{equation}
H = - L (\bK) \sum_{j = 1}^{N} \sum_{\mathclap{\substack{\ell \geq 2 \\ \mathrm{even}}}} \,\, \sum_{\mathclap{m = - \ell}}^{\ell} \mJ_{\ell} \big[ \bK , \bK_{j} \big] \, Y_{\ell m} (\hbL) \, Y_{\ell m} (\hbL_{j} (t)) ,
\label{rewrite_H}
\end{equation}
where ${ \hbL (t) }$ stands for the instantaneous orbital orientation
of the test particle, and ${ \hbL_{j} (t) }$ for that  of the bath particles.
Similarly, $\bK$ is the conserved parameter of the test particle,
and $\bK_{j}$ the ones of the bath particle.
In Eq.~\eqref{rewrite_H}, we also introduced the norm of the angular momentum
as ${ L (\bK) = m \sqrt{G \MBH a (1 - e^{2})} }$.
The coupling coefficients, ${ \mJ_{\ell} [ \bK , \bK_{j} ] }$,
depend only on the stars' conserved parameters,
and are constant throughout the \VRR\ dynamics.
Their detailed expressions are recalled in Appendix~\ref{sec:Coupling}.
Finally, in Eq.~\eqref{rewrite_H}, we also introduced the real spherical harmonics,
${ Y_{\ell m} (\hbL) }$, defined with the normalisation
${ \!\int\! \rd \hbL \, Y_{\ell m} Y_{\ell^{\prime} m^{\prime}} \!=\! \delta_{\ell \ell^{\prime}} \delta_{m m^{\prime}} }$.

The instantaneous state of the background bath
is fully described by the discrete \DF\
\begin{equation}
\fb (\hbL , \bK , t) = \sum_{j = 1}^{N} \deltaD (\hbL - \hbL_{j} (t)) \, \deltaD (\bK - \bK_{j}) .
\label{def_phib}
\end{equation}
It can naturally be expanded in spherical harmonics as
\begin{equation}
\fb (\hbL , \bK , t) = \fb_{\alpha} (\bK , t) \, Y_{\alpha} (\hbL) ,
\label{expansion_vphi}
\end{equation}
where the sum over ${ \alpha = (\ell_{\alpha} , m_{\alpha}) }$ is implied,
and we wrote
\begin{equation}
\fb_{\alpha} (\bK , t) = \sum_{j = 1}^{N} \deltaD (\bK - \bK_{j}) \, Y_{\alpha} (\hbL_{j} (t)) .
\label{decomposition_bath}
\end{equation}
As already derived in Eq.~{(9)} of~\cite{FouvryBarOr2019},
the harmonics of the bath evolve according to
\begin{equation}
\frac{\partial \fb_{\alpha} (\bK , t)}{\partial t} \!=\! - \!\! \int \!\! \rd \bKp \, \mJ_{\gamma} \big[ \bK , \bKp \big] E_{\alpha \gamma \delta} \fb_{\gamma} (\bKp , t) \fb_{\delta} (\bK , t) .
\label{evol_eq}
\end{equation}
In that expression, the sum over the harmonic indices ${ \gamma , \delta }$ is implied,
recalling that the coupling coefficient, $\mJ_{\gamma}$, only depends on $\ell_{\gamma}$.
We also introduced the (constant) Elsasser coefficients, $E_{\alpha \gamma \delta}$,
whose properties are presented in Appendix~\ref{sec:Elsasser}.
Equation~\eqref{evol_eq} is the fundamental equation of \VRR\
and describes exactly that dynamics.
The complexity of Eq.~\eqref{evol_eq} comes from the fact
that it is a quadratic matrix differential equation for fields.

\subsection{Noise in VRR}
\label{sec:Noise_VRR}

Rather than describing the exact fate of all the bath particles,
we aim at characterising the statistical properties
of the coefficients ${ \fb_{\alpha} (\bK , t) }$.
This was done in~\cite{FouvryBarOr2019},
and we recall here the key equations.

As they are self-consistently generated by ${ N \gg 1 }$ particles,
we can assume via the central limit theorem that
${ \fb_{\alpha} (\bK , t) }$ is a Gaussian random field.
If we also assume that the bath's evolution is stationary in time,
these stochastic fields are fully characterised by their correlation function
\begin{equation}
\Cb_{\alpha \beta} (\bK , \bKp , t - \tp) = \big\langle \fb_{\alpha} (\bK , t) \, \fb_{\beta} (\bKp , t) \big\rangle , 
\label{def_Cb}
\end{equation}
where ${ \langle \, \cdot \, \rangle }$ stands for the ensemble average over realisations,
i.e.\ over the initial conditions of the bath.
\cite{FouvryBarOr2019} showed that these correlation functions can be well approximated
by temporal Gaussians of the form
\begin{equation}
\Cb_{\alpha\beta} (\bK , \bKp , t \!-\! \tp) \!=\! \delta_{\alpha \beta} \deltaD (\bK \!-\! \bKp) n (\bK) \re^{- \frac{A_{\ell_{\alpha}}}{2} (t / \Tc (\bK))^{2}} ,
\label{res_Cb}
\end{equation}
with the coefficient ${ A_{\ell} \!=\! \ell (\ell \!+\! 1) }$.
In that expression, it was also assumed that the background bath is, on average,
isotropic, and we introduced the \DF\
of the stars' parameters, ${ n (\bK) }$,
with the normalisation convention ${ \!\int\! \rd \hbL \rd \bK n(\bK) = N }$.
This \DF\ is the quantity we aim to constrain using VRR.
 
Equation~\eqref{res_Cb} fully characterises the statistical properties
of the potential fluctuations generated by the background particles.
We note that the amplitude of its temporal correlation is proportional
to the background's stellar density, ${ n (\bK) }$.
As the system is isotropic on average, these correlations
are diagonal when expanded in spherical harmonics,
and only depend on the index $\ell$.
Finally, in Eq.~\eqref{res_Cb}, we also introduced the coherence time of the noise,
${ \Tc (\bK) }$, that follows from Eq.~{(19)} of~\cite{FouvryBarOr2019}
and reads
\begin{equation}
\frac{1}{\Tc^{2} (\bK)} = \!\! \int \!\! \rd \bKp \, n (\bKp) \sum_{\ell} B_{\ell} \, \mJ_{\ell}^{2} \big[ \bK , \bKp \big] ,
\label{def_Tc}
\end{equation}
with the constant coefficient ${ B_{\ell} \!=\! \ell (\ell \!+\! 1) (2 \ell \!+\! 1) / (8 \pi) }$.
The coherence time characterises the time one has to wait
for the bath to reshuffle enough so that its fluctuations
become statistically independent.
As highlighted in the coming section,
this timescale will also set the typical timescale
on which neighbouring test particles will be able to drift
away from one another.

\section{neighbour separation}
\label{sec:NeighCalc}

In this section, we now tackle the problem of describing
the simultaneous separation of two test stars sharing
similar orientations and parameters.

\subsection{Dynamics of test particles}
\label{sec:DynTest}

\begin{figure*}
\centering
\includegraphics[width=0.22 \textwidth]{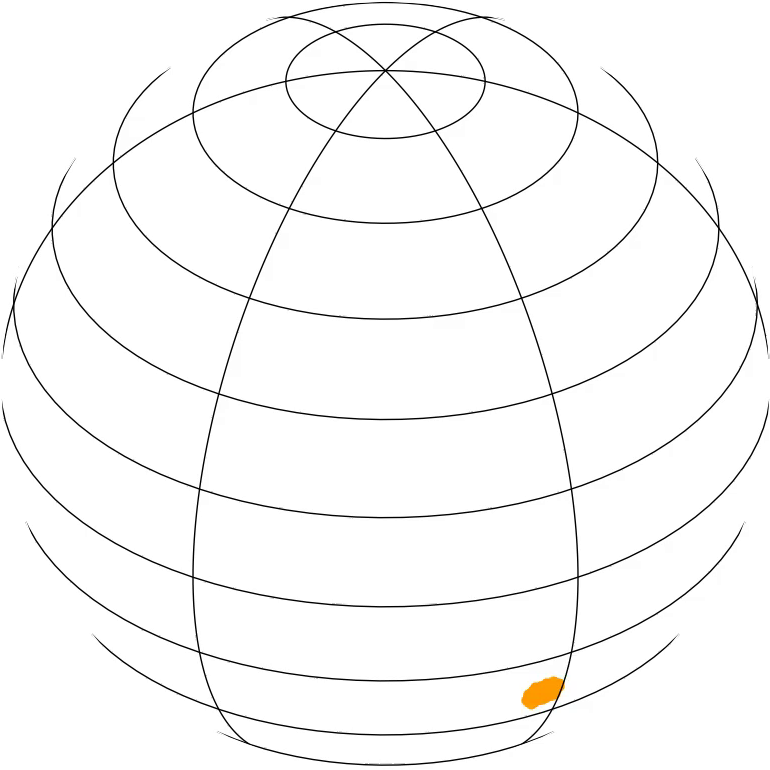}
\hspace{0.01\textwidth}
\includegraphics[width=0.22 \textwidth]{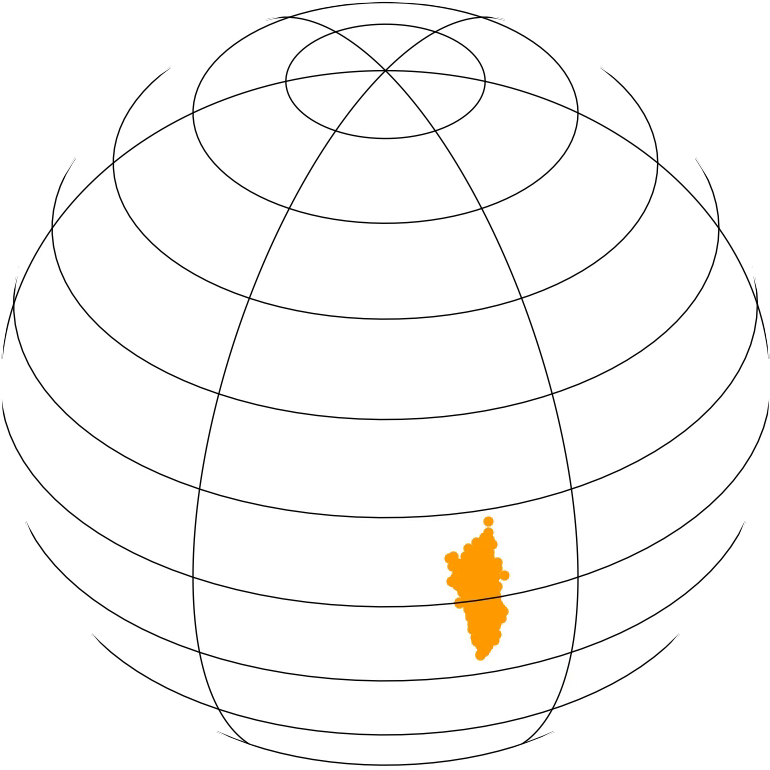}
\hspace{0.01\textwidth}
\includegraphics[width=0.22 \textwidth]{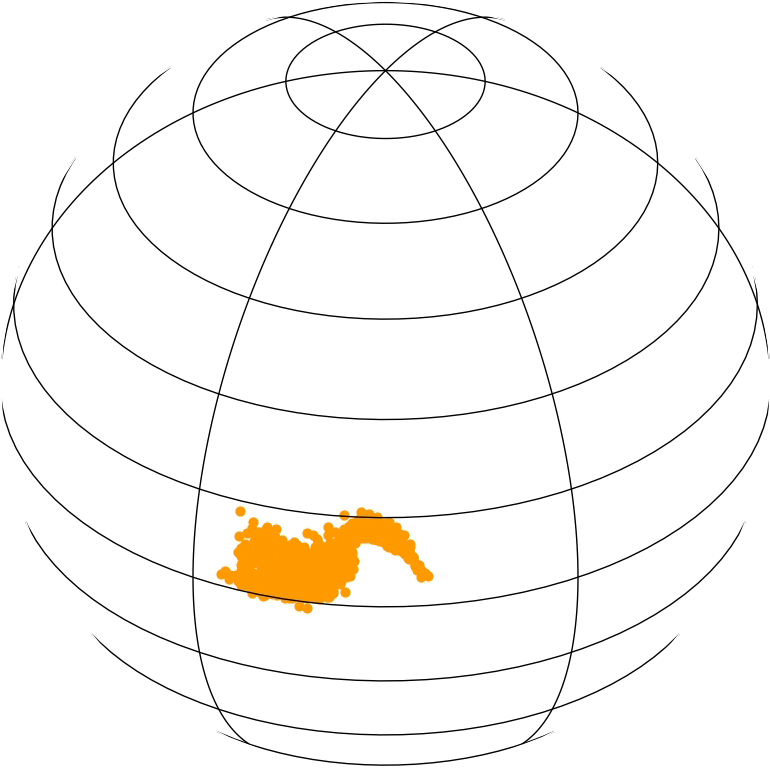}
\hspace{0.01\textwidth}
\includegraphics[width=0.22 \textwidth]{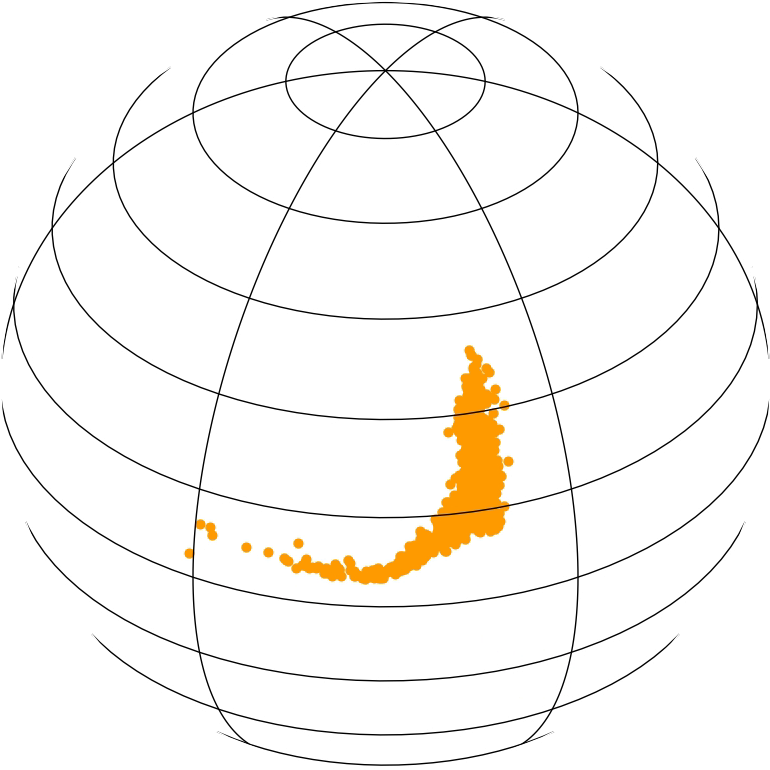}\\
\vspace{0.01\textwidth}
\includegraphics[width=0.22 \textwidth]{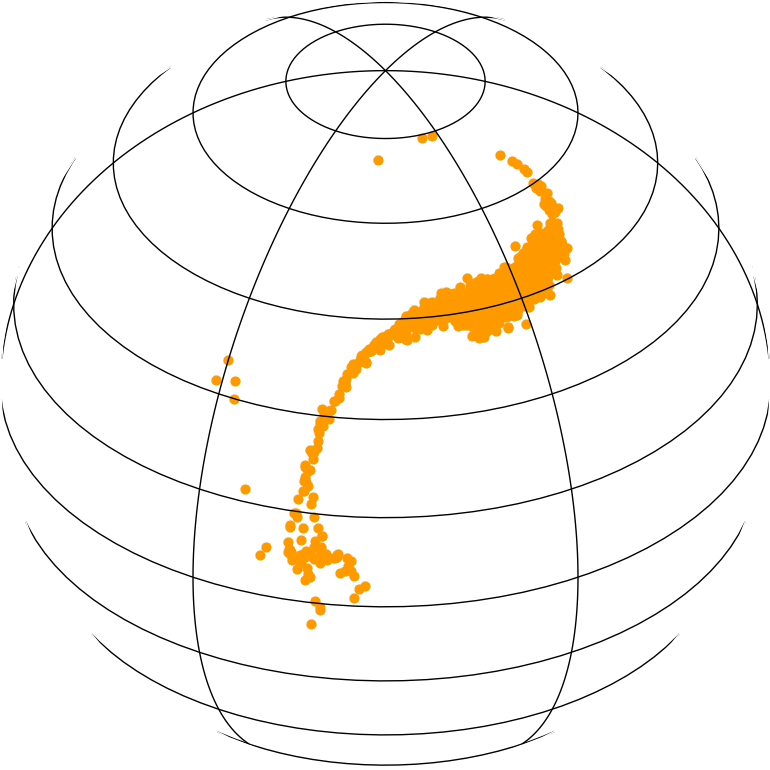}
\hspace{0.01\textwidth}
\includegraphics[width=0.22 \textwidth]{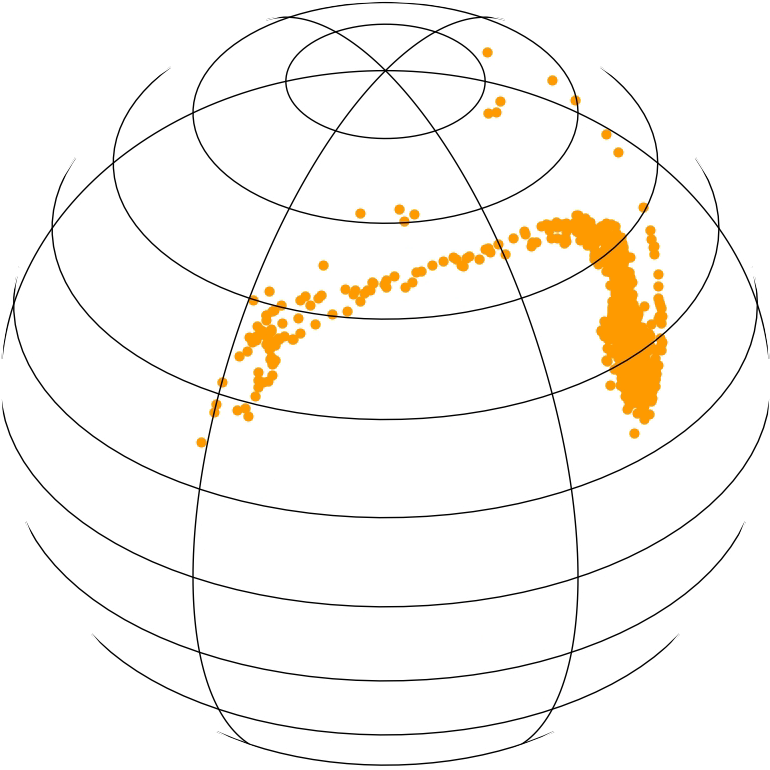}
\hspace{0.01\textwidth}
\includegraphics[width=0.22 \textwidth]{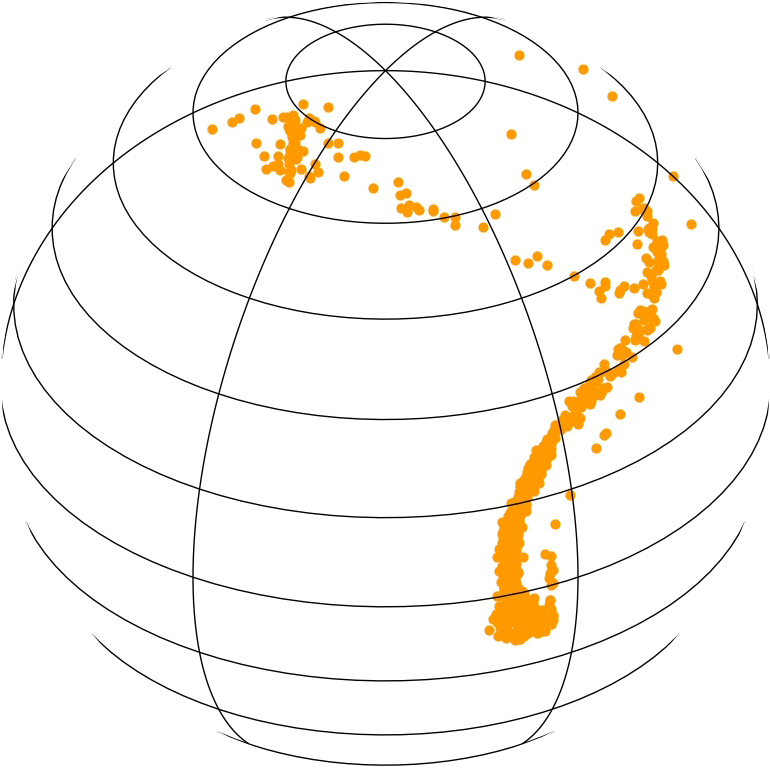}
\hspace{0.01\textwidth}
\includegraphics[width=0.22 \textwidth]{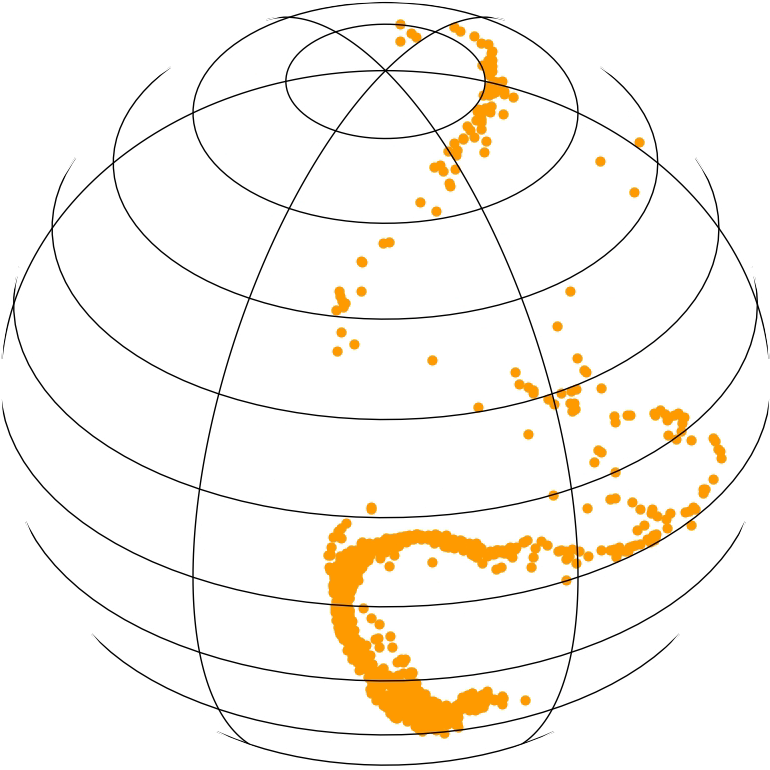}      
\caption{Illustration of the separations between neighbouring tracer particles,
differing initially both in orientations and orbital parameters.
Details for this simulation are spelled out in Appendix~\ref{sec:Simulations}.
This time sequence (from left to right and top to bottom)
shows that a patch of orbital planes initially close to each other 
diffuses while interacting with the background unresolved cluster.
The patch becomes quite elongated,
as a result of the phase space shearing generated
by the instantaneous potential fluctuations.
Similarly, the patch also moves somewhat
faster than it spreads.
Eventually, the test particles will become
distributed over the whole sphere.
   }
   \label{fig:Dilution}
\end{figure*}

Let us consider therefore two test (or tracer) particles,
resp.\ indexed by `$1$' and `$2$'. They represent here  stars
which are bright enough to be observed,
while the background bath corresponds to the
unresolved old stellar cluster (possibly involving black holes),
too dim to be directly imaged.
We place ourselves in the test particle limit,
i.e.\ we assume that the motions of the test particles
are fully imposed by the background bath
(in the so-called zero mass limit). 
As such, we neglect any back-reaction of these particles on the bath,
and we neglect any self-gravity between the two test particles.
This effectively halves the Hamiltonian and allows for the statistics given
by Eq.~\eqref{res_Cb} to be used in our subsequent calculations.
As a result, the background bath is therefore fully self-gravitating,
i.e.\ follows the dynamics imposed by the Hamiltonian from Eq.~\eqref{H_full},
while the test particles are only probing the gravitational potential
in the cluster but do not contribute to it.
Our goal is now to constrain the efficiency with which
a young stellar cluster (i.e.\ the test particles)
can dissolve as a result of the potential fluctuations
generated by the old unresolved stellar cluster (i.e.\ the background bath).
In Fig.~\ref{fig:Dilution},
we illustrate one example of such a dilution
in a numerical simulation.

We denote the parameters of the test particles with ${ \bK_{i} }$
(with ${ i = 1,2 }$)
and their current orientation with ${ \hbL_{i} (t) }$.
Similarly to Eq.~\eqref{def_phib},
each test particle is fully characterised by its \PDF\
\begin{equation}
f^{i} (\hbL , t) = \deltaD (\hbL - \hbL_{i} (t)) .
\label{def_phi_test}
\end{equation}
It can naturally be expanded in spherical harmonics as
${ f^{i} (\hbL , t) = f^{i}_{\alpha} (t) Y_{\alpha} (\hbL) }$,
where the sum over $\alpha$ is implied.
Here, the harmonics decomposition simply reads
\begin{equation}
f^{i}_{\alpha} (t) = Y_{\alpha} (\hbL_{i} (t)) .
\label{def_harm_neigh}
\end{equation}
Because the test particles do not contribute to the system's instantaneous potential,
their individual dynamics follow from Eq.~\eqref{evol_eq}
and take the simpler form
\begin{equation}
\frac{\partial f_{\alpha}^{i} (t)}{\partial t} = - \!\! \int \!\! \rd \bK \, \mJ_{\gamma} \big[ \bK_{i} , \bK \big] \, E_{\alpha \gamma \delta} \, \fb_{\gamma} (\bK , t) \, f^{i}_{\delta} (t) ,
\label{evol_test}
\end{equation}
where, once again, the sums over ${ \gamma , \delta }$ are implied.
In order to better highlight its properties, we can finally rewrite Eq.~\eqref{evol_test}
as
\begin{equation}
\frac{\partial f_{\alpha}^{i} (t)}{\partial t} = - Q_{\alpha \delta}^{i} (t) \, f^{i}_{\delta} (t) ,
\label{evol_test_short}
\end{equation}
where ${ Q_{\alpha \delta}^{i} (t) \!=\! \!\int\! \rd \bK \mJ_{\gamma} [\bK_{i} , \bK] E_{\alpha \gamma \delta} \fb_{\gamma} (\bK , t) }$ is a stochastic matrix 
because ${ \fb_{\gamma} (\bK , t) }$ is a stochastic field,
whose statistical properties were already spelled out
in Eq.~\eqref{res_Cb}.
Equation~\eqref{evol_test_short}
now takes the form of a set of coupled \textsl{linear} differential equations,
sourced by the fluctuations of the background bath
through the time-dependent matrix ${ \bQ_{i} (t) }$.
Let us emphasise that the matrices ${ \bQ_{i} (t) }$
are indexed by the test particles' index, $i$,
as they depend on their conserved parameters $\bK_{i}$.
Yet, ${ \bQ_{i} (t) }$ and ${ \bQ_{j} (t) }$
are highly correlated one with another,
as they both depend on the same instantaneous fluctuations
${ \fb_{\gamma} (\bK , t) }$ generated by the bath.
It is essential to accurately capture these correlations
in order to describe the  process of neighbour separation.

\subsection{Correlation of separation}
\label{sec:CorrelTest}

Having written in Eq.~\eqref{evol_test} the exact evolution equation
for the test particles, we can now address the computation
of their rate of separation.

\begin{figure}
    \centering
   \includegraphics[width=0.35 \textwidth]{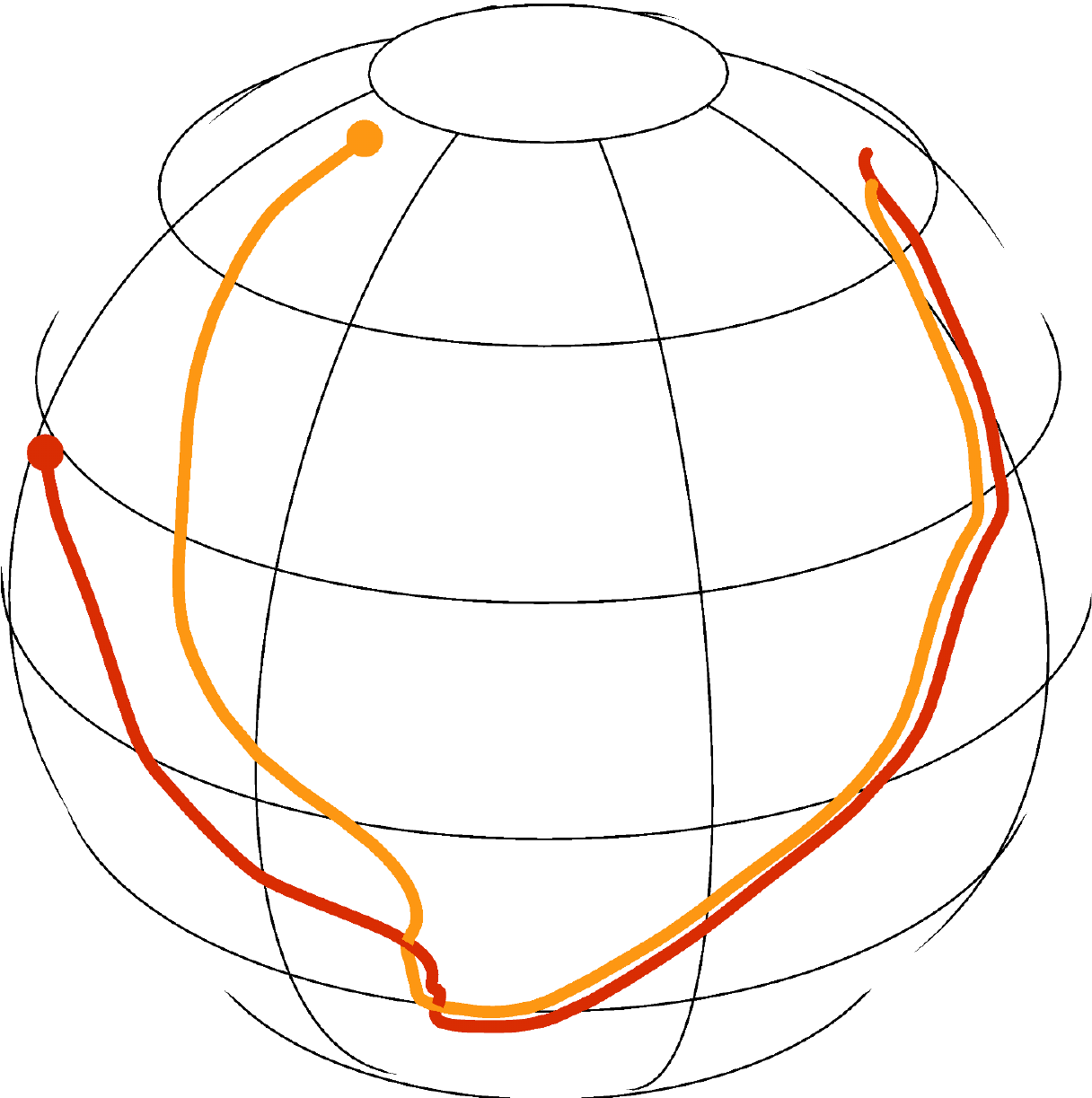}
   \caption{Illustration of separation between
   two neighbouring tracer particles.
The two paths initially follow each other closely, but eventually
diverge from one another.
This behaviour is typical of \VRR\ near a supermassive \BH\@,
and happens on timescales of the order of $\Tc$,
as defined in Eq.~\eqref{def_Tc}.
   }
   \label{fig:NeighbourSeparation}
\end{figure}

\begin{figure}
    \centering
   \includegraphics[width=0.45 \textwidth]{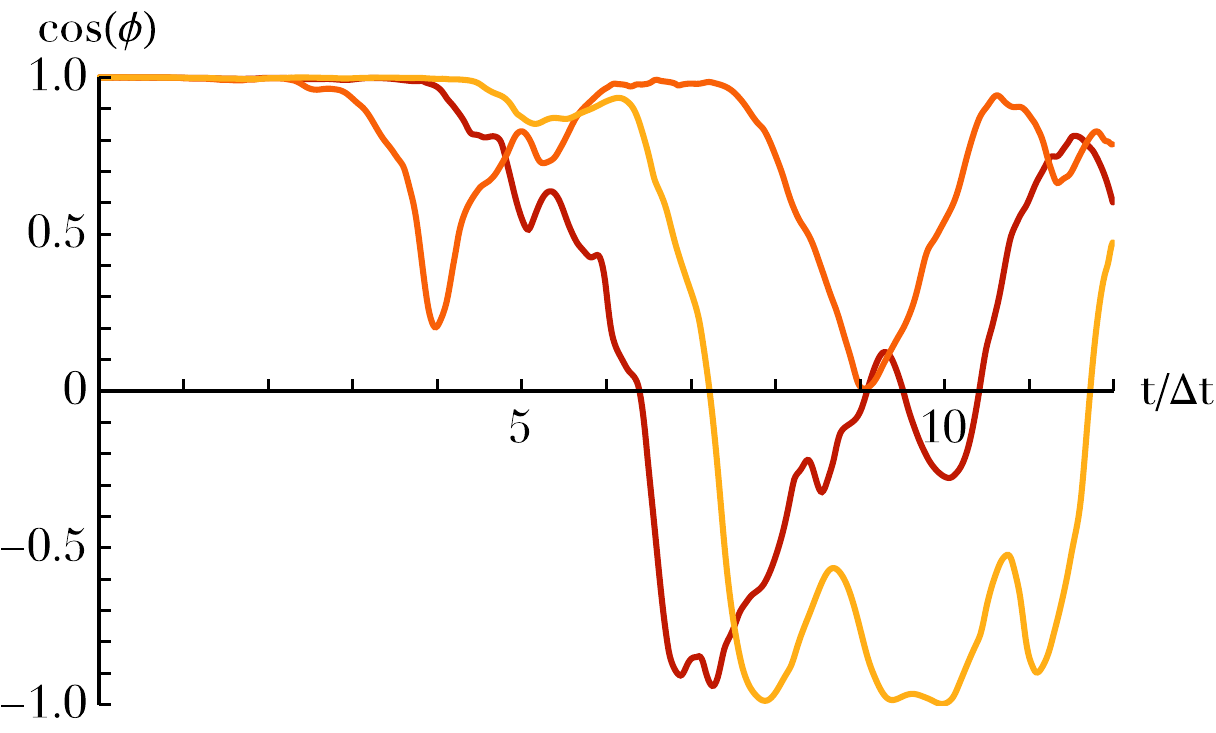}
   \caption{Illustration of typical random walks
   of the angle ${ \cos (\phi) }$ between three pairs
   of test particles (with the same orbital parameters)
   extracted from the numerical simulations
   described in Appendix~\ref{sec:Simulations}.
   Here, we can note that the two orientations
   typically decorrelate after a few ${ \Delta t }$,
   with ${ \Delta t }$ defined as in Eq.~\eqref{choice_Deltat}.
   }
   \label{fig:RandomWalksNBody}
\end{figure}

We are interested in the efficiency with which two nearby particles
diffuse away from each other.
We illustrate such a separation in Fig.~\ref{fig:NeighbourSeparation}.
Here, the main difficulty stems from the fact that these two neighbour particles
are not independent, as they evolve within the same background noise.
As a result, in contrast to Eq.~\eqref{def_Cb},
we are not interested anymore in the correlation of an harmonics at different times,
but rather in the correlation of the harmonics of two different particles,
at the same time.
We therefore define the correlation function
\begin{equation}
C_{\alpha\beta} (t) = \big\langle f^{1}_{\alpha} (t) \, f^{2}_{\beta} (t) \big\rangle ,
\label{def_C_neigh}
\end{equation}
where  the two harmonics are computed at the exact same time.
In practice, for an isotropic noise
and isotropic initial conditions
(as will be the case, e.g.\@, in Eq.~\eqref{calc_C_neigh}),
the correlation function from Eq.~\eqref{def_C_neigh}
has a simple interpretation.
Indeed, owing to the addition theorem
for spherical harmonics, one has
\begin{align}
C_{\alpha \beta} (t) = \frac{\delta_{\alpha\beta}}{4 \pi} \big\langle P_{\ell_{\alpha}} (\cos (\phi (t))) \big\rangle ,
\label{interp_C}
\end{align}
with $P_{\ell}$ the Legendre polynomials,
and ${ \phi (t) }$ the instantaneous angle between the two particles.
Since ${ P_{1} (\cos (\phi)) = \cos (\phi) }$,
the case ${ \ell_{\alpha} = 1 }$ of Eq.~\eqref{def_C_neigh}
is of prime importance as it directly informs us
on the evolution of the angular separation
between the two test particles
on the unit sphere.
Higher order spherical harmonics
are similarly directly connected to higher order moments
of ${ \cos (\phi) }$.
Characterising the separation of neighbours
in the \VRR\ process
amounts, in particular, to characterising the statistical properties
of the random walk undergone by ${ \phi (t) }$,
i.e.\ characterising the statistical properties of the angular separation between the two test particles.
In Fig.~\ref{fig:RandomWalksNBody},
we illustrate examples of such random walks
in numerical simulations.

To characterise this stochastic dynamics,
we follow a method inspired from section~{4} of~\cite{FouvryBarOr2019},
and note that the dynamics of each individual
test particle can be solved formally
using Magnus series~\citep{Blanes2009}.
One of the main advantages of such a solution is that it guarantees
a good behaviour at late times.
Following Eq.~{(29)} of~\cite{FouvryBarOr2019},
the time evolution of the test particles can be solved as
\begin{equation}
f^{i}_{\alpha} (t) = \big[ \re^{\bO_{i} (t)} \big]_{\alpha \delta} \, f^{i}_{\delta} (0) ,
\label{sol_test}
\end{equation}
where the sum over $\delta$ is implied.
In that expression, ${ \bO_{i} (t) }$ is the Magnus matrix,
which to second order in the perturbative expansion reads
\begin{equation}
\bO_{i} (t) = - \!\! \int_{0}^{t} \!\!\! \bQ_{i} (\tau) \, \rd \tau + \frac{1}{2} \!\! \int_{0}^{t} \!\! \rd \tau \!\! \int_{0}^{\tau} \!\!\! \rd \taup \, \big[ \bQ_{i} (\tau) , \bQ_{i} (\taup) \big] ,
\label{def_Omega}
\end{equation}
with ${ [ \bA , \bB ] = \bA \bB - \bB \bA }$ the matrix commutator.

We can now use the formal solution from Eq.~\eqref{sol_test}
to rewrite the correlation of the neighbours of Eq.~\eqref{def_C_neigh} as
\begin{equation}
C_{\alpha \beta} (t) = \big\langle \big[ \re^{\bO_{1} (t)} \big]_{\alpha \gamma} \, \big[ \re^{\bO_{2} (t)} \big]_{\beta \delta} \big\rangle \, \big\langle f^{1}_{\gamma} (0) \, f^{2}_{\delta} (0) \big\rangle ,
\label{rewrite_C_neigh}
\end{equation}
where the sums over $\gamma$ and $\delta$ are implied,
and we have solved for the time evolution of both test particles.
In that expression, we assumed that the statistics of the background noise
is independent of the statistics of the initial locations of the test particles.
As a result, in the l.h.s.\ of Eq.~\eqref{rewrite_C_neigh},
the first ensemble average is over realisations of the background,
while the second average is over the initial location of the test particles.

In Appendix~\ref{sec:InitStat}, we compute the initial statistics of the test particles.
In particular, we show that for isotropic initial conditions,
i.e.\ while the location of the two test particles are correlated one with another,
the distribution of any given test particle on the unit sphere is statistically uniform,
the second average from Eq.~\eqref{rewrite_C_neigh}
can generically be written as
\begin{equation}
\big\langle f^{1}_{\gamma} (0) \, f^{2}_{\delta} (0) \big\rangle = \frac{\delta_{\gamma \delta}}{4 \pi} \, D_{\gamma} ,
\label{def_D}
\end{equation}
In that expression, by isotropy, the diagonal matrix ${ \bD }$
sees its entries $D_{\gamma}$ only depend on $\ell_{\gamma}$.
In the particular case where the two test particles are launched
with the exact same initial orientations, one has ${ \bD = \bI }$,
the identity matrix.

Having characterised the test particles' initial conditions,
we can now go back to Eq.~\eqref{rewrite_C_neigh} which can be written as 
\begin{equation}
C_{\alpha\beta} (t) = \big\langle \big[ \re^{\bO_{1} (t)} \, \bD \, \re^{-\bO_{2} (t)} \big]_{\alpha \beta} \big\rangle ,
\label{rewrite_C_neigh_II}
\end{equation}
where we used the fact that the matrix $\bQ^{i}$ is skew-symmetric,
i.e.\ ${ \bQ_{i} = - \bQ_{i}^{\rt} }$,
so that ${ [ \re^{\bO_{i}} ]^{\rt} = \re^{- \bO_{i}} }$.
At this stage, there are two origins to the separation
undergone by the test particles:
(i) their initial angular separation,
as captured by the matrix $\bD$;
(ii) their difference in orbital parameters,
as captured by ${ \bO_{1} \neq \bO_{2} }$.
The main difficulty in the computation of the ensemble average
from Eq.~\eqref{rewrite_C_neigh} comes from the non-commutativity
of the matrix exponential.
This is the challenging part of the present calculation.

In Appendix~\ref{sec:CompNeigh},
performing a truncation at second-order in the bath fluctuations,
we compute explicitly the ensemble average from Eq.~\eqref{rewrite_C_neigh_II},
and obtain
\begin{equation}
C_{\alpha\beta} (t) = \delta_{\alpha\beta} \, D_{\alpha} \, C_{\alpha}^{\bO} (t) \, C_{\alpha}^{\bD} (t) .
\label{calc_C_neigh}
\end{equation}
In that expression, ${ C_{\alpha}^{\bO} (t) }$ (resp.\ ${ C_{\alpha}^{\bD} (t) }$)
captures the separation of the two test particles sourced by their differences
in conserved parameters (resp.\ in initial orientations).
They generically read
\begin{equation}
C_{\alpha}^{\bO} (t) =  \exp \bigg[ \frac{1}{2} \big\langle \big[ \big( \bO_{1} - \bO_{2} \big)^{2} \big]_{\alpha \alpha} \big\rangle \bigg] ,
\label{def_COmega}
\end{equation}
and
\begin{equation}
C_{\alpha}^{\bD} (t) = \exp \bigg[ \sum_{\gamma} \frac{D_{\alpha} - D_{\gamma}}{D_{\alpha}} \big\langle \bO^{1}_{\alpha\gamma} \, \bO^{2}_{\gamma\alpha} \big\rangle \bigg] .
\label{def_CD}
\end{equation}
We emphasise that Eq.~\eqref{def_CD} is symmetric
w.r.t.\ ${ (1 \!\leftrightarrow\! 2) }$,
owing to the skew-symmetries of ${ \bO^{i} }$.
Having written Eqs.~\eqref{def_COmega} and~\eqref{def_CD}
as exponentials
will guarantee a physically admissible behaviour at late times,
where correlations must tend to $0$.

As detailed in Appendix~\ref{sec:CompNeigh},
one can push further the calculation
using Eq.~\eqref{res_Cb} that provides us with an explicit expression
for the correlation of the bath's fluctuations.
Following this route,
the associated expressions for ${ C_{\alpha}^{\bO} (t) }$
and ${ C_{\alpha}^{\bD} (t) }$
are spelled out explicitly in terms of $n(\bK)$
in Eqs.~\eqref{CbO} and~\eqref{CbD}.

These generic expressions become particularly enlightening
in the limit of small initial angular separations.
Assuming that the two test particles are initially
separated by a small angle $\phi_0$,
as shown in Eq.~\eqref{calc_C_neigh_limit},
one can write Eq.~\eqref{calc_C_neigh} as
\begin{align}
C_{\ell}^{\bO} (t) & \, = \exp \!\big[\! - \tfrac{1}{2} A_{\ell} \Psi^{-} (\bK_{1} , \bK_{2} , t) \big] ,
\nonumber
\\
C_{\ell}^{\bD} (t) & \, = \exp \!\big[\! - \tfrac{1 - \cos (\phi_{0})}{2} A_{\ell} \Psi^{+} (\bK_{1} , \bK_{2} , t) \big] .
\label{shortC_limit}
\end{align}
In these expressions,
we introduced the two auxiliary functions
${ \Psi^{-} }$ and ${ \Psi^{+} }$ as
\begin{align}
\Psi^{-} ( \bK_1, \bK_2 , t ) & \, = \sum_\ell B_\ell \!\!\int\!\! \rd \bK \, n(\bK) \big( \mJ_\ell \big[ \bK_1,\bK \big] \!-\! \mJ_\ell \big[ \bK_2,\bK \big] \big)^2
\nonumber
\\ 
\times & \, \frac{2 \Tc^{2} (\bK)}{A_{\ell}} \chi \big[ \sqrt{A_\ell /2 }(t/ \Tc(\bK)) \big] ,
\label{def_PsiMinus}
\end{align}
and
\begin{align}
\Psi^{+} ( \bK_1, \bK_2 , t ) & \, = \sum_\ell B_\ell \!\!\int\!\! \rd \bK \, n(\bK) \mJ_\ell \big[ \bK_1,\bK \big] \mJ_\ell \big[ \bK_2,\bK \big]
\nonumber
\\
\times & \, \frac{2 \Tc^{2} (\bK)}{A_{\ell}} (A_{\ell} \!-\! 2) \, \chi \big[ \sqrt{A_\ell /2 }(t/\Tc(\bK)) \big] ,
\label{def_PsiPlus}
\end{align}
In Eqs.~\eqref{def_PsiMinus} and~\eqref{def_PsiPlus},
the only temporal dependence
is in the universal dimensionless function ${ \chi (\tau) }$ that reads
\begin{align}
\chi (\tau) & \, = \!\! \int_{0}^{\tau} \!\!\! \rd \tau_{1} \!\! \int_{0}^{\tau} \!\!\! \rd \tau_{2} \, \re^{- (\tau_{1} - \tau_{2})^{2}}
\nonumber
\\
& \, = \re^{- \tau^{2}} - 1 + \sqrt{\pi} \, \tau \, \erf(\tau) .
\label{def_chi}
\end{align}
This function directly stems from the double time integral
of the Gaussian bath fluctuations from Eq.~\eqref{res_Cb}.
In particular, it follows the asymptotic behaviour
${ \chi (\tau) \propto \tau^{2} }$ (resp.\ ${ \propto \tau} $)
for ${ \tau \ll 1 }$ (resp.\ ${ \tau \gg 1 }$)
which corresponds to the ballistic
(resp.\ diffusive) part of the \VRR\ dynamics.

Equation~\eqref{shortC_limit}
offers a simple interpretation of the dynamical
mechanisms driving the separation of neighbours.
In that expression,
a first source of separation stems from
the different in the test particles' parameters,
as captured by the function $\Psi^{-}$ from Eq.~\eqref{def_PsiMinus}.
This can be seen from the coupling factor
${ (\mJ_{\ell} [\bK_{1} , \bK] \!-\! \mJ_{\ell} [\bK_{2} , \bK])^{2} }$
in Eq.~\eqref{def_PsiMinus},
which highlights that both test particles couple to the bath differently.
The closer $\bK_{1}$ and $\bK_{2}$,
the slower the separation of the particles.
As expected, we recover that $C_{\alpha}^{\bO}$
becomes one in Eq.~\eqref{def_PsiMinus},
when ${ \bK_{1} \!=\! \bK_{2} }$,
i.e.\ when the underlying annuli
of the two test particles are identical.

On top of this first effect,
a second source of separation
originates from any initial misalignment
in the test particles' orientations.
This is captured by the function $\Psi^{+}$
from Eq.~\eqref{def_PsiPlus},
which does not vanish
even when the two test particles have the same orbital parameters.
In addition, as highlighted in Eq.~\eqref{shortC_limit},
the smaller the initial separation of the two test particles,
i.e.\ the smaller ${ (1 \!-\! \cos (\phi_{0})) }$,
the longer it takes for the neighbours to get separated.
In particular, we note that ${ C_{\ell}^{\bD} }$
becomes one when ${ \cos (\phi_{0}) \to 1 }$,
i.e.\ when the two test particles share the exact
same initial orientation.
Note that Eq.~\eqref{def_PsiPlus}
involves an extra ${ (A_\ell \!-\! 2) \!=\! (\ell(\ell \!+\! 1) \!-\! 2) }$ 
which corresponds to a Laplacian, reflecting the fact 
that pair separations are only sensitive to tides, not forces.
Moreover,
should the harmonics ${ \ell = 1 }$ have been able 
to drive the \VRR\ dynamics,
this particular harmonics would not have been able
to drive any neighbour separation
through orientations mismatches,
as highlighted by the vanishing factor
${ A_{\ell = 1} \!-\! 2 \!=\! 0 }$ in Eq.~\eqref{def_PsiPlus}.

Let us further note that the prediction
from Eq.~\eqref{shortC_limit} is self-similar,
i.e.\ the only dependence on the considered harmonics $\ell$
is carried by ${ A_{\ell} }$,
which is factored in the exponent.
For a given pair of test stars
(that is, fixing the shape of $\Psi^+$),
the prediction from Eq.~\eqref{shortC_limit}
only depends on the product ${ A_{\ell} (1 \!-\! \cos (\phi_0)) }$.
Since ${ 1/\sqrt{A_{\ell}} }$ is the characteristic scale
of the $\ell^{\mathrm{th}}$ harmonics,
this product compares the test particles' separation
with that of the considered harmonics.
In particular, on the one hand,
harmonics whose scale is larger than the particles' separation,
i.e.\ such that ${ A_{\ell} (1 \!-\! \cos (\phi_0)) \ll 1 }$
do not decorrelate efficiently.
This is because the potential generated by
these bath harmonics
is roughly constant on the scale of the particles' separation.
We note that even for these large scale harmonics,
there is still an unavoidable separation stemming
from the non-vanishing contribution of $C_{\ell}^{\bO}$,
which, in that limit, captures the effect of phase mixing,
i.e.\ the frequency shearing of test particles
with different orbital parameters.
On the other hand, higher order harmonics,
i.e.\ such that
${ A_{\ell} (1 \!-\! \cos (\phi_0)) \gtrsim 1 }$,
are much more efficient at driving neighbour separation
since they vary on angular scales similar to or smaller
than the initial separation of the test particles.

Finally, we emphasise that both functions
$\Psi^{-}$ and $\Psi^{+}$
only depend on the test particles' parameters,
$\bK_{1}$ and $\bK_{2}$,
as well as on the background cluster's parameters, ${ n(\bK) }$.
As highlighted in Eq.~\eqref{shortC_limit},
these functions are independent
of the harmonics $\ell$ of the considered correlation function,
as well as of the statistics of the test particles' initial separation,
given by ${ \cos (\phi_0) }$.
These functions will prove very useful
in Section~\ref{sec:Piecewise}
to construct our piecewise prediction.

Equations~\eqref{calc_C_neigh} and~\eqref{shortC_limit}
are the main results of this section.
In practice, one is not limited to only considering two test particles,
but can rather consider an arbitrary population of test particles
that follow initially a given smooth distribution.
It is straightforward to expand Eq.~\eqref{calc_C_neigh}
to such a population,
as detailed in Appendix~\ref{sec:Population}.
In short, one has
\begin{equation}
\big \langle f_{\alpha}(\bK,t) \, f_{\beta}(\bK',t) \big \rangle \simeq  p(\bK) \, p(\bKp) \, C_{\alpha \beta} (\bK,\bKp,t),
\label{Fokker_X_bignMT}
\end{equation}
where ${ p (\bK) }$ is the PDF of
 the orbital parameters of the test particles and $ C_{\alpha \beta} (\bK,\bKp,t)$ is given by 
 Eq.~\eqref{calc_C_neigh},
 assuming that the orbital parameters of the two test particles
 are resp.\ $\bK$ and $\bKp$.
 
Let us emphasise once again the generality of Eq.~\eqref{calc_C_neigh}
that captures jointly three physical contributions
modulating the efficiency of the neighbours separation:
(i) the orbital distribution of the background particles,
via ${ n(\bK) }$;
(ii) the orbital differences of the two test particles,
via $\bK_{1}$ and $\bK_{2}$,
and further through the \PDF\ ${ p(\bK) }$
appearing in Eq.~\eqref{Fokker_X_bignMT};
(iii) the difference in the initial orientation,
via $\bD$.
In all these expressions,
time is measured in units of ${ \Tc (\bK) }$,
the coherence time of the bath's fluctuations,
which defines the typical timescale for that process.
As such, Eq.~\eqref{calc_C_neigh} is a very general result,
that can be applied to a wide variety of physical processes.
For instance, in Section~\ref{sec:Applications} we will use it to put
constraints on ${ n(\bK) }$,
which describes the unresolved old stellar cluster.

\begin{figure}
    \centering
   \includegraphics[width=0.45 \textwidth]{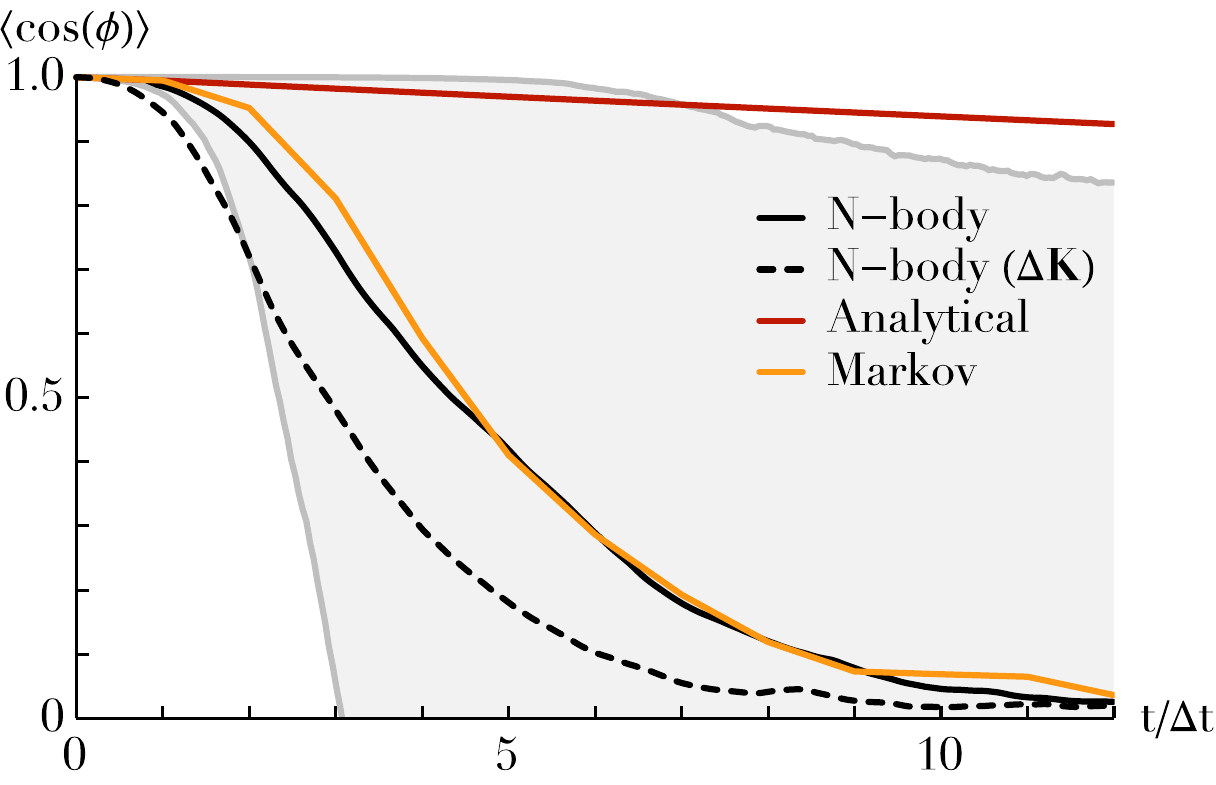}
   \caption{
   Correlation of the separation of a population of test particles
   with the same orbital parameters,
   as measured in numerical simulations
   (see Appendix~\ref{sec:Simulations} for details)
   and predicted by Eq.~\eqref{calc_C_neigh}.
   Here, the timescale ${ \Delta t }$ was chosen
   following Eq.~\eqref{choice_Deltat}.
   The `Analytical' prediction corresponds to the direct application
   of Eq.~\eqref{calc_C_neigh},
   while the `Markov' prediction corresponds
   to the improved prediction
   following the approach from Appendix~\ref{app:Virtual}.
 Because of the very slow separation of nearby particles,
 the prediction from Eq.~\eqref{calc_C_neigh}
 does not provide a good match to the late-time decorrelation
 of test particles that start with very similar initial orientations.
Finally, the curve `${ \Delta \bK }$' illustrates the correlation
measured in numerical simulations where the test particles
do not share the exact same orbital parameters,
which further accelerates their dilution.
   }
   \label{fig:CorrelationSameK}
\end{figure}

We may now test the predictions of Eq.~\eqref{calc_C_neigh}
against some tailored numerical simulations.
These simulations are similar to those
of~\cite{FouvryBarOr2019},
and we detail our exact setup in Appendix~\ref{sec:Simulations}.
Our main result is illustrated in Fig.~\ref{fig:CorrelationSameK}.
Here, we compare the measurements of particle's dilution
in numerical simulations
with the analytical predictions from Eq.~\eqref{calc_C_neigh}.
First, in that figure, we represented two numerical measurements of correlations,
depending on whether or not the test particles
also differ in their conserved orbital parameters.
As expected, the separation in parameters,
as captured by the term $C_{\alpha}^{\bO}$
from Eq.~\eqref{calc_C_neigh},
contributes to further accelerating the particles' separation.
Unfortunately, in Fig.~\ref{fig:CorrelationSameK},
we note that the `Analytical' prediction,
i.e.\ the prediction from Eq.~\eqref{calc_C_neigh},
does not manage to accurately capture the
late-time decay of the particles' separation.
However, we do note that on the coherence time,
i.e.\ for ${ t \lesssim \Delta t }$
(see the definition of ${ \Delta t }$ in Eq.~\eqref{choice_Deltat}),
the analytical prediction indeed matches the numerical measurements.
This is expected from the fact that the prediction from Eq.~\eqref{calc_C_neigh}
was obtained through a Taylor series around ${ t = 0 }$.
Yet, in its present form,
the prediction from Eq.~\eqref{calc_C_neigh} is not able to describe
the unavoidable separation
of the test particles sharing very similar orientations
and orbital parameters,
as this prediction follows a plateau
that only decays on very long timescales.
Fortunately, this apparent failure of the analytical prediction
can be alleviated by using an appropriate Markovian approach,
leading to the improved `Markov' prediction
in Fig.~\ref{fig:CorrelationSameK},
that quantitatively matches with the numerical measurements.
This is what we explore in the next section.

\section{Piecewise Markovian prediction}
\label{sec:Piecewise}

The results from the previous section,
in particular Figure~\ref{fig:CorrelationSameK},
show that the straight application of the
theoretical prediction from Eq.~\eqref{calc_C_neigh}
is not enough to compute the rate of separation
of the test particles when they are too similar,
i.e.\ when both their initial orientations
and their conserved orbital parameters
are very similar.
In these cases,
the correlations measured in the $N$-body simulations
present a plateau in ${ t \!=\! 0 }$.
This suggests that the perturbative expansion
used to obtain Eq.~\eqref{calc_C_neigh}
might not contain enough information about the dynamics
to accurately model the particles' separation.
Our goal is now to show how one can improve
the analytical prediction from Eq.~\eqref{calc_C_neigh}
so that it could also apply to cases
where the separation of the neighbours
is very slow.

In practice, in Appendix~\ref{sec:CompNeigh}
we performed a second-order Taylor expansion,
so that the analytical prediction from Eq.~\eqref{calc_C_neigh}
only matches the first two derivatives at ${ t \!=\! 0 }$
of the numerically measured correlation.
As soon as a plateau appears in Fig.~\ref{fig:CorrelationSameK},
i.e.\ as soon as the contribution from the first derivatives
is small compared to the contribution of higher-order derivatives,
this perturbative development breaks down.
This happens in particular 
when the two test particles
share very similar orientations.
Indeed, since the potential field generated by the background bath
is large-scale and continuous,
the two neighbours feel a very similar potential,
and as such need a lot of time
(compared to $\Tc$,
the coherence time of the noise)
to separate one from another.

Fortunately, regardless of the initial similarity between
the test particles,
Eq.~\eqref{calc_C_neigh} works well for short timescales,
that is when ${ t \lesssim \Tc (\bK) }$.
As a result, a reasonable way of fixing the late-time behaviour
of our prediction
is to construct a sequence of short-time predictions,
following for each of them Eq.~\eqref{calc_C_neigh}.
Let us therefore pick a timelapse ${ \Delta t }$
(whose precise value will be picked later on),
and construct a piecewise prediction of ${ \langle \cos (\phi) \rangle }$,
splitting the prediction from Eq.~\eqref{calc_C_neigh}
in timelapses of duration ${ \Delta t }$.
Doing so, we therefore construct a sequence
${ \cos (\phi_{0}) , ... ,  \cos(\phi_{n}) \!=\! \cos(\phi (t \!=\! n \Delta t)) }$ of angular separations
between the two test particles.
In order to construct the prediction for a given timelapse,
say ${ n \Delta t \to (n+1) \Delta t }$,
we follow Eq.~\eqref{calc_C_neigh},
and use ${ \cos (\phi_n) }$ as the initial angular separation between
the two test particles
and ${ t = \Delta t }$ as the time duration
during which Eq.~\eqref{calc_C_neigh} is pushed forward in time.

Such a piecewise protocol
is not equivalent to making a single prediction for
the whole time series using
only ${ \cos (\phi_{0}) }$ as the initial separation
in Eq.~\eqref{calc_C_neigh}.
Indeed, there are two main differences with the present
piecewise approach.
First, when making the analytical prediction in Fig.~\ref{fig:CorrelationSameK},
the only angular information used in the equations was
the statistics of the initial angular separation, ${ \cos (\phi_0) }$.
Here, the current value of the angular separation, ${ \cos (\phi_n) }$,
is used at the start of each timelapse of duration ${ \Delta t }$.
Second, because the current angular separation, ${ \cos (\phi_n) }$,
is now used in Eq.~\eqref{calc_C_neigh}
as an initial condition of the $n^{\mathrm{th}}$ timelapse,
the piecewise approach neglects any correlation
that might exist in the background noise
between the various timesteps.
Indeed, 
when deriving the prediction from Eq.~\eqref{calc_C_neigh},
we had to assume that the initial conditions of the test particles
are independent from the state and statistics of the background bath (as highlighted in Eq.~\eqref{rewrite_C_neigh}).
In the present piecewise case,
since we proceed by successive timesteps,
we neglect any such correlations.
This is our Markovian assumption,
a key ingredient of the piecewise prediction.

Hence the ideas behind this piecewise approach
are very similar to the explicit Euler method
used to solve differential equations.
Rather than limiting ourselves to approximating
the solution with a single perturbative expansion around its initial conditions,
we construct the solution  of Eq.~\eqref{evol_test_short} step by step.
There is however one difference with traditional step-by-step methods,
which is the fact that Eq.~\eqref{evol_test_short} is stochastic,
so that the noise driving the separation of neighbours
is time-correlated.
As a result, by proceeding by successive timelapses,
we unavoidably neglect some part of that correlation.
As such, if we were to take ${ \Delta t }$ arbitrarily small,
as is usually done in non-stochastic cases to increase their accuracy,
we would be actually replacing the \VRR\ fluctuations
by a noise uncorrelated in time.
In that limit, the reconstructed motion would be Brownian,
which drastically differs from the large scale
gravitationally-driven motion imposed by \VRR\@.

The choices for ${ \Delta t }$ are therefore limited.
One the one hand, in order to capture most of the noise correlation,
one needs ${ \Delta t \gtrsim \Tc }$,
with $\Tc$ (see Eq.~\eqref{def_Tc}) an estimate
of the noise coherence time.
On the other hand, in order for each of the timelapses to be
accurately predicted,
one needs to take ${ \Delta t }$ as small as possible.
Given these two constraints,
a natural choice is to take ${ \Delta t }$
of the order of $\Tc$.

In practice, one can note from Eq.~\eqref{res_Cb}
that the coherence time of the noise
generated by the bath, ${ \Tc (\bK) }$,
depends both on $\bK$
(i.e.\ the bath has a whole range
of decorrelation timescales),
as well as on $\ell$
(i.e.\ different harmonics
separate neighbours on different physical scales).
For two test particles having identical orbital parameters, $\bK$,
we choose to define the timestep ${ \Delta t }$ as
\begin{equation}
\Delta t = \Tc (\bK) .
\label{choice_Deltat}
\end{equation}
Here, returning to Eq.~\eqref{res_Cb},
we note that the ${ \ell = 2 }$ harmonics of the noise,
which has the longest correlation time
and the largest scale correlation length,
decays on a timescale of the order ${ \sqrt{2/A_{2}} \Tc \!=\! \Tc / \sqrt{3} }$,
which justifies our choice for ${ \Delta t }$.
Furthermore, in Eq.~\eqref{choice_Deltat},
we chose to evaluate the coherence time
for the orbital parameters $\bK$,
following the observation that
test particles mainly interact with bath particles
that have similar parameters,
as can be seen in the dependence of the coupling coefficients,
${ \mJ_{\ell} [\bK , \bKp ] }$, in Fig.~\ref{fig:ShapeJl}.
In the general case where the two test particles
do not share the same orbital parameters,
i.e.\ ${ \bK_{1} \neq \bK_{2} }$,
we opt for the most conservative choice.
We therefore take the maximum of ${ \Delta t }$
obtained for both particles.

When considering a population of more than two test particles,
we have shown in Eq.~\eqref{Fokker_X_bignMT}
that the dilution is given by the average
of the two-point correlation functions
of pairs of neighbours.
As a consequence, to apply the piecewise approach
to a population of test particles,
one only has to deal with each pair of neighbour separately,
and then average over them.
Note that, following Eq.~\eqref{choice_Deltat}
different timesteps ${ \Delta t }$
can be used for each pair of test particles.
Since the coherence time $\Tc$
can vary substantially between pairs of particles,
this allows for the use of a somewhat optimal timestep
for each pair.

\subsection{Direct evolution of the correlation}

Having decided upon a timestep ${ \Delta t }$
in Eq.~\eqref{choice_Deltat},
let us now apply the previous piecewise protocol
to find a better estimate of the separations
measured in Fig.~\ref{fig:CorrelationSameK}.
Since our goal is to construct
typical sequences ${ \cos(\phi_0) \rightarrow ... \rightarrow \cos(\phi_n) }$
of angular separations between two neighbours,
we must estimate the statistics
of the transition from one angle to the following,
i.e.\ the statistics of the transition ${ \cos(\phi_{i}) \rightarrow \cos(\phi_{i+1}) }$.
Of course,  these transitions are stochastic
so that the separations ${ \cos(\phi_i) }$
are themselves random variables
that depend on the realisations of the background noise.
Specifically we aim to compute the average properties
of these separations,
and, as in Fig.~\eqref{fig:CorrelationSameK},
predict ${ \langle \cos (\phi) \rangle }$.

We therefore need
to compute the expectation of ${ \cos (\phi_{i+1}) }$
conditionally to the value of ${ \cos(\phi_{i}) }$.
Indeed, within the Markovian approximation,
it is only the angular separation of the test particles
at the start of a given timelapse, i.e.\ ${ \cos (\phi_{i}) }$,
that matters
for its subsequent evolution during that same timestep.
As a result, we can naturally write
\begin{equation}
\langle \cos (\!\phi_{i+1}\!) \rangle \!= \!\!\int\!\! \rd ( \cos(\!\phi_i\!)) \, \rho_i( \cos(\!\phi_i\!))  \langle \cos (\!\phi_{i+1}\!) | \cos(\!\phi_i) \rangle,
\label{AG_conditionalE}
\end{equation}
In that expression, ${ \langle \cos (\phi_{i + 1}) | \cos(\phi_{i}) \rangle }$
follows from Eq.~\eqref{calc_C_neigh},
evaluated after a time ${ t = \Delta t }$
for two test particles systematically separated
by the constant angle ${ \cos (\phi_i) }$
at the start of the timelapse
(see Eq.~\eqref{D_Fixed} for the associated
coefficients ${ D_{\ell} }$).
In Eq.~\eqref{AG_conditionalE},
we also formally introduced ${ \rho_i }$ as the \PDF\ of ${ \cos(\phi_i) }$.
In practice, that \PDF\@ is not known,
so that without further approximation,
the integral from Eq.~\eqref{AG_conditionalE}
cannot be explicitly computed.

The simplest way around
is to rely on a first-order development
of ${ \langle \cos (\phi_{i + 1}) | \cos(\phi_{i}) \rangle }$
near ${ \cos (\phi_{i}) \simeq 1 }$,
i.e.\ in the limit of  small angular separations.
As detailed in Appendix~\ref{sec:piecewiseAppendix},
in that limit one obtains a linear relationship between
the initial condition ${ \cos (\phi_i) }$
and the conditional expectation
${ \langle \cos (\phi_{i+1}) | \cos (\phi_{i}) \rangle }$,
reading
\begin{equation}
\langle \cos (\phi_{i+1}) | \cos(\phi_{i}) \rangle = \xi_{0} + \xi_{1} \cos (\phi_i) .
\label{AG_linearMomentsRel}
\end{equation}
The coefficients $\xi_{k}$ follow from
the prediction of Eq.~\eqref{calc_C_neigh},
and their detailed values are given in Eq.~\eqref{def_xi}.
In particular, we emphasise that these coefficients
are independent of the test particles' current angular separation,
and depend only on their orbital parameters, $\bK_{1}$ and $\bK_{2}$,
as well as on the background's \DF\@, ${ n (\bK) }$.

Following Eq.~\eqref{AG_conditionalE},
let us now compute the average
of Eq.~\eqref{AG_linearMomentsRel}
to obtain   
\begin{equation}
\langle \cos (\phi_{i+1}) \rangle = \xi_{0} +  \xi_{1} \, \langle \cos (\phi_i) \rangle .
\label{AG_linearMomentsRel_avg}
\end{equation}
This is the key relation of the piecewise approach,
as we have been able to obtain a relation
between the successive values ${ \langle \cos (\phi_i) \rangle }$
and ${ \langle \cos (\phi_{i+1}) \rangle }$.
Equation~\eqref{AG_linearMomentsRel_avg}
is an arithmetic-geometric relation.
As a consequence, given an initial condition ${ \langle \cos (\phi_0) \rangle }$,
it uniquely defines a sequence of expectations
for all the subsequent timesteps, ${ t = i \Delta t , \, i \geq 0 }$.
We refer to Eq.~\eqref{PWApp_FirstExpansion_II}
for the explicit solution of Eq.~\eqref{AG_linearMomentsRel_avg}.
This is our piecewise prediction.

\begin{figure}
    \centering
   \includegraphics[width=0.45 \textwidth]{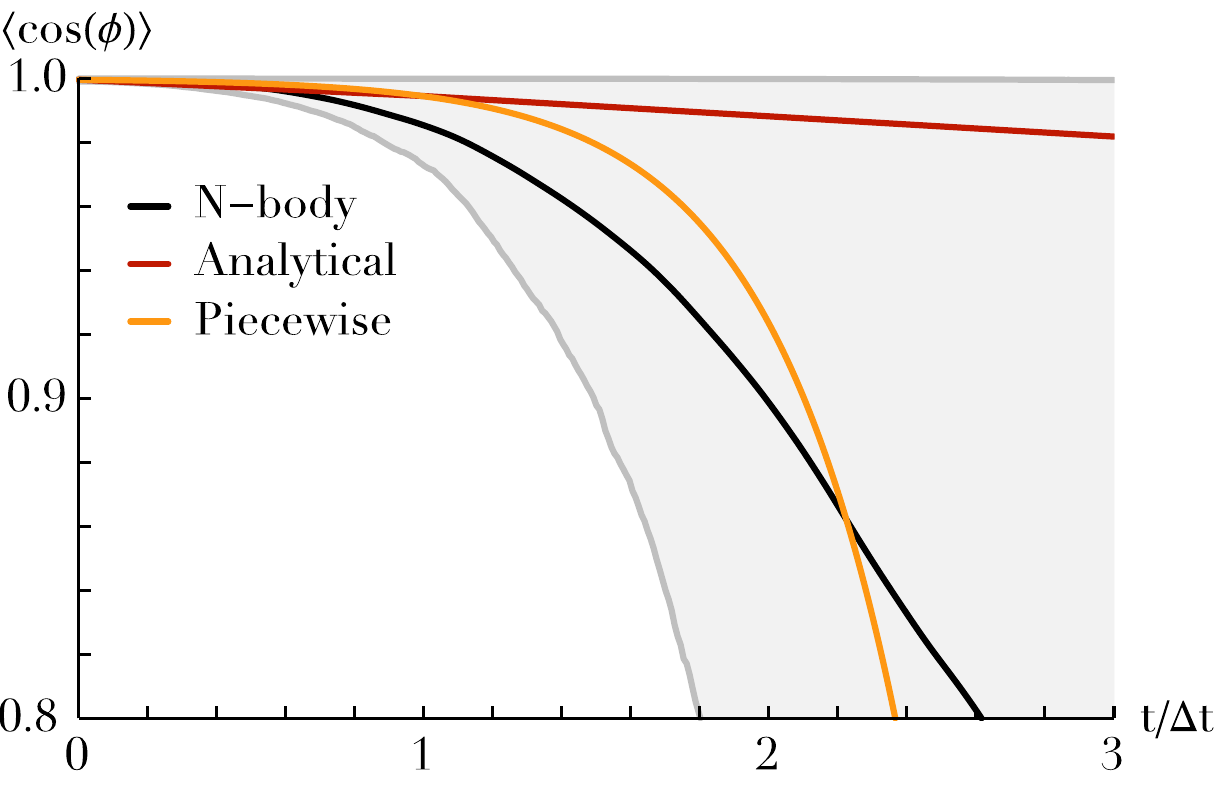}
   \caption{Initial correlation of the separation
   of a population of test particles with the same orbital parameters,
   using the exact same numerical simulations
   as in Fig.~\ref{fig:CorrelationSameK}.
   The `Analytical' prediction corresponds
   to the direct application of Eq.~\eqref{calc_C_neigh},
   while the `Piecewise' prediction corresponds
   to the use of the piecewise Markovian approach
   of Eq.~\eqref{AG_linearMomentsRel_avg}.
   This second prediction is able to describe
   the slow initial separation of the particles
   and, as a result, matches the numerical measurements
   much better.
   }
   \label{fig:PiecewiseSameK}
\end{figure}

In Fig.~\ref{fig:PiecewiseSameK},
we compare the piecewise prediction from 
Eq.~\eqref{AG_linearMomentsRel_avg}
against the numerical measurements
from Fig.~\ref{fig:CorrelationSameK}.
While the match between the numerical measurements
and the piecewise prediction is not ideal,
the piecewise prediction presented in Fig.~\ref{fig:PiecewiseSameK},
can describe the first stages of (slow)
separation of the test particles,
while the straightforward application
of the analytical prediction from Eq.~\eqref{calc_C_neigh}
failed at it.

In practice, the perturbative expansion
performed in Eq.~\eqref{AG_linearMomentsRel}
is only valid for small angular separations.
As a consequence, a piecewise sequence
generated with the protocol from Eq.~\eqref{AG_linearMomentsRel_avg}
will only match the numerical measurements
as long as the test particles remain sufficiently close to one another.
This is, however, not a problem
since these first steps of (very) slow initial separation
are the hardest ones to predict,
as already illustrated in Fig.~\ref{fig:CorrelationSameK}.
Indeed, once ${ \langle \cos (\phi_i) \rangle }$ has slightly decreased,
i.e.\ once the test particles have been slightly stirred away
from one another through the \VRR\ fluctuations,
a straightforward use of Eq.~\eqref{calc_C_neigh}
would be enough to predict the rest of the time series.
This would allow us to extend our prediction for the separation
to arbitrarily large times.
In practice, we find that the piecewise prediction behaves properly
for ${ \langle \cos (\phi) \rangle \gtrsim 0.8 }$,
i.e.\  ${ \phi \lesssim 37^{\circ} }$.
This is enough for most astrophysical applications,
e.g.\@, the possible dilution of SgrA*'s clockwise stellar disc
whose typical angular separation is ${16^{\circ}}$~\citep{Gillessen2017}.

\begin{figure}
    \centering
   \includegraphics[width=0.45 \textwidth]{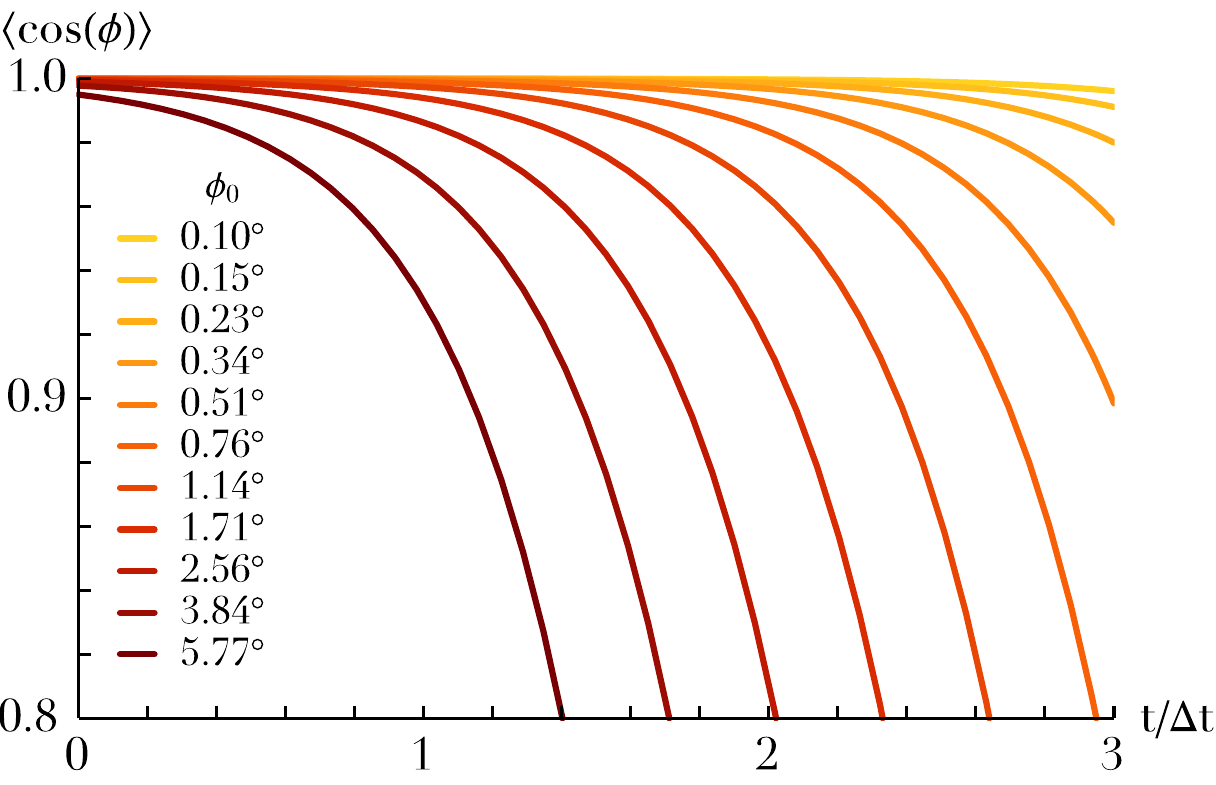}
   \\
   \includegraphics[width=0.45\textwidth]{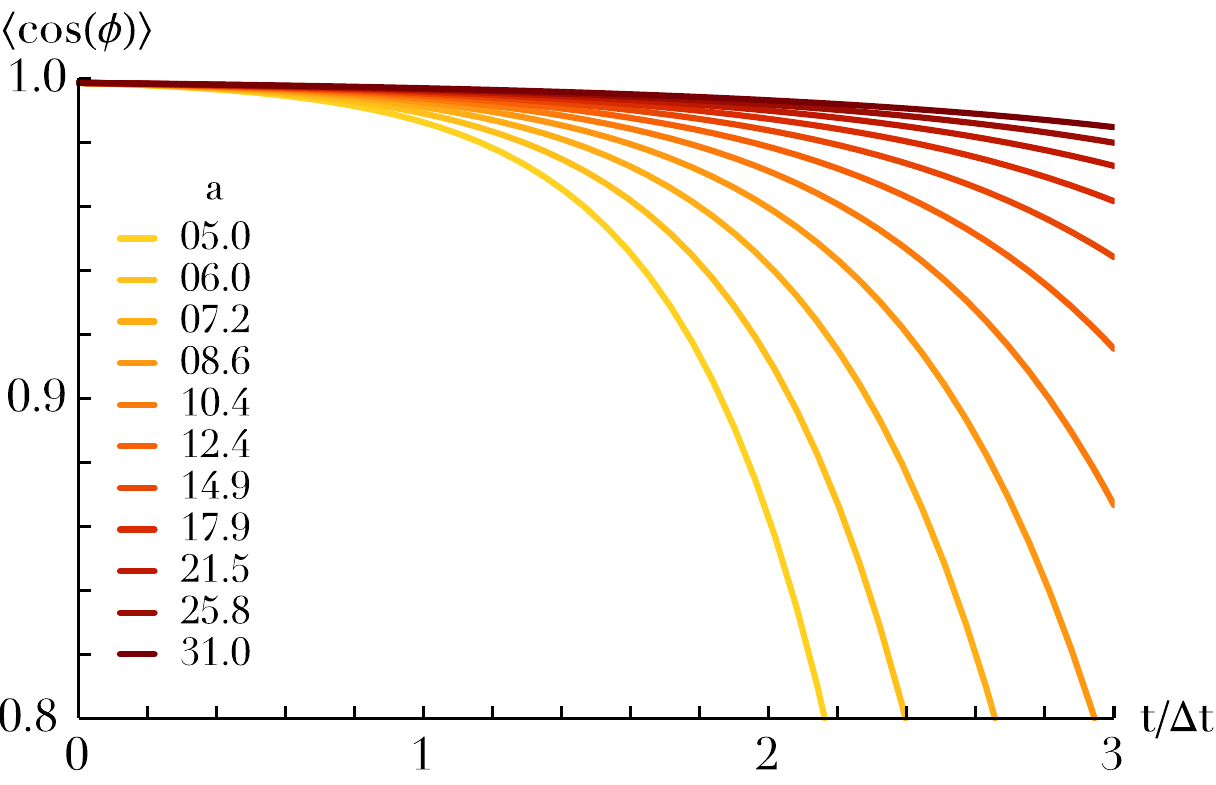}
   \caption{Illustration of the rate of neighbour separation
   as predicted by the piecewise approach from Eq.~\eqref{AG_linearMomentsRel_avg}.
   Here, the background bath follows
   the same statistics as in Fig.~\ref{fig:PiecewiseSameK},
   while the test particles share the same orbital parameters,
   ${ (a,e) \simeq (10,0.21) }$.
  \textsl{Top panel}: Correlation of the separation
   as one varies the initial angle $\phi_0$ between the particles.
   The closer the two particles, the slower the dilution.
   \textsl{Bottom panel}: Same correlations for test particles
   initially separated by ${ \phi_0 = 3^{\circ} }$,
   as one varies their shared semi-major axis, $a$.
      In that panel, the rescaled time was picked for the smallest 
   ${ \Delta t (a \!=\! 5) }$.
   The larger $a$, the slower the dilution.
   }
   \label{fig:PiecewiseVaryingPhiandSMA}
\end{figure}

Following Eq.~\eqref{AG_linearMomentsRel_avg},
determining the timescale
associated with the dilution of an initial patch
of test particles
then only amounts to computing once the two coefficients, $\xi_{k}$,
and determining at which time ${ \langle \cos (\phi) \rangle }$
gets below a given threshold, e.g.\@, 0.95.
Since this is such a simple calculation,
it can then be used to very efficiently explore
the parameters of the test particle's distribution.
This is illustrated in Fig.~\ref{fig:PiecewiseVaryingPhiandSMA},
where we show how the efficiency of the dilution
varies as one changes the initial angular separation
of the test particles or their conserved orbital parameters.
As expected, 
the smaller the initial patch, the slower the separation.
Similarly, the larger the semi-major axes of the test particles,
the larger their $\Tc$ (see Eq.~\eqref{def_Tc}),
and therefore the slower their dilution.
In Section~\ref{sec:Applications},
we will further illustrate the versatility of this approach
by presenting some first applications
of Eq.~\eqref{AG_linearMomentsRel_avg}
to estimate the efficiency of the dilution
of the clockwise disc surrounding SgrA*.

As highlighted in Eq.~\eqref{AG_linearMomentsRel_avg},
the piecewise prediction can only predict
the expectation of ${ \langle \cos (\phi) \rangle }$
throughout the dilution of the test particles.
It cannot predict higher order moments of ${ \langle \cos (\phi) \rangle }$,
nor can it be used to produce effective random
walks of the stochastic variable ${ \cos (\phi) }$.
Relying on the use of a restricted ${ \ell = 2 }$ toy model
(see Appendix~\ref{sec:ToyModel}),
this is what we explore in Appendix~\ref{app:Virtual}
to generate virtual dilutions.
It is this particular approach that was used
in Fig.~\ref{fig:CorrelationSameK}
to produce a Markovian prediction that matches
the numerical measurements even at late times.

\section{Application:  cusp properties from dilution}
\label{sec:Applications}

Let us now consider a background stellar cusp distribution
similar to SgrA*'s. We will now show how our 
 results allow us to 
probe the underlying kinematic properties of the 
unresolved nucleus when requesting consistency
with level of neighbours dilutions.
We will not aim to be very realistic nor
match any specific data, as this will be the topic of an upcoming 
investigation.   

Following~\cite{Gillessen2017},
we take the mass of the central \BH\ to be
${ \MBH = 4.3 \!\times\! 10^{6} \Msun }$.
For simplicity, we first assume that the background
old stellar cluster is made of a single-mass stellar population
of individual mass ${ \mstar = 1 \Msun }$.
We assume that the stars' eccentricities follow a thermal
distribution, ${ f_{e} (e) = 2 e }$~\citep{Merritt2013},
and that the number of stars per unit $a$ follows a
power-law distribution of the form ${ n_{a} (a) \propto a^{2 - \gamma} }$.
The detailed normalisations for that setup are all summarised
in Appendix~\ref{sec:PowerLaw}.
In such a configuration,~\cite{FouvryBarOr2019}
(see Eq.~{(48)} therein)
have shown that the coherence time
of the fluctuations, ${ \Tc (\bK) }$
(see Eq.~\eqref{def_Tc}),
follows the simple dependence
\begin{equation}
\Tc (a , e) \simeq 1.4 \!\times\!  \frac{P (a)}{\sqrt{N( \!<\! a)}} \, \frac{\MBH}{\sqrt{ \langle m^{2} \rangle}} \, \sqrt{1 \!-\! e^{2}} ,
\label{expression_Tc_cusp}
\end{equation}
where ${ P (a) \!=\! 2 \pi (a^{3} / (G \MBH))^{1/2} }$
is the (fast) Keplerian period,
${ N ( \!<\! a) \!\propto\! a^{3 - \gamma} }$
is the number of stars physically within a sphere
of radius $a$ from the centre,
and ${ \sqrt{\langle m^{2} \rangle} }$
captures the mass spectrum of the background cluster.
In particular,
the larger the spread in mass,
the larger the eccentricity,
the lighter the central black hole,
the shorter the coherence time.
Equation~\eqref{expression_Tc_cusp}
will prove useful to interpret
some of the trends of the upcoming figures.

We now consider a population of test particles
 mimicking the stars belonging to the clockwise disc~\citep{Gillessen2017}.
To simplify, we assume that all the test stars share the same orbital parameters,
so that we take, as an example, ${ a_{\rt} = 50 \, \mathrm{mpc} }$
and ${ e_{\rt} = 0.1 }$.
As such, we are not accounting
for any separation stemming from differences
in the test particles' orbital parameters,
see Eq.~\eqref{def_PsiMinus},
which would further reduce
the timescales predicted here.
We assume that the test stars are born
with an average angular separation given by ${ \langle \cos (\phi_{0}) \rangle }$,
and that their age are somewhat similar to those  of the inner S-stars~\citep{Habibi2017},
so that ${ t_{\star} \simeq 10 \, \mathrm{Myr} }$.
Having specified the parameters of the disc's stars,
 the prediction from Eq.~\eqref{calc_C_neigh}
(in particular its piecewise version from
Eq.~\eqref{AG_linearMomentsRel_avg})
can be used
to estimate the typical time required for such a stellar disc to dilute.
Following~\cite{Gillessen2017},
the current angular dispersion
of the clockwise disc is approximately given by
${ \langle \phidisc \rangle \simeq 16^{\circ} }$,
This corresponds to an average separation
${ \langle \cos (\phidisc) \rangle \simeq 0.96 }$,
which is well within the regime
of applicability of the piecewise prediction,
see Fig.~\ref{fig:PiecewiseSameK}.
For a given model of the background cluster
and a given initial condition,
we then define the diffusion time, $\Tdiff$,
as the time required for ${ \langle \cos (\phi) \rangle }$
to reach ${ \langle \cos (\phidisc) \rangle }$.
Once the average angular separation between the test stars
has reached such a large value,
one may consider that the stellar disc has been effectively
dissolved by the \VRR\ fluctuations.

\begin{figure}
    \centering
   \includegraphics[width=0.45 \textwidth]{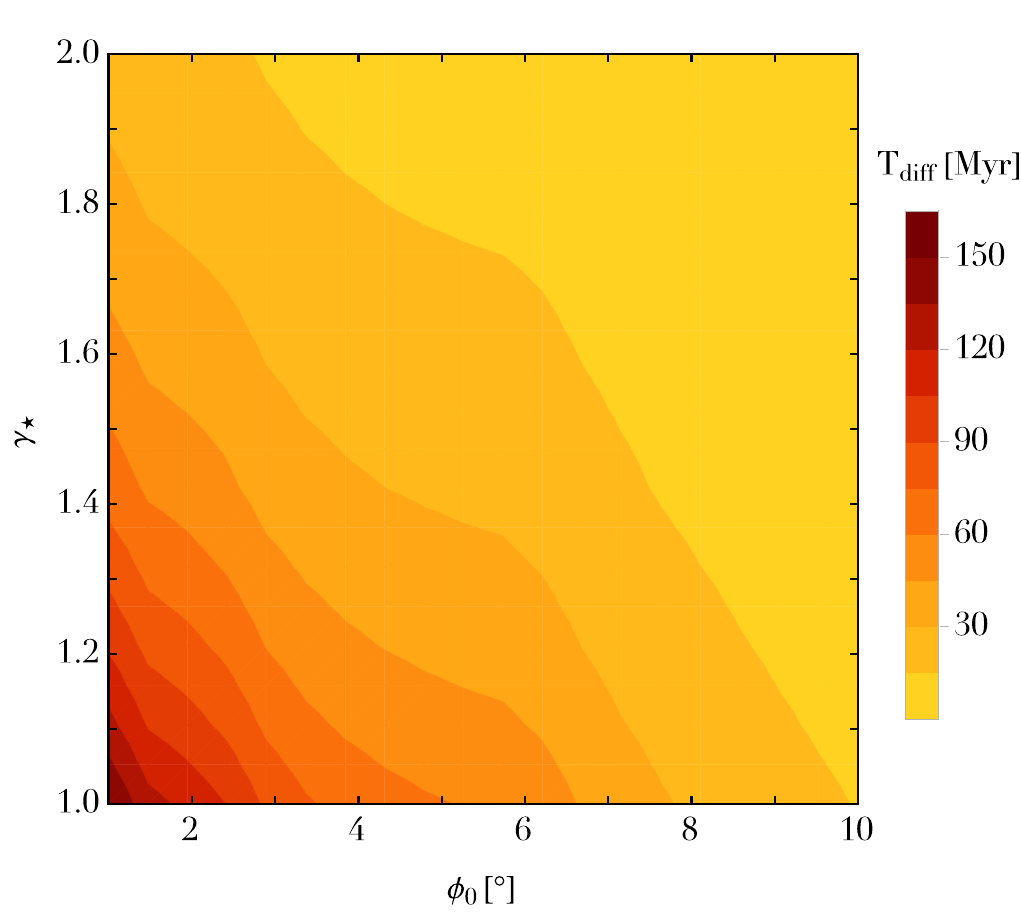}
   \caption{Illustration of the dependence
   of the dilution time, $\Tdiff$, of the stellar disc,
   as a function of the disc's initial angular separation, $\phi_{0}$,
   and the cusp index of the background stellar cluster.
   Light colours correspond to fast dilution times.
   In particular, we recover that the larger the initial
   angular dispersion and the larger the cusp index,
   the faster the dilution.
   }
   \label{fig:AstroGammaAndPhi0}
\end{figure}

Figure~\ref{fig:AstroGammaAndPhi0} illustrates the variations
of the dilution time
as one varies the power index
of the background stellar cusp, ${ \gamma_{\star} }$,
as well as the initial angular dispersion,
${ \langle \cos (\phi_{0}) \rangle }$.
As already highlighted in Fig.~\ref{fig:PiecewiseVaryingPhiandSMA},
the smaller the initial patch of stars,
the longer it takes for the initially coherent stellar disc
to dissolve.
Figure~\ref{fig:AstroGammaAndPhi0} also shows that the cuspier
the density profile,
the faster the dilution.
This dependence is a direct consequence
of the factor ${ 1/\sqrt{N( \!<\! a)} }$
in the expression of ${ \Tc (\bK) }$
from Eq.~\eqref{expression_Tc_cusp}:
the larger $\gamma_{\star}$,
the larger ${ N ( \!<\! a) }$,
therefore the smaller
the coherence time, ${ \Tc (\bK) }$,
and hence the faster the dilution of the disc.
While the numerical values used here
are in some sense ad hoc,
and would definitely require more careful selections,
Fig.~\ref{fig:AstroGammaAndPhi0} shows
how the present
formalism could be used
to place constraints on the parameters of the unresolved
background stellar cluster (here through its index $\gamma_{\star}$)
as well as on the formation channels of stars in galactic nuclei
(here through the size of their initial angular patch $\phi_{0}$).

Let us finally assume that the background
cluster is composed
not only of old stars,
but also of \IMBHs\@.
For simplicity, let us assume that the total enclosed mass
remains the same,
and that the \IMBHs\ follow the same ${ (a,e) }$ distribution
than the stars,
with an individual mass given by ${ \mbh = 100 \, \Msun }$.
The details of our normalisation are given
in Appendix~\ref{sec:PowerLaw}.
In the top panel of Fig.~\ref{fig:AstroIMBH},
we illustrate the dependence
of the dilution time
as one varies the mass fraction of the \IMBHs\@,
as well as their shared power index, $\gamma$.
\begin{figure}
    \centering
   \includegraphics[width=0.45 \textwidth]{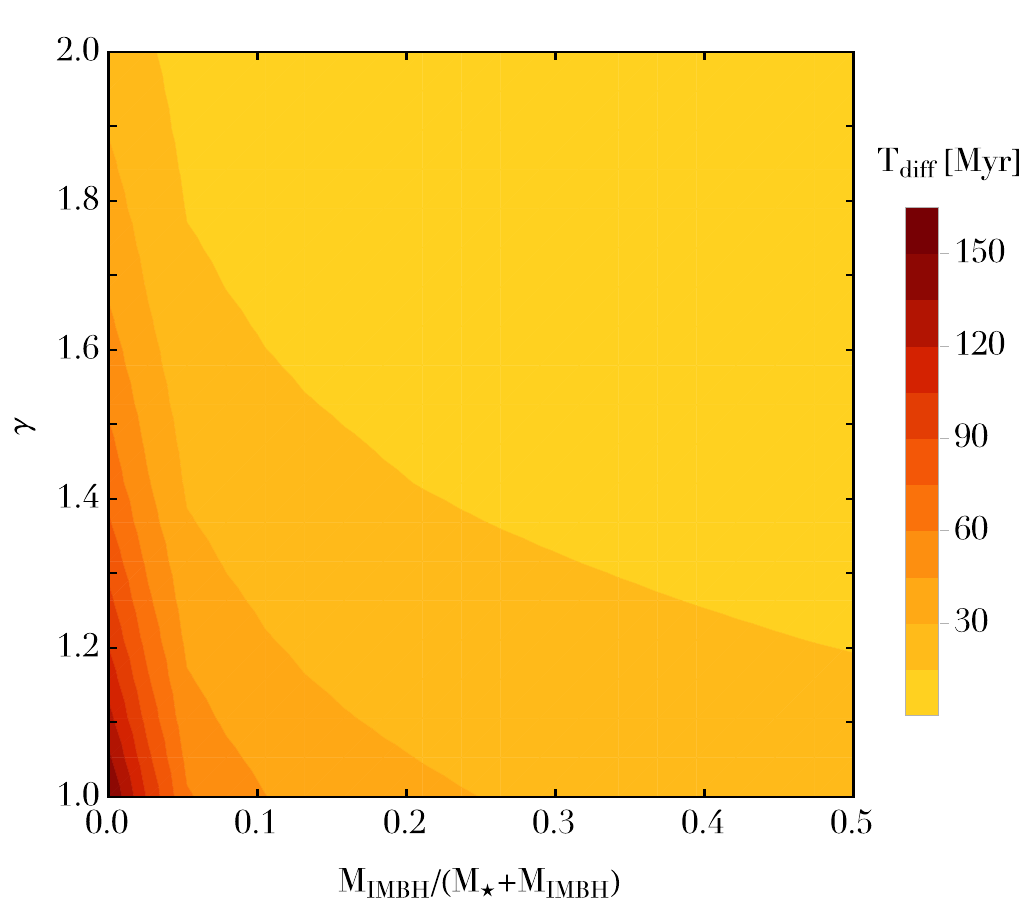}
   \\
   \includegraphics[width=0.45\textwidth]{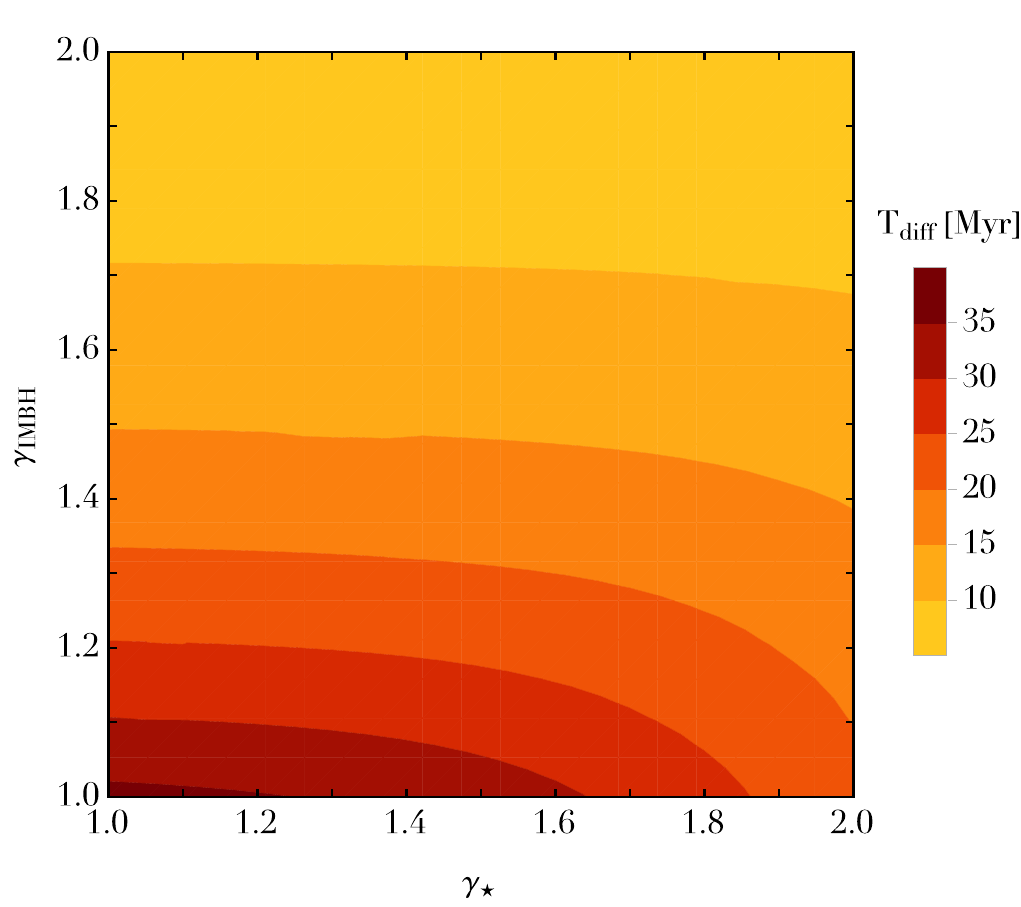}
   \caption{Illustration of the dependence
   of the dilution time, $\Tdiff$, of the stellar disc,
   when the background cluster also
   contains some \IMBHs\@,
   assuming an initial angular separation
   given by ${ \phi_{0} = 3^{\circ} }$.
   \textsl{Top panel}: As a function
   of the \IMBHs\@' mass fraction
   and their shared cusp index.
   \textsl{Bottom panel}: As a function
   of the old stars' and \IMBHs\@' cusp indices,
   assuming an \IMBH\ mass fraction of ${ M_{\mathrm{IMBH}}/(M_{\mathrm{IMBH}} \!+\! M_{\star}) \!=\! 20\% }$.
   As expected, the higher the fraction
   of \IMBHs\@,
   and the higher the cusps' indices,
   the faster the dilution of the stellar disc.   
   }
   \label{fig:AstroIMBH}
\end{figure}
As expected, one recovers that a larger fraction
of \IMBHs\ leads to a faster dilution of the disc stars.
This is a direct consequence
of the factor ${ 1/\sqrt{\langle m^{2} \rangle} }$
present in Eq.~\eqref{expression_Tc_cusp}:
the larger the fraction of \IMBHs\@,
the larger ${ \langle m^{2} \rangle }$,
and therefore the smaller ${ \Tc (\bK) }$.
In practice, this acceleration of the dilution
through the mass spectrum is slightly
dampened by the factor ${ 1/\sqrt{N( \!<\! a)} }$
from Eq.~\eqref{expression_Tc_cusp}
which increases when the mass fraction of \IMBHs\
increases as the total number of
background particles decreases overall.

In practice, within the old unresolved stellar cluster,
\NR\ has had the time to lead to mass segregation,
so that one expects
that the old stars and the \IMBHs\
not to share the same power index,
the heavier particles being
on more cuspy distributions~\citep{BahcallWolf1977}.
This is briefly explored in the bottom panel of Fig.~\ref{fig:AstroIMBH},
where for a fixed mass fraction in \IMBHs\@,
we explore the dependence of the dilution time
as one varies independently the power indices
of both populations.
In that figure, we recover that cuspier profiles
lead to more efficient separations,
the effect being  most visible for the \IMBH\@'s population.

All in all, even though the typical timescale of \VRR\
is given by $\Tc$,
Figs.~\ref{fig:AstroGammaAndPhi0} and~\ref{fig:AstroIMBH}
show that the timescale for particle separation
can be significantly longer than $\Tc$.
For the range of parameters explored here,
we approximately find
\begin{equation}
\frac{\Tdiff}{\Tc} \simeq 1 \, \hbox{---} \, 15 ,
\end{equation}
depending mainly on the disc's initial
angular dispersion.

Of course, the explorations from
Figs.~\ref{fig:AstroGammaAndPhi0} and~\ref{fig:AstroIMBH}
are only preliminary applications
in the  context of SgrA*.
In particular, here we only focused
on the dynamical constraint associated
with the observation of the clockwise disc
on the scale ${ a \simeq 50 \, \mathrm{mpc} }$.
But a dual constraint also comes from the fact
that in more internal regions, ${ a \simeq 5 \mathrm{mpc} }$,
the observed S-cluster has a seemingly isotropic
distribution of orientations.
As such, to be physically admissible
any model for SgrA*'s cluster
must ensure both a fast enough dilution time of stellar discs
in its center,
and a slow enough dilution time further out.
These joint constraints will eventually allow us to bracket 
the physical parameters of the cluster.

\section{Conclusions}
\label{sec:Conclusion}

This paper illustrated how to quantitatively describe
the two-point statistical properties of neighbour separation
induced by vector resonant relaxation.
Section~\ref{sec:VRR}
assumed the limit of an isotropic distribution of background stars,
whose potential fluctuations slowly stir neighbour particles away from one another.
A key result was obtained in Eq.~\eqref{calc_C_neigh},
which highlighted the two main effects sourcing the separation
of nearby particles, namely their difference in orientations
and their differences in conserved orbital parameters.
Relying on a Markovian assumption,
this estimator was subsequently improved 
in Section~\ref{sec:Piecewise}
and Appendix~\ref{app:Virtual}
through the construction of both a piecewise prediction
and generations of virtual realisations.
This approach allowed us in particular to describe
the (very) slow separation of highly correlated neighbour stars,
e.g.\@, test particles starting with very close orientations.
Throughout the text, all our predictions were compared
with tailored numerical simulations,
which led to a quantitative agreement
on the rate of neighbour separation.
Finally,  Section~\ref{sec:Applications}
 presented a first application of this formalism
to determine the efficiency with which young stellar discs,
such as the one observed within SgrA*,
can spontaneously dissolve under the effects
of the stochastic \VRR\ dynamics.
In particular, we showed
how the initial distribution of angular separation of the disc,
the profile of the background unresolved  cluster,
as well as the possible presence of \IMBHs\
all influence the efficiency with which discs
spontaneously dissolve in galactic nuclei.

This paper only addresses some aspect of what a 
complete theory of \VRR\@ should achieve.
Here is a list possible avenues for future development.

First, the toy model for SgrA* presented in Section~\ref{sec:Applications}
is only a preliminary validation and illustration
of the computation of the dissolution time
of SgrA*'s clockwise disc.
Indeed, two observational constraints must be
satisfied by any model
of the underlying background stellar cluster:
(i) the most inner S-stars seem to have
an isotropic distribution of orientations,
while (ii) some of the outer stars seem
to belong to a disc-like structure~\citep{Gillessen2017}.
As a consequence, the separation of neighbours sourced by \VRR\
must be on the one hand efficient enough to mix the inner stars, 
and on the other hand inefficient enough to allow for the outer disc
to survive up to the present time.
Using jointly these observational requirements,
one should therefore
be in a position to place constraints
on the properties of the underlying stellar cluster
(e.g.\@, possible existence of \IMBHs\@)
and of the formation mechanism of the S-stars
(e.g.\@, through an episode of star formation in a disc).
Such detailed explorations of parameter space
are made possible by the present formalism
because it allows very easily for variations
in the bath's \DF\ (via ${ n(\bK) }$),
the test particles' \DF\ (via ${ p (\bK) }$),
as well as their initial separation
(via ${ \langle \phi_{0} \rangle }$).
This will be the subject of a future work.

The present framework
relied extensively on the isotropic assumption,
both for the background potential fluctuations,
but also for the test particles' initial conditions.
As a result,  any effects associated
with possible anisotropic clusterings in orientation~\citep{SzolgyenKocsis2018} was neglected.
Indeed, such non-spherical structures
stemming from the \VRR\ long-term
thermodynamical equilibria
can undoubtedly affect
the statistical properties of the fluctuations in the system
(in particular for the most massive background particles),
and as a result may also change
the efficiency of neighbour separation.

Here, the  limit of test particles was assumed,
so that  any self-gravity among neighbouring particles was neglected.
This assumption is only (reasonably) valid
in the limit where these pairwise interactions are indeed
negligible in front of the bulk of interactions sourced
by the background stellar cluster.
In practice,
owing to the sharpness of the coupling coefficients, $\mJ_{\ell}$,
see Fig.~\ref{fig:ShapeJl},
a given star mainly interacts with stars that share
similar parameters (in particular semi-major axes),
and similar orientations.
All in all, this tends to enhance the gravitational coupling
between two neighbour particles,
making our present assumption of test particles less valid.
For example, the importance of self-gravity among the particles from the same
disc was already highlighted in Figs.~{6} and~{7} of~\cite{KocsisTremaine2011},
where one notes that self-gravity increases the coherence
of the disc, and reduces the efficiency with which it can dissolve.
Accounting for this self-gravitating component is no easy task,
and definitely deserves further theoretical investigations.

The present analysis focused on an isotropic description of the dilution process.
Strikingly, simulations  such as those shown in Fig.~\ref{fig:Dilution} suggest that this process is significantly non-isotropic, 
with a clear elongation of the tracer distribution. It would therefore be interesting to quantify,
e.g.\@, through the three-point correlation function,
the rate at which these elongation arise.
This could then be used as a
supplementary observational constraint
to leverage all the information
coming from future observations of SgrA*~\citep{Do2019}.
Similarly, it could also prove useful to further expand
the analytical calculations from Appendix~\ref{sec:CompNeigh}
to compute the theoretical predictions up to fourth-order
in the fluctuations.
This could be of importance in particular to better describe very slow separations,
to improve the quality of the piecewise predictions,
or to be able to generate virtual dilutions
also in the case of test particles
with different orbital parameters.

Finally, this paper focused  on systems dominated
by a central mass.
But, as long as one updates accordingly the coupling coefficients,
${ \mJ_{\ell} }$, the same \VRR\ process also happens
in globular clusters~\citep{MeironKocsis2018}.
In that context, one could  investigate the efficiency
of the mixing (in orientations) of neighbouring stars.
In the light of the exquisite GAIA data,
this dynamical process could prove important to understand how co-eval
stars can mix in the crowded stellar environments of globular clusters.

\section*{Acknowledgements}

We thank B.\ Bar-Or, F.\ Vincent, G.\ Perrin and K.\ Tep
for insightful conversations.
This work is partially supported by the grant
Segal ANR-19-CE31-0017
of the French Agence Nationale de la Recherche.
We thank St\'ephane Rouberol for the smooth running of the
Horizon Cluster, where the simulations were performed.

\appendix

\section{Coupling coefficients}
\label{sec:Coupling}

Let us detail the expression of the coupling coefficients,
${ \mJ_{\ell} \big[ \bK_{i} , \bK_{j} \big] }$,
that characterise the strength of the pairwise \VRR\ interaction
in the Hamiltonian from Eq.~\eqref{rewrite_H}.
Following Eq.~{(51)} of~\cite{FouvryBarOr2019},
they stem from the multipole expansion
of the Newtonian interaction and read
\begin{equation}
\mJ_{\ell} \big[ \bK_{i} , \bK_{j} \big] \!=\! \frac{G m_{i} m_{j}}{L (\bK_{i})} \frac{4 \pi P_{\ell}^{2} (0)}{2 \ell + 1}
\!\! \int_{0}^{\pi} \!\!\! \frac{\rd M_{i}}{\pi} \!\! \int_{0}^{\pi} \!\!\! \frac{\rd M_{j}}{\pi} \, \frac{\Min [ r_{i} , r_{j} ]^{\ell}}{\Max [ r_{i} , r_{j} ]^{\ell + 1}} ,
\label{def_Jl}
\end{equation}
with ${ L (\bK) = m \sqrt{G \MBH a (1 \!-\! e^{2})} }$
the norm of the angular momentum.
Note the presence of the Legendre polynomials, ${ P_{\ell} (0) }$,
which ensures that these coefficients are non-zero only for even values of $\ell$.
Finally, in Eq.~\eqref{def_Jl} we introduced $M$ as the mean anomaly
associated with a given Keplerian orbit.
This equation already shows that more radial orbits (${ e \simeq 1 }$) will have larger 
$\mJ_{\ell}$, since they have smaller angular momentum $L$.
As a result, they are more easily reoriented.

In order to highlight the scale-invariance
of these coupling coefficients,
let us rewrite the orbit-average integral
as an integral over eccentric anomalies,
that we denote with $\phi$.
To do so, we also introduce `$\mathrm{out}$'
(resp.\ `$\mathrm{in}$') as the index $i$ or $j$
with the larger (resp.\ smaller) semi-major axis,
and introduce the ratio ${ \alpha = \ain / \aout \leq 1 }$.
Given the relation ${ \rd M / \rd \phi = r / a }$,
and the explicit expression of the radius
${ r \!=\! a (1 \!-\! e \cos (\phi)) }$,
Eq.~\eqref{def_Jl} finally  reads
\begin{equation}
\mJ_{\ell} \big[ \bK_{i} , \bK_{j} \big] = \frac{G m_{i} m_{j}}{\aout} \frac{1}{L (\bK_{i})} \, s_{\ell} \big[ \alpha , \ein , \eout \big] ,
\label{rewrite_Jl_with_sl}
\end{equation}
where the dimensionless coefficients ${ s_{\ell} \big[ \alpha , \ein , \eout \big] }$ read
\begin{align}
s_{\ell} \big[ \alpha , \ein , & \, \eout \big] = \frac{4 \pi P_{\ell}^{2} (0)}{2 \ell + 1} \, \frac{1}{\alpha} \!\! \int_{0}^{\pi} \!\! \frac{\rd \phi}{\pi} \!\! \int_{0}^{\pi} \!\! \frac{\rd \phip}{\pi}
\nonumber
\\
\times & \, \frac{\Min \big[ \alpha (1 \!-\! \ein \cos (\phi)) , (1 \!-\! \eout \cos (\phip)) \big]^{\ell + 1}}{\Max \big[ \alpha (1 \!-\! \ein \cos (\phi)) , (1 \!-\! \eout \cos (\phip)) \big]^{\ell}} .
\label{def_sl}
\end{align}
Such a dimensionless writing is an appropriate form to deal
with infinite power-law distributions
as in Section~\ref{sec:Applications}.

\begin{figure}
    \centering
   \includegraphics[trim = 0.cm 0cm 0cm 1.3cm , clip , width=0.45 \textwidth]{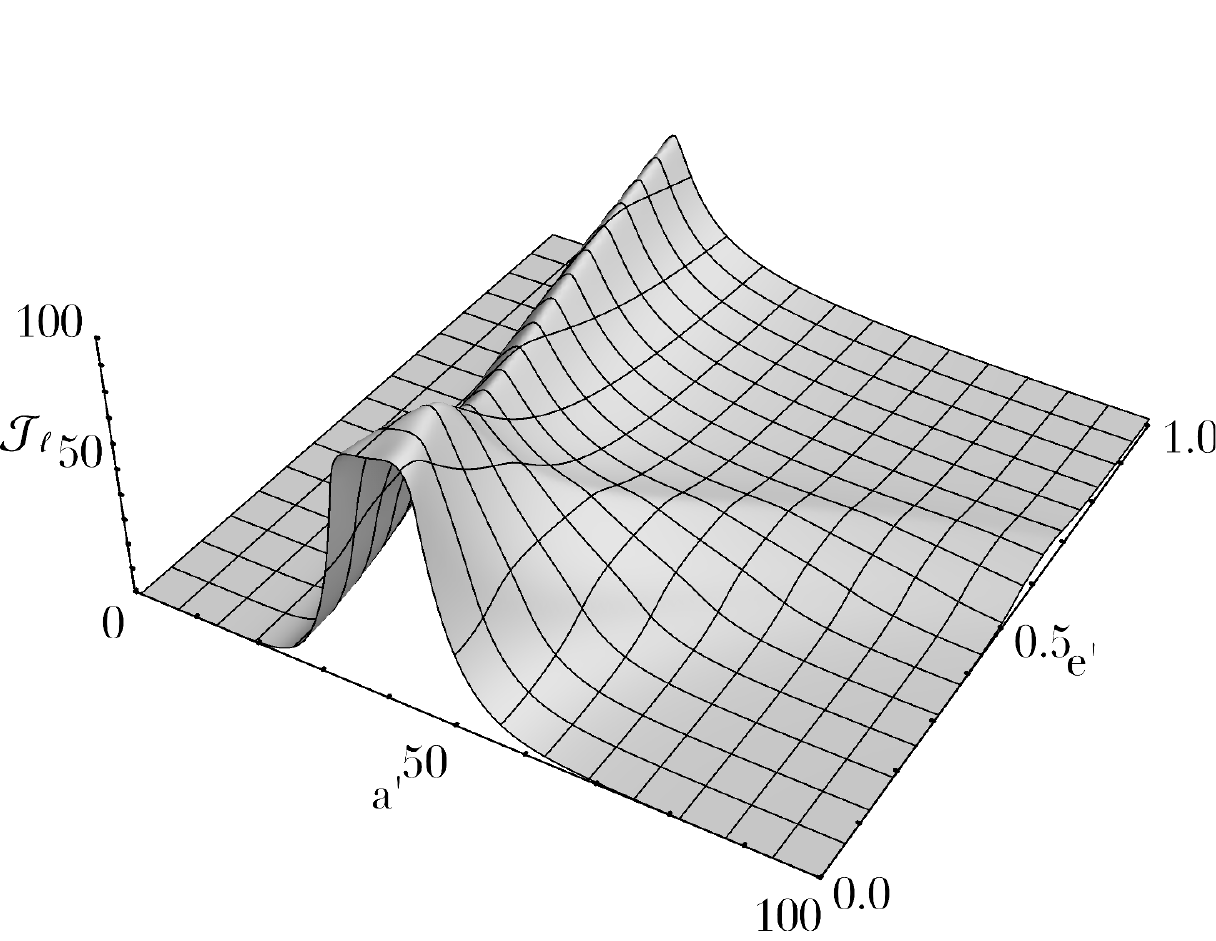}
   \includegraphics[trim = 0.cm 0cm 0cm 1.3cm , clip , width=0.45 \textwidth]{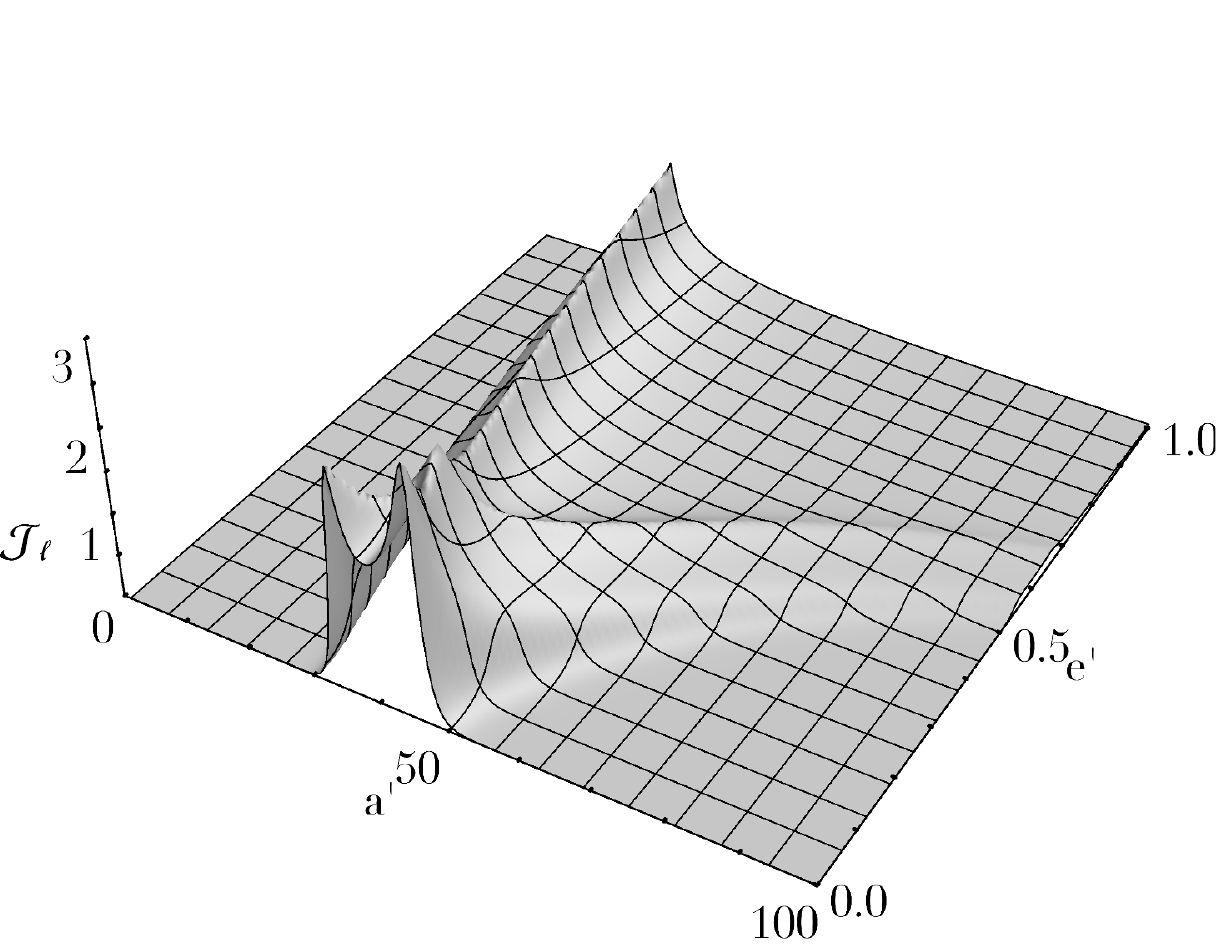}
   \caption{
   Illustration of the coupling coefficients ${ \mJ_{\ell} \big[ \bK , \bKp \big] }$
   for ${ \ell = 10 }$ (top) and ${ \ell = 40 }$ (bottom).
   Here, we chose ${ (a , e) = (40, 0.15) }$.
   The vertical axes have been arbitrarily rescaled.
   As highlighted in the text, a given orbit mostly couples with 
   orbits with similar semi major axes.
   }
   \label{fig:ShapeJl}
\end{figure}

Figure~\ref{fig:ShapeJl} illustrates the typical shape
of the coupling coefficients.
In that figure, we note that a given particle is strongly coupled
to particles that share similar conserved parameters,
in particular similar semi-major axes.
Moreover, we note that as $\ell$ increases,
the amplitude of the coupling coefficient, $\mJ_{\ell}$,
drastically reduces.

\section{The Elsasser coefficients}
\label{sec:Elsasser}

Let us detail some contraction rules satisfied by the Elsasser coefficients used throughout the paper.
Complementary definitions and expressions can be found in Appendix~{B} of~\cite{FouvryBarOr2019}. 
The Elsasser coefficients can be decomposed as
\begin{equation}
E_{\alpha \gamma \delta} = E^{L}_{\ell_{\alpha} \ell_{\gamma} \ell_{\delta}} \, E^{M}_{\alpha \gamma \delta} ,
\label{decomposition_Elsasser}
\end{equation}
where $E^{L}_{\ell_{\alpha} \ell_{\gamma} \ell_{\delta}}$ only depends
on ${ (\ell_{\alpha} , \ell_{\gamma} , \ell_{\delta} ) }$,
while $E^{M}_{\alpha \gamma \delta}$ also depends
on ${ (m_{\alpha} , m_{\gamma} , m_{\delta}) }$.
These coefficients satisfy various exclusion rules.
In particular, to be non-zero, one has to satisfy
\begin{align}
\text{(C1)}: & \,\, |m_{\alpha}| \leq \ell_{\alpha}; \;\; |m_{\gamma}| \leq \ell_{\gamma}; \;\; |m_{\delta}| \leq \ell_{\delta} ,
\nonumber
\\
\text{(C2)}: & \,\, \ell_{\alpha} + \ell_{\gamma} + \ell_{\delta} \; \text{ is odd},
\nonumber
\\
\text{(C3)}: & \,\, |\ell_{\alpha} - \ell_{\gamma}| < \ell_{\delta} < \ell_{\alpha} + \ell_{\gamma} \; \text{ (strict triangular inequality)} ,
\nonumber
\\
\text{(C4)}: & \,\, \text{all pairs } (\ell_{\alpha} , m_{\alpha}), (\ell_{\gamma} , m_{\gamma}), (\ell_{\delta} , m_{\delta}) \; \text{ are different}.
\label{exclusion_rules}
\end{align}
In addition, the Elsasser coefficients also follow the symmetry
relation ${ E_{\alpha \gamma \delta} \!=\! E_{\delta \alpha \gamma} \!=\! - E_{\alpha \delta \gamma} }$.
These coefficients also satisfy various contraction rules.
As already used in~\cite{FouvryBarOr2019},
one has
\begin{align}
& \, \sum_{m_\gamma,m_\delta} E^M_{\alpha\gamma\delta} E^M_{\beta \gamma \delta} = \delta_{\alpha \beta} \frac{1}{2\ell_\alpha +1} ,
\nonumber
\\
& \, \sum_{\ell_\delta} \frac{1}{2\ell_\alpha+1} \big( E^L_{\ell_\alpha \ell_\gamma \ell_\delta} \big)^2 = A_{\ell_\alpha} B_{\ell_\gamma} ,
\label{Elm_FouvryContractions}
\end{align}
where the first relation only holds when ${ (\ell_{\alpha} , \ell_{\gamma} , \ell_{\delta}) }$
satisfy the exclusion rules from Eq.~\eqref{exclusion_rules}.
In the previous formula,
we also introduced  ${ A_\ell = \ell(\ell+1) }$
as well as ${ B_\ell = \ell(\ell+1)(2\ell+1)/ (8\pi) }$.
To these two rules, we finally add the following one
\begin{equation}
\sum_{\ell_\delta} \frac{1}{2\ell_\alpha+1} A_{\ell_\delta} \big( E^L_{\ell_\alpha \ell_\gamma \ell_\delta} \big)^2 \!=\! A_{\ell_\alpha} B_{\ell_\gamma} ( A_{\ell_\alpha} \!+\! A_{\ell_\gamma}\!-\! 2 ) ,
\label{Elm_SqContraction}
\end{equation}
which will prove useful in Appendix~\ref{sec:CompNeigh}
to compute the effect of the separation in orientation
in the limit of very close particles.

\section{Initial statistics}
\label{sec:InitStat}

Let us detail the initial statistics of the test particles' separation,
occurring in the second-expectation from Eq.~\eqref{rewrite_C_neigh}.
Throughout the paper, we will consider two different initial statistics,
namely a sampling with a fixed angular separation
between two test particles,
and a smooth sampling following the Von Mises-Fisher statistics.

\subsection{Fixed angle separation}
\label{sec:FixedAngle}

A simple route to sample two particles on the unit sphere
is to draw them with a constant and given angular separation.
To do so, one may draw the orientation of the first particle, $\hbL_{1}$,
uniformly on the sphere.
Then, the second orientation, $\hbL_{2}$, is sampled uniformly on the circle
such that ${ \hbL_{1} \!\cdot\! \hbL_{2} = \cos (\phi_{0}) }$,
with ${ \cos (\phi_0) \in [-1,1] }$,
the fixed angular separation between the two particles.
The joint \PDF\ of such a process is given by
\begin{equation}
P (\hbL_{1} , \hbL_{2}) = \frac{1}{8 \pi^{2}} \deltaD (\hbL_{1} \cdot \hbL_{2} - \cos (\phi_{0})) .
\label{PDF_Fixed}
\end{equation}
As required by Eq.~\eqref{def_D},
one can compute the statistics of the test particles' initial separation.
It follows from the addition theorem for spherical harmonics,
and one gets
\begin{equation}
D_{\ell} = P_{\ell} (\cos (\phi_{0})) ,
\label{D_Fixed}
\end{equation}
with $P_{\ell}$ the Legendre polynomial.

In the limit of small angular separation,
i.e.\ ${ \cos (\phi_0) \to 1 }$,
the following Taylor expansion holds
\begin{equation}
P_{\ell} (\cos (\phi_0)) \simeq 1 - \tfrac{1}{2} \ell (\ell + 1) (1 - \cos (\phi_0)) .
\label{DL_Legendre}
\end{equation}
In particular, one recovers that for a vanishing angular separation,
 ${ P_{\ell} (\cos (\phi_0)) \to 1 }$,
i.e.\ ${ \bD = \bI }$.
Such a distribution will prove useful  to construct
the Markovian piecewise prediction
presented in Section~\ref{sec:Piecewise}.

\subsection{Von Mises-Fisher distribution}
\label{sec:VonMises}

The Von Mises-Fisher distribution~\citep{Wood1994} is characterised by the \PDF\
\begin{equation}
P (\hbL) = \frac{\kappa}{4 \pi \sinh (\kappa)} \, \re^{\kappa \hbL \cdot \hbL_{0}} .
\label{PDF_VonMises}
\end{equation}
In that expression, $\hbL_{0}$ stands for a preferred direction,
and $\kappa$ for the concentration of the \PDF\@.
The larger $\kappa$, the smaller is the spread of the \PDF\ on the unit sphere.
As such, Eq.~\eqref{PDF_VonMises} is the analog of a Gaussian distribution on the sphere.
Sampling the test particles according to that statistics amounts
therefore to sampling once $\hbL_{0}$ uniformly on the sphere,
as required by isotropy,
and then sampling the test particles according to Eq.~\eqref{PDF_VonMises}.

\begin{figure}
    \centering
   \includegraphics[width=0.35 \textwidth]{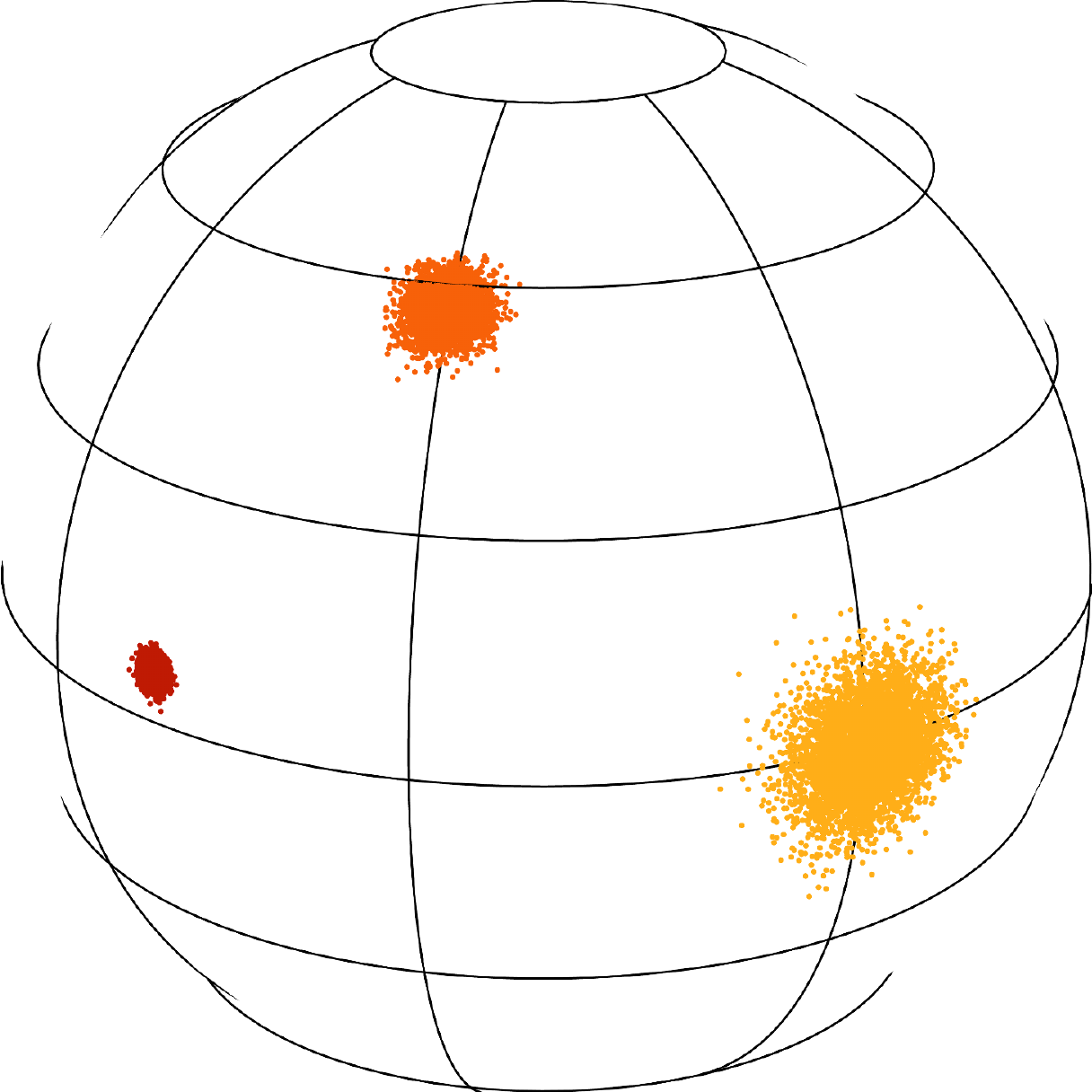}
   \caption{Illustration of the Von Mises-Fisher  distribution for
 three initial patches with 
     ${ \kappa = 200}$ (yellow), ${1000 }$ (orange) and $5000$ (red) resp\@.
     The larger $\kappa$ the narrower the distribution
     of initial orientations.
     Such a patch is meant to represent a set of co-eval stars
     which were born with comparable
     orbital parameters (including orientation).
     The smaller the initial patch the longer it takes for the \VRR\@-driven dilution to operate.
   }
   \label{fig:VMF}
\end{figure}

Following Eq.~\eqref{def_D}, it is then straightforward to compute
the statistics of the test particles' initial separation.
One gets
\begin{equation}
D_{\ell} = \frac{\pi \kappa}{2 \sinh^{2} (\kappa)} \, \big( I_{\ell + 1/2} (\kappa) \big)^{2} ,
\label{D_VonMises}
\end{equation}
with $I_{\ell}$ the modified Bessel function of the first kind.
In the limit of small angular separations,
i.e.\ ${ \kappa \to + \infty }$,
one has asymptotically
\begin{equation}
D_{\ell} \simeq 1 - \frac{1}{\kappa} \ell (\ell + 1) .
\label{DL_VMF}
\end{equation}
We recover therefore that for ${ \kappa \to + \infty }$,
one has ${ D_{\ell} \to 1 }$,
i.e.\ ${ \bD = \bI }$,
so that the initial patch of test particles
on the unit sphere is a Dirac delta.
Figure~\ref{fig:VMF} illustrates some examples
of initial samplings according the \PDF\ from Eq.~\eqref{PDF_VonMises}.

\section{Expectation for separation}
\label{sec:CompNeigh}

Let us detail the calculations
leading from Eq.~\eqref{rewrite_C_neigh_II} to Eq.~\eqref{calc_C_neigh}.
We start by performing a second-order Taylor expansion
of the matrix exponentials in Eq.~\eqref{rewrite_C_neigh_II},
so that
\begin{align}
 \re^{\bO_1} \bD \, \re^{- \bO_2} & \, \simeq \big( \bI + \bO_1+\tfrac{1}{2} \bO_1^2 \big) \, \bD \, \big( \bI - \bO_2 + \tfrac{1}{2} \bO_2^2 \big)
\label{exp_PertDev}
 \\
 & \simeq \bD \!+\! \bO_1 \bD\! -\! \bD \bO_2 \!+\! \tfrac{1}{2} \bO_1^2 \bD + \tfrac{1}{2} \bD \bO_2^2 - \bO_1 \bD \bO_2 .
\nonumber
\end{align}
Before computing the ensemble average of that expression,
let us first determine the average of $\bO_{i}$.
Returning to Eq.~\eqref{def_Omega},
we have
\begin{align}
 \big \langle \bO_i \big \rangle = & \, - \!\!\int_{0}^{t} \!\! \rd \tau \big \langle \bQ_{i}(\tau) \big \rangle
 \nonumber
 \\
 & \, + \frac{1}{2} \!\! \int_{0}^{t} \!\! \rd \tau \!\!\int_{0}^{\tau} \!\! \rd \taup \big\langle \big[ \bQ_i (\tau) , \bQ_i(\taup) \big] \big\rangle .
 \label{mean_Omega_Expand}
\end{align}
In that expression, the linear term in $\bQ_i$
is proportional to the harmonics of the bath.
Since the latter is isotropic, these harmonics vanish on average,
so that one has ${ \langle \bQ_{i} \rangle = 0 }$.
Let us now show that the quadratic term also vanishes.
Indeed, for indices $i,j$ one has
\begin{align}
 &\big\langle \big[ \bQ_i(\tau) \, \bQ_j(\taup) \big]_{\alpha \beta} \big\rangle = \sum_{\gamma,\delta,\sigma} E_{\alpha \gamma \sigma} E_{\sigma \delta \beta}
\label{mean_QQ}
\\
& \, \times \!\!\int\!\! \rd \bK \rd \bKp \, \mJ_\gamma \big[ \bK_i,\bK \big] \mJ_\delta \big[ \bK_j , \bKp \big] \big \langle \fb_\gamma(\bK,\tau) \fb_\delta (\bKp,\taup) \big\rangle .
\nonumber
\end{align}
Using the Gaussian ansatz from Eq.~\eqref{res_Cb}
for the bath correlations, 
as well as the contraction rules for the Elsasser coefficients
(see Eq.~\eqref{Elm_FouvryContractions}),
we can write
\begin{align}
 \big\langle \big[ & \bQ_i(\tau) \, \bQ_j(\tau') \big]_{\alpha \beta} \big\rangle = - \delta_{\alpha \beta} A_{\ell_\alpha} 
 \label{mean_QQ_diag}
 \\
 &\times\sum_\ell B_\ell \!\!\int\!\! \rd \bK \, \mJ_\ell \big[ \bK_i,\bK \big] \mJ_\ell \big[ \bK_j,\bK \big] \Cb_{\ell} (\bK,\tau-\tau').
\nonumber
\end{align}
Note that this expression is symmetric
both w.r.t.\ ${ i \!\leftrightarrow\! j }$
and ${ \tau \!\leftrightarrow\! \taup }$,
so that the two matrices ${ \bQ_{i} }$ and $\bQ_{j}$
commute on average.
As a result, the commutators in Eq.~\eqref{mean_Omega_Expand}
vanish on average, and so does that whole expression,
i.e.\ we have ${  \big \langle \bO_i \big \rangle = 0}$.
Hence the terms linear
in $\bO_{i}$ in Eq.~\eqref{exp_PertDev} also vanish on average.
Following these simplifications,
and owing to the fact that $\bD$ is a diagonal matrix,
we get
\begin{align}
 \big\langle \big[ \re^{\bO_1} \bD \, \re^{- \bO_2} \big]_{\alpha \beta} \big\rangle & = \delta_{\alpha \beta} D_\alpha + \tfrac{1}{2} \delta_{\alpha \beta} D_\alpha \big\langle \big[ \bO_1^2 + \bO_2^2 \big]_{\alpha \alpha} \big\rangle
 \nonumber
 \\
 & - \sum_{\gamma} D_{\gamma} \big\langle \big [\bO_1\big]_{\alpha \gamma} \big[ \bO_2 \big ]_{\gamma \beta}  \big\rangle ,
 \label{mean_quadratic_Index}
\end{align}
where in the first line we used Eq.~\eqref{mean_QQ_diag}
to ensure that only the diagonal coefficients of ${ \langle \bO^{2}_{i} \rangle }$
are non-zero on average.
Given that all these terms are already quadratic in $\bO_i$,
in Eq~\eqref{def_Omega},
we may keep only the terms linear in $\bQ_i$.
Subsequently, using Eq.~\eqref{mean_QQ_diag}
(which remains true for distinct ${i,j}$),
we get that the $\bQ$ matrices commute on average,
and therefore so do the $\bO$ matrices.
As a consequence, we can then write
\begin{align}
\big\langle \big[ \re^{\bO_1} \bD \, \re^{- \bO_2} & \big]_{\alpha \beta} \big\rangle  = \delta_{\alpha \beta} D_\alpha\! +\! \tfrac{1}{2} \delta_{\alpha \beta} D_\alpha \big\langle \big[ \big( \bO_1 - \bO_2 \big)^2 \big]_{\alpha \alpha} \big\rangle
\nonumber
\\ 
& - \sum_\gamma \! \big( D_\gamma \!-\! \delta_{\alpha \beta} D_\alpha \big) \big\langle \big[ \bO_1 \big]_{\alpha \gamma} \big[ \bO_2 \big]_{\gamma \beta} \big\rangle .
\label{mean_quadratic_commuting}
\end{align}
Using once again the contraction rules of the Elsasser coefficients
(as was done in Eq.~\eqref{mean_QQ_diag}),
we can rewrite Eq.~\eqref{mean_quadratic_commuting}
under the isotropic form
\begin{align}
 \big\langle \big[ \re^{\bO_1} \bD \, \re^{- \bO_2} \big]_{\alpha \beta} \big\rangle &= \delta_{\alpha \beta} \big( D_\alpha + \tfrac{1}{2} D_\alpha \big\langle \big[ \big( \bO_1 - \bO_2 \big)^2 \big]_{\alpha \alpha} \big\rangle
 \nonumber
 \\
 &- \sum_{\gamma} \big( D_\gamma \!-\! D_\alpha \big) \big\langle \big[ \bO_1 \big]_{\alpha \gamma} \big[ \bO_2 \big]_{\gamma \alpha} \big\rangle \big) .
 \label{mean_quadratic_TaylorFinal}
\end{align}
At this stage, since Eq.~\eqref{mean_quadratic_TaylorFinal}
was obtained through a Taylor expansion,
any functional ansatz satisfying this initial expansion is as valid.
As it is a polynomial, Eq.~\eqref{mean_quadratic_TaylorFinal} is not bounded,
and therefore cannot model the correlation's behaviour for ${ t \!\rightarrow\! \infty }$.
To circumvent this issue, we now choose the same ansatz
as in~\cite{FouvryBarOr2019}
and rely on an exponential function.
As all the terms are already quadratic in $t$
and the matrix is diagonal,
that exponential simply reads
\begin{align}
\big\langle \big[ \re^{\bO_1} \bD \, \re^{- \bO_2} & \big]_{\alpha \beta} \big\rangle =  \delta_{\alpha \beta} \, D_\alpha \exp\!\big[ \tfrac{1}{2} \big\langle \big[ \big( \bO_1 - \bO_2 \big)^2 \big]_{\alpha \alpha} \big\rangle \big]
 \nonumber
 \\
 & \!\!\!\! \times \exp \bigg[ \sum_{\gamma}\! \frac{D_{\alpha} \!-\! D_{\gamma}}{D_{\alpha}} \big\langle \big[ \bO_1 \big]_{\alpha \gamma} \big[ \bO_2 \big]_{\gamma \alpha} \big\rangle \bigg] .
 \label{mean_quadratic_Exponential}
\end{align}
In Eq.~\eqref{mean_quadratic_Exponential},
we have recovered Eq.~\eqref{calc_C_neigh}
presented in the main text.
Finally, one only has to plug in the expression for $\bO_i$
from Eq.~\eqref{def_Omega},
and,
following Eq.~\eqref{mean_QQ_diag}
substitute the bath correlation function by its Gaussian expression
from Eq.~\eqref{res_Cb} to get
\begin{align}
C_{\alpha}^{\bO}& (t) \, = \exp \bigg[ - \frac{1}{2} A_{\ell_{\alpha}} \sum_{\ell} B_{\ell} \!\! \int \!\! \rd \bK \, n (\bK)
\label{CbO}
\\
\times & \, \big( \mJ_{\ell} \big[ \bK_{1} , \bK \big] \!-\! \mJ_{\ell} \big[ \bK_{2} , \bK \big] \big)^{2} \,\frac{2 \Tc^{2} (\bK)}{A_\ell} \chi \big[ t \sqrt{A_{\ell} / 2}/ \Tc (\bK) \big] \bigg] ,
\nonumber
\end{align}
and
\begin{align}
& \, C_{\alpha}^{\bD} (t)= \exp \bigg[ \sum_{\ell , \ell_{\gamma}} \frac{D_{\ell_{\alpha}} - D_{\ell_{\gamma}}}{(2 \ell_{\alpha} + 1) D_{\ell_{\alpha}}} \big( E_{\ell_{\alpha} \ell \ell_{\gamma}}^{L} \big)^{2}
\label{CbD}
\\
& \! \times \!\!\int\!\!\! \rd \bK n (\bK) \mJ_{\ell} \!\big[ \bK_{1} ,\! \bK \big] \! \mJ_{\ell} \!\big[ \bK_{2} ,\! \bK \big] \frac{2 \Tc^{2} (\bK)}{A_\ell} \chi [ t \sqrt{A_{\ell} / 2} / \Tc (\bK) ] \bigg] ,
\nonumber
\end{align}
with the dimensionless function $\chi$ defined in Eq.~\eqref{def_chi}.
In Eq.~\eqref{CbD}, we also introduced the isotropic component
of the Elsasser coefficients ${ E_{\ell_{\alpha} \ell \ell_{\gamma}}^{L} }$,
whose main properties are briefly recalled in Appendix~\ref{sec:Elsasser}.

To conclude this Appendix,
let us now give a simplified form of Eq.~\eqref{mean_quadratic_Exponential}
when the stars share initially
very similar orientations.
We assume that the initial orientations
of the two test particles
is drawn with a fixed angular separation, ${ \cos (\phi_{0}) }$.
In Eq.~\eqref{D_Fixed},
we already characterised such a statistics,
and showed that ${ D_{\ell} \!=\! P_{\ell} (\cos (\phi_0)) }$.
Let us now assume that this separation is small
compared to any other relevant scale of the problem.
In practice, this amounts to requiring that
${ \ell (\ell + 1) \ll {1}/{(1 - \cos (\phi_0))} }$
for any $\ell$ that significantly contributes to the dynamics.

Relying on the development of $D_{\ell}$
from Eq.~\eqref{DL_Legendre},
we can the use the contraction rule of the Elsasser
coefficients from Eq.~\eqref{Elm_SqContraction},
to rewrite Eq.~\eqref{calc_C_neigh} as
\begin{align}
C_{\alpha \beta} (t) = & \, \delta_{\alpha \beta} \, P_{\ell_{\alpha}} (\cos (\phi_{0})) \, \exp \!\big[\! - \tfrac{1}{2} A_{\ell_{\alpha}} \Psi^{-} ( \bK_{1} , \bK_{2} , t ) \big]
\nonumber
\\
\times & \, \exp \!\big[\! - \tfrac{1 - \cos (\phi_{0})}{2} A_{\ell_{\alpha}} \Psi^{+} ( \bK_{1} , \bK_{2} , t ) \big] ,
\label{calc_C_neigh_limit}
\end{align}
where $\Psi^{-}$ and $\Psi^{+}$
are given by Eqs.~\eqref{def_PsiMinus} and~\eqref{def_PsiPlus}.
We refer to the main text for a discussion
of the effects associated with each of these terms.

\section{Dilution of a population}
\label{sec:Population}

Let us briefly show how one can naturally
expand the expression from Eq.~\eqref{calc_C_neigh}
to a population with an arbitrary number of test particles.

Throughout Section~\ref{sec:NeighCalc}, we focused on the description
of the separation of two particles
through their two-point correlation functions
(see Eq.~\eqref{def_C_neigh})
written with spherical harmonics.
In that case, all our predictions effectively depend only on ${ \cos (\phi) }$,
with $\phi$ the angular separation between the two particles.
Fortunately, such a result
can straightforwardly be used
to describe the anisotropies that appear 
when a larger population of test particles
diffuses on the unit sphere,
e.g.\@, as illustrated in Fig.~\ref{fig:Dilution}.
Indeed, owing to the large scale potential fluctuations,
an initially symmetric patch of test particles
can reach very complex, cramped, and anisotropic shapes.
In the present appendix, we show how one can easily generalise Eq.~\eqref{calc_C_neigh}
to a population of test particles.
This is in particular useful to interpret the simulations
from Appendix~\eqref{sec:Simulations}.

Let us consider a population of $n$ test particles
described its discrete \DF\
\begin{equation}
f (\hbL,\bK,t) = \frac{1}{n} \sum_{i=1}^n \deltaD (\bK-\bK_i) \, \deltaD (\hbL-\hbL_i(t)) ,
\label{Fokker_DDF}
\end{equation}
which satisfies the normalisation convention ${ \!\int\! \rd \bK \rd \hbL \, f = 1 }$.
This \DF\ can naturally be expanded in spherical harmonics as
\begin{equation}
f_\alpha (\bK,t) = \frac{1}{n} \sum_{i=1}^n \deltaD (\bK-\bK_i) Y_\alpha (\hbL_i(t)).
\label{Fokker_harmonics}
\end{equation}

Following Eq.~\eqref{def_C_neigh},
we are interested in the two-point correlation functions
of these spherical harmonics.
It reads
\begin{align}
f_\alpha (\bK,t) f_\beta (\bKp , t) = \frac{1}{n^{2}} \sum_{i,j=1}^n & \deltaD(\bK-\bK_i) \, \deltaD(\bKp - \bK_j)\nonumber
\\  & \times Y_\alpha (\hbL_i(t)) Y_\beta (\hbL_j(t)).
\label{Fokker_X_def}
\end{align}
Let us now compute the ensemble average of the previous expression.
We assume that the orbital parameters of the test particles
are distributed according to some \PDF\@,
${ p (\bK) }$, normalised so that ${ \!\int\! \rd \bK p (\bK) = 1 }$,
independent of their initial orientations.
Paying attention to the cases ${ i = j }$ and ${ i \neq j }$,
we can rewrite Eq.~\eqref{Fokker_X_def} as
\begin{align}
\big\langle f_\alpha (\bK,t) f_\beta(\bKp , t) \big\rangle = & \frac{n(n \!-\! 1)}{n^{2}} \, p(\bK) \, p(\bKp) \, C_{\alpha \beta} (\bK,\bKp,t) \nonumber\\
+ & \frac{1}{4 \pi} \frac{1}{n} \, \deltaD(\bK-\bKp) \, p(\bK) ,
\label{Fokker_X_avg}
\end{align}
In that expression, we introduced ${ C_{\alpha \beta} (\bK , \bKp , t) }$
as the two-point correlation of two test particles
with orbital parameters $\bK$ and $\bKp$,
as already defined in Eq.~\eqref{def_C_neigh}
For ${ n \gg 1 }$, Eq.~\eqref{Fokker_X_avg}
naturally becomes
\begin{equation}
\big \langle f_{\alpha}(\bK,t) f_{\beta}(\bK',t) \big \rangle \simeq  p(\bK) \, p(\bKp) \, C_{\alpha \beta} (\bK,\bKp,t).
\label{Fokker_X_bign}
\end{equation}
Equation~\eqref{Fokker_X_bign}
shows therefore that the correlation function
of a given population of test particles
essentially corresponds to the average correlation of particles' pairs,
averaged over realisation of the orbital parameters
(that are distributed according to the \PDF\ ${ p (\bK) }$).

To conclude,
following Eq.~\eqref{Fokker_harmonics},
for a population of test particles,
a natural observable to consider is therefore
\begin{align}
\frac{1}{n^{2}} \!\!\sum_{i , j = 1}^{n}\!\!\! \big\langle Y_{\alpha} (\hbL_{i} (t)) & \, Y_{\beta} (\hbL_{j} (t)) \big\rangle
\label{Fokker_observable}
\\
& \, \simeq \!\! \int \!\! \rd \bK \rd \bKp \, p (\bK) \, p(\bKp) \, C_{\alpha \beta} (\bK , \bKp , t) ,
\nonumber
\end{align}
where we used Eq.~\eqref{Fokker_X_bign}.
In practice, we are mostly interested
in the typical angular size of the patch,
i.e.\ the case ${ \ell_{\alpha} = \ell_{\beta} = 1 }$.
In that case, Eq.~\eqref{Fokker_observable} becomes
\begin{align}
\frac{1}{n^{2}} \!\!\sum_{i , j = 1}^{n}\!\!\! \big\langle Y_{\alpha} (\hbL_{i} (t)) Y_{\beta} (\hbL_{j} (t)) \big\rangle
& \, = \frac{1}{4 \pi} \frac{1}{n^2} \sum_{i,j=1}^n \langle \cos (\phi_{ij}(t))\rangle
\nonumber
\\
& \,  = \frac{1}{4 \pi} \bigg| \frac{1}{n} \sum_{i=1}^{n} \hbL_i (t) \bigg|^{2},
\label{Fokker_angularSize}
\end{align}
with ${ \phi_{ij} (t) }$ the angular separation
between the particles ${ i , j }$.
We finally emphasise that even though each individual orientation
is a unit vector,
their average, appearing in Eq.~\eqref{Fokker_angularSize},
is not.
More precisely, this average tends to be unitary
when the patch of test particles is very localised,
and vanishes when the patch becomes isotropic
on the unit sphere.

\section{Numerical simulations}
\label{sec:Simulations}

Let us briefly present the numerical simulations
to which our analytical results are compared.
The code used to integrate that system is the exact same
as that used in~\cite{FouvryBarOr2019},
and detailed in Appendix~{C} therein.
In a nutshell, at every timestep, the code computes
the particles' instantaneous magnetisations.
It has an overall complexity scaling like ${ \mO (N^{2} \ellmax^{2}) }$,
with $N$ the total number of particles,
and ${ \ellmax }$ the maximum harmonic number considered
in the pairwise interaction.
Once the velocity vectors are computed,
particles are displaced forward in time
using a fourth-order Runge-Kutta scheme,
with a constant global timestep.
The present method can benefit from parallelisation when computing
the magnetisations,
that are seen as contractions of large matrices and vectors.

The simulations used throughout the text are composed as such.
The background bath is made of ${ N = 10^{3} }$ stars.
Their conserved quantities ${ \bK_{i} = (m_{i}, a_{i}, e_{i}) }$
satisfy ${ m = \mmin }$, ${ \amin \leq a \leq \amax }$,
and ${\emin \leq e \leq \emax}$.
We pick our units so that ${ G = \mmin = \amin = 1 }$,
and we consider the ranges ${ \amax / \amin = 100 }$,
${ \emin = 0 }$, and ${ \emax = 0.3 }$.
Each of these parameters are drawn independently from one another,
according to \PDFs\ proportional to
${ (\deltaD (m - \mmin) , a^{1/2} , e) }$.
Finally, the initial orientations of the background stars
are drawn uniformly on the sphere,
and interactions are truncated at ${ \ellmax = 50 }$.
The background particles are fully self-gravitating,
i.e.\ are all coupled to each other.
The timestep of the integration is picked using the exact same method
and parameters as in~\cite{FouvryBarOr2019}.

On top of that, in order to investigate the separation of neighbour test particles,
we also add to the simulation ${ n = 10^{3} }$ test particles.
These are test particles, i.e.\ they do not contribute to the mean potential,
but only probe its instantaneous value.
The initial parameters of the test particles,
${ \bKt = (\mt , \at , \et) }$,
are picked
with ${ \mt = \mmin }$,
${ 9 \leq \at \leq 11 }$,
and ${ \emin \leq \et \leq \emax }$.
Within that range, the parameters are drawn
with the same \PDFs\ as for the background particles.
For simulations where the test particles have all the same orbital parameters,
we used the median value of these distributions,
that correspond to ${ (\at , \et) \simeq (10,0.21) }$.
For each realisations,
the initial orientation of the test particles
is drawn according
to the Von Mises-Fisher distribution
from Eq.~\eqref{PDF_VonMises}.
Adding the test particles to the dynamics does not drastically
increase the complexity of the code, as it then scales
like ${ \mO (N (N \!+\! n) \ellmax^{2}) }$.

In practice, we performed two sets of simulations of that system:
(i) the test particles all have the same orbital parameters,
and an initial dispersion in orientation characterised by ${ \kappa = 5000 }$,
as defined in Eq.~\eqref{PDF_VonMises};
(ii) the test particles have an initial distribution of orbital parameters,
and an initial orientation also given by ${ \kappa = 5000 }$.
The main interest of the first set of simulations is to allow
us to investigate the effect of the separation in orientations,
in the absence of any separation stemming from differences in parameters.
For each of these simulations,
we considered 200 different realisations
to perform the ensemble averages.

\section{Piecewise prediction}
\label{sec:piecewiseAppendix}

Let us now detail the calculations
leading to the piecewise prediction
from Section~\ref{sec:Piecewise}, and
illustrated in Fig.~\ref{fig:PiecewiseSameK}.
The goal is to give a simple prediction
for the expectation ${ \langle \cos (\phi_i) \rangle }$
corresponding to the timestep
${ t_i = i \Delta t }$ with ${ i \geq 0 }$.
Thanks to the Markovian assumption,
the statistics of ${ \cos (\phi_{i+1}) }$
is fully determined by the statistics of the previous
angle separation ${ \cos (\phi_{i}) }$
and the properties of the background stochastic noise.
As such, we can write
\begin{equation}
\langle \cos (\phi_{i+1}) \rangle = \!\!\int\!\! \rd ( \cos(\phi_i)) \, \rho_i (\cos (\phi_i)) \, \langle \cos (\phi_{i+1}) | \phi_i \rangle ,
\label{PWApp_MasterEq}
\end{equation}
where $\rho_i$ is the PDF of ${ \cos(\phi_i) }$,
i.e.\ the statistics of the initial angular separation
for the timelapse ${ \cos (\phi_i) \to \cos (\phi_{i+1}) }$.
In Eq.~\eqref{PWApp_MasterEq},
we also introduced
${ \langle \cos (\phi_{i+1}) | \cos (\phi_i) \rangle }$
as the conditional expectation of ${ \cos (\phi_{i+1}) }$
given the value of ${ \cos (\phi_i) }$,
i.e.\ the expectation of the new angular separation
${ \cos (\phi_{i+1}) }$ after a time ${ \Delta t }$
for test particles initially separated by ${ \cos (\phi_{i}) }$.
As such, in that expectation ${ \cos (\phi_{i}) }$
is taken to be an initial condition
rather than a random variable.

Let us therefore assume that the two test particles
are initially separated by ${ \cos (\phi_{i}) }$.
In Appendix~\ref{sec:FixedAngle},
we have already characterised the statistics 
of such fixed angular separations.
Following Eq.~\eqref{D_Fixed},
we can therefore describe the initial separation
of the test particles
via the coefficients ${ D_{\ell} = P_{\ell} (\cos (\phi_{i})) }$.
We can then use the prediction from Eq.~\eqref{calc_C_neigh}
to obtain, after a time ${ t = \Delta t }$,
the conditional expectation for the test particles'
new separation.
For any harmonics $\ell$, we write
\begin{equation}
\langle P_{\ell} (\cos (\phi_{i+1}\!)) | \cos (\phi_{i}) \rangle \!=\! P_{\ell} (\cos (\phi_{i})) \, C_{\ell}^{\bO} \!(\Delta t) C_{\ell}^{\bD} \!(\Delta t) ,
\label{PWApp_cond}
\end{equation}
with ${ C_{\ell}^{\bO} (\Delta t) }$ and ${ C_{\ell}^{\bD} (\Delta t) }$
given by Eq.~\eqref{calc_C_neigh}.
Since we have ${ P_{1} (\cos(\phi_{i+1})) \!=\! \cos (\phi_{i+1}) }$,
we are mainly interested
in the case ${ \ell = 1 }$ of Eq.~\eqref{PWApp_cond}.

Even if the r.h.s.\ of Eq.~\eqref{PWApp_cond}
is fully known,
performing explicitly the integral from
Eq.~\eqref{PWApp_MasterEq} is impossible.
Indeed,  the \PDF\ of ${ \rho_{i} (\cos (\phi_{i})) }$ is unknown,
nor can we hope to compute that integral
with the intricate integrand
from Eq.~\eqref{PWApp_cond}.
To proceed further, we rely on a first-order perturbative expansion
of the r.h.s.\ of Eq.~\eqref{PWApp_cond}
for small angular separations,
i.e.\ for ${ \cos (\phi_{i}) \simeq 1 }$.
At this stage, the only dependences on ${ \cos (\phi_{i}) }$
in the r.h.s.\ of Eq.~\eqref{PWApp_cond}
are through the coefficients ${D_{\ell} = P_{\ell} (\cos (\phi_{i})) }$.
In particular, these coefficients only appear
in the expression of ${ C_{\ell}^{\bD} (\Delta t) }$,
as given by Eq.~\eqref{CbD}.

In that limit, Eq.~\eqref{PWApp_cond}
becomes
\begin{align}
\langle \cos (\phi_{i+1}) | \cos (\phi_i)  \rangle &= \xi_0 + \xi_1 \cos (\phi_i) ,
\label{PWApp_FirstExpansion}
\end{align}
where the coefficients $\xi_{k}$ are given by
\begin{align}
\xi_{0} & \, = \langle \cos (\phi_{i+1}) | \cos (\phi_{i}) \rangle \big|_{\cos (\phi_{i}) = 1} - \xi_{1} ,
\nonumber
\\
\xi_{1} & \, = \frac{\partial }{\partial \cos (\phi_{i})} \bigg[ \langle \cos (\phi_{i+1}) | \cos (\phi_{i}) \rangle \bigg]_{\cos (\phi_{i}) = 1} .
\label{generic_xi}
\end{align}
Fortunately, in the limit ${ \cos (\phi_{i}) \to 1 }$,
we can follow the exact same development
as in Eq.~\eqref{calc_C_neigh_limit}
to write
\begin{equation}
\langle \cos (\phi_{i+1}) | \cos (\phi_{i}) \rangle \simeq \cos (\phi_{i}) \, \re^{- \Psi^{-}} \re^{- \Psi^{+} (1 - \cos (\phi_{i}))} 
\label{approx_expect}
\end{equation}
where we used the fact that ${ A_{\ell = 1}/2 = 1 }$,
and where the functions $\Psi^{-}$ and $\Psi^{+}$
were introduced in Eq.~\eqref{def_PsiMinus}
and~\eqref{def_PsiPlus},
and are evaluated in ${ t = \Delta t }$.
Following the definition from Eq.~\eqref{generic_xi},
we finally obtain
\begin{equation}
\xi_{0}
 \, = - \, \re^{- \Psi^{-}} \, \Psi^{+} , \;\;\;
\xi_{1}
 \, = \re^{- \Psi^{-}} (1 + \Psi^{+}) .
\label{def_xi}
\end{equation}
At this stage, since the coefficients $\xi_{0}$
and $\xi_{1}$ do not depend anymore on the test
particles' separation, ${ \cos (\phi_{i}) }$,
we can then take the average
of the expansion from Eq.~\eqref{PWApp_FirstExpansion}
w.r.t.\ ${ \cos (\phi_{i}) }$,
to get the expectation ${ \langle \cos (\phi_{i + 1} ) \rangle }$,
as defined in Eq.~\eqref{PWApp_MasterEq}.
We get
\begin{equation}
\langle \cos (\phi_{i+1}) \rangle = \xi_0 + \xi_1 \langle \cos (\phi_i) \rangle , 
\label{PWApp_FirstExpansion_II}
\end{equation}
This inductive definition can be solved explicitly as
\begin{align}
\langle \cos (\phi_i) \rangle = q + \xi_1^i \big( \langle \cos (\phi_0) \rangle - q \big) ,
\label{PWApp_AGSolved}
\end{align}
with ${ q = \xi_0 / (1 - \xi_1) }$,
and ${ \langle \cos (\phi_{0}) \rangle }$
standing for the average angular separation
of the test particles at the initial time.
In Eq.~\eqref{PWApp_AGSolved},
 note that for test particles with identical parameters,
one has ${ \Psi^{-} = 0 }$, so that ${ q = 1 }$,
while in the generic case, one has ${ q \!\geq\! 1 }$.
Finally, even if Eq.~\eqref{PWApp_AGSolved}
depends on the discrete time index, $i$,
it can still be viewed  as a continuous function of time,
provided one makes the replacement ${ i \to t / \Delta t }$.
Such continuous predictions are  plotted
in Figs.~\ref{fig:PiecewiseSameK} and~\ref{fig:PiecewiseVaryingPhiandSMA}.

Note that the prediction from Eq.~\eqref{PWApp_AGSolved}
does not rely on the precise shape
of the \PDFs\ ${ \rho_{i} (\cos (\phi_i)) }$,
as we are ultimately integrating over them.
In addition, since the coefficients $\xi_{k}$ only depend
on the coupling coefficients, ${ \mJ_{\ell} [\bK_{i} , \bK] }$,
they do not need to be recalculated when
the initial angular separation, ${ \langle \cos (\phi_0) \rangle }$,
is changed,
but only when the background bath,
i.e.\ ${ n(\bK) }$,
or the parameters of the test particles,
i.e.\ $\bK_{1}$ and $\bK_{2}$,
are changed.

Because it relies on the perturbative expansion
from Eq.~\eqref{PWApp_FirstExpansion},
when ${ \cos (\phi_i) }$ decreases,
Eq.~\eqref{PWApp_AGSolved} becomes inaccurate,
and even diverges to ${ - \infty }$ at late times.
In practice, however, this is not a problem,
since Eq.~\eqref{PWApp_FirstExpansion} has good behavior
up to ${ \langle \cos (\phi) \rangle \gtrsim 0.8 }$
(see, e.g.\@, Fig.~\ref{fig:PiecewiseSameK}).
As a result, using this piecewise prediction,
we are in a position to properly describe
the early separation of neighbours up to ${ \phi \lesssim 37^{\circ} }$.
This is more than enough for astrophysical applications.
Moreover, should one wish to extend this prediction
to even larger angular separations,
the present piecewise approach can in practice
be extended with the analytical prediction from Eq.~\eqref{calc_C_neigh},
that rightfully works in regimes where the angular separation
between the test particles is large.

\section{Restricted $\ell=2$ toy model}
\label{sec:ToyModel}

The toy model presented in this Appendix
serves two main purposes:
(i) to illustrate quantitatively the impact of the bath
on the test particles in some simplified framework;
(ii) to allow for the tuning of the transition \PDF\
on which the Markovian model of Appendix~\ref{app:Virtual}
is based.

Similarly to~\cite{Roupas2017,Hamers2018},
we construct our toy model
by restricting the \VRR\ interaction 
to the sole ${ \ell = 2 }$ harmonics,
i.e.\ only the coupling coefficients, ${ \mJ_{\ell} }$,
with ${ \ell = 2 }$ are taken to be non-zero.
To further simplify the setup, we assume that background cluster
is composed only of identical stars.
As a result, in Eq.~\eqref{evol_test},
there is a single coupling coefficient,
which we denote $\mJ_{2}$
(which can however vary between
different test particles.)
Similarly, following Eq.~\eqref{def_Tc},
there is a single coherence time for the bath,
which we denote $\Tc$.
In the case ${ \ell = 2 }$,
the equations of motion for the test particle
can then be rewritten as
\begin{equation}
\frac{\rd \hbL}{\rd t} = \hbL \times \big( \bM (t) \cdot \hbL \big),
\label{eq_toy}
\end{equation}
where the time dependent matrix ${ \bM (t) }$ reads
\begin{equation}
\bM (t) =
\begin{pmatrix}
\eta_2 & \eta_{-2} & -\eta_1 \\
\eta_{-2} & -\eta_2 & -\eta_{-1} \\
-\eta_1 & - \eta_{-1} & \sqrt{3} \eta_0
\end{pmatrix}, \hfill
\label{M_def_toy}
\end{equation}
where we have set
\begin{equation}
\eta_m (t) = \sqrt{B_{2}} \mJ_{2} \fb_{2m}(t)\,.
\label{eta_def_toy}
\end{equation}
Here, the stochastic matrix ${ \bM (t) }$ contains
all the information about the potential fluctuations generated
by the background bath.

At any given time, ${ \bM (t) }$, which is a symmetric matrix,
has three orthogonal eigenvectors,
that we denote ${ \be_{k} }$ (with ${ k = 1,2,3 }$).
These vectors define three orientations,
and therefore six `poles' on the unit sphere.
For any particle standing on one of these poles,
one has ${ \bM(t) \!\cdot\! \hbL \propto \hbL }$,
so that these are equilibrium points of the dynamics
as the cross product from Eq.~\eqref{eq_toy} vanishes.
Decomposing any generic ${ \hbL (t) }$ over that same basis,
as ${ \hbL (t) = \sum_{k=1}^{3} \Lambda_{k} \be_{k} }$,
we can rewrite Eq.~\eqref{eq_toy} under the simpler form
\begin{equation}
\frac{\rd \bm{\Lambda}}{\rd t} = \pm \, \bm{\Lambda} \times \big( \bD (t) \cdot \bm{\Lambda} \big) ,
\label{D_eq_toy}
\end{equation}
where ${ \bD (t) }$ is the diagonal matrix,
whose entries are the three eigenvalues of ${ \bM (t) }$,
and the global sign depends on the relative order of the eigenvectors.
Of course, Eq.~\eqref{D_eq_toy} remains a quadratic (and stochastic) differential equation
hard to generically solve.

To better understand the characteristics
of that dynamics, let us assume that 
${ \hbL (t) }$ is close to one of the eigenvectors,
say ${ \be_{a} }$.
Assuming ${ \Lambda_{a} \gg \Lambda_{b} , \Lambda_{c} }$,
we can then linearise Eq.~\eqref{D_eq_toy}
to obtain simple linear evolution equations for the two remaining coordinates,
${ k = b , c }$, as
\begin{equation}
\frac{\rd^{2} \Lambda_{k}}{\rd t^{2}} = - \Lambda_{a}^{2} \, (d_{b} - d_{a}) (d_{c} - d_{a}) \, \Lambda_{k} ,
\label{eqToy_linearise}
\end{equation}
with $d_{k}$ the eigenvalues of $\bD$.
From the structure of Eq.~\eqref{eqToy_linearise},
we can note that the stability of the six equilibrium points
depends on the eigenvalue
with which they are associated~\citep{Roupas2017}.
Indeed, the equilibrium points associated
with the largest and smallest eigenvalues
are stable, while the ones associated with the
intermediate eigenvalue are unstable.

Following this property, Eq.~\eqref{eqToy_linearise} therefore defines
six dynamical regions on the sphere,
four of them enforce stable rotating trajectories around the stable points,
while the two other unstable regions
drive diverging trajectories.
We finally note from the non-linearity of Eq.~\eqref{eqToy_linearise}
that the frequency for these dynamics depends on the distance to the equilibrium points.
This will naturally enforce a phase mixing
of particles close to these stable points,
as a result of this shear in frequency.
In addition to this orientation-induced phase mixing,
test particles with different orbital parameters will also shear apart,
as their orbital frequencies depend linearly
on the coupling coefficients.

\begin{figure}
\centering
\includegraphics[width=0.35 \textwidth]{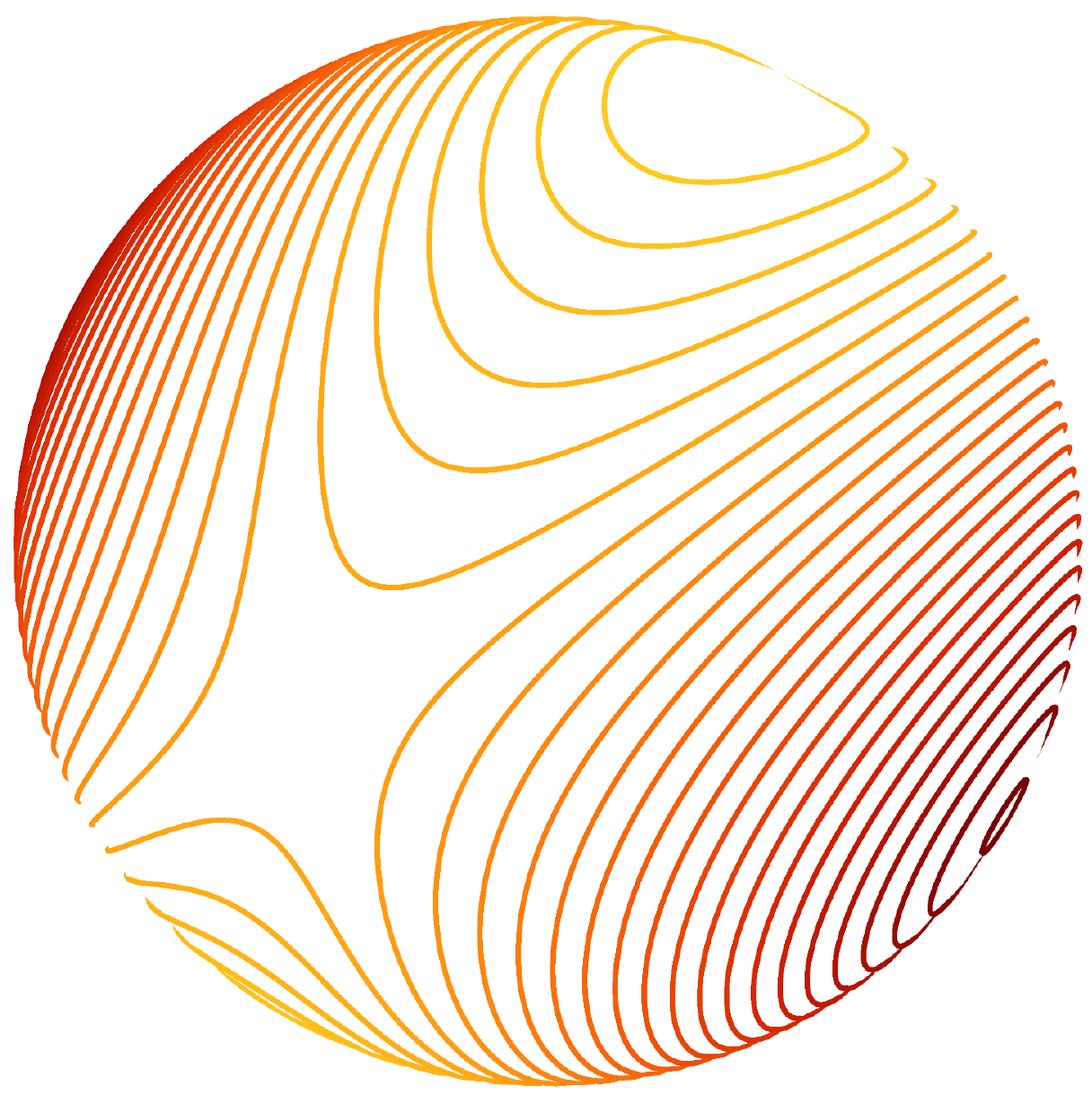}   
\caption{Illustration of trajectories in the toy model
from Eq.~\eqref{eq_toy}
for a time-independent background noise,
i.e.\ for ${ \bM (t) \!=\! \mathrm{cst.} }$
One can clearly note the presence of attraction points
(e.g.\@, near the pole) and separatrices
where the separation of neighbours will be
the most efficient.
   }
   \label{fig:TrajToyModel}
\end{figure}

On top of this phase mixing,
the most efficient way to separate nearby test particles
appears in the vicinity of the unstable points
which define separatrices on the unit sphere,
where test particles get swiftly separated.
In Fig.~\ref{fig:TrajToyModel},
we illustrate the typical orbits
of the present toy model.
For such a model,
orbits are generically the intersections
of the unit sphere with an ellipsoid
having the same centre~\citep{Roupas2017}.
In that figure, one can clearly note the presence of
stable regions that source phase mixing,
as well as unstable regions
associated with separatrices.

At this stage, it is important to emphasise
that the previous discussion has been made without taking
into account the explicit time-dependence of ${ \bM (t) }$,
i.e.\ it is only valid for timescales ${ t \ll \Tc }$.
Because of the stochastic changes of ${ \bM(t) }$,
the test particles' dynamics
becomes obviously more intricate
on longer timescales.
Indeed, as the matrix changes, the eigen-directions,
and therefore the six dynamical regions,
move around the sphere,
so that no point on the sphere remains an equilibrium point.
Eventually, the test particles continuously pass from one dynamical
region to the other,
which keeps distorting their trajectories
and stirring them away from one another.

In practice, this reshuffling of the eigen-directions
turns out to be the main dynamical process driving
the growth of the angular separation between test particles.
Indeed, if the eigen-directions were constant,
two particles launched with similar
directions would likely find themselves within the same dynamical region.
Their trajectories would therefore be close to each other,
and they would only drift apart through phase mixing.
In that case, only if the two particles were initially separated
by a separatrix will they have very different trajectories.
The closer initially the two particles,
the less likely 
for them sit on two sides of a separatrix.
However, once one accounts for the time-dependence of ${ \bM (t) }$,
these separatrices start moving around the sphere.
It gets then much more likely that at some point a separatrix
will get in between the two test particles.
From that moment onwards, the separation between the particles
is drastically accelerated, as their trajectories become very different.
Hence, it is this succession of rapid stirring induced by the stochastic motion
of the separatrices that is the main dynamical driver
of the separation of neighbours in the \VRR\ dynamics.

From this discussion, it appears that an essential quantity
is therefore the ratio between
the ballistic timescale
(i.e.\ the dynamical time for the motion of a test particle
in a given and fixed noise)
and the coherence timescale
(i.e.\ the time required for a significant modification
of the background noise).
The ballistic timescale depends primarily on the typical amplitude
of the background fluctuations and on the coupling strength
 between the test particles and the bath.
The coherence timescale is given by the correlation time,
i.e.\ by $\Tc$ as defined in Eq.~\eqref{def_Tc}.
If one assumes that the test particles share the same orbital
parameters as that of the  bath, following Eq.~\eqref{res_Cb},
one can show that the ratio of the ballistic and coherence timescales
equals $\sqrt{3}$.
This implies that both timescales are
such that the motion of the test particles and the reshuffling
of the eigen-directions happen simultaneously.
This further complicates the dynamics,
as the lack of  timescale separation
requires that the two effects are accounted for simultaneously.

\begin{figure}
\centering
\includegraphics[width=0.35 \textwidth]{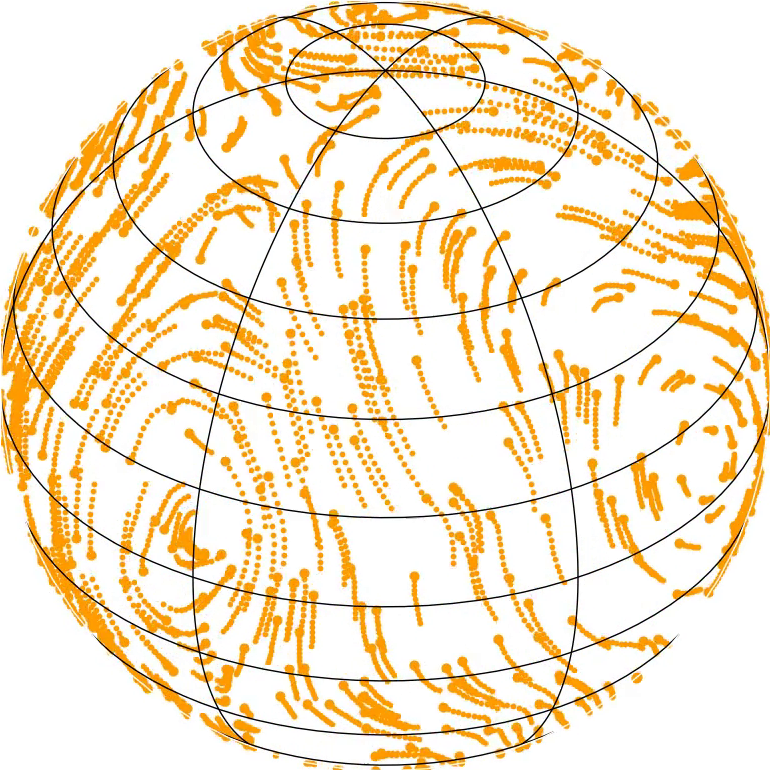}   
\caption{Illustration of the test particles' trails at late time
in numerical simulations,
where all the test particles have the same set of orbital parameters
(see Appendix~\ref{sec:Simulations}).
Stars strikingly seem to rotate for some time around some 
attraction orientations related to those described in detail in Fig~\ref{fig:TrajToyModel}
for the toy model.
When higher order harmonics contribute to the potential,
at any given time,
the unit sphere is segmented by many separatrices
stemming from the (correlated) potential fluctuations
generated by the background stellar cluster.
A patch of test particles which happens to pass near
one such (transient) separatrix
will diffuse apart more swiftly.
   }
   \label{fig:Vortices}
\end{figure}

Even though this toy model is a useful and insightful tool to understand
the dynamics that drive neighbour separation,
it is only an approximation of the general \VRR\ dynamics
where all harmonics are taken into account.
As illustrated in Fig.~\ref{fig:Vortices},
higher-order harmonics allow for the creation
of even more separatrices,
which further boost the efficiency of particles' separation.

In practice, to implement numerically the toy model from Eq.~\eqref{eq_toy},
we proceeded as follows.
For a given realisation, the stochastic matrix, ${ \bM (t) }$,
is generated by sampling time-correlated processes, ${ \eta_{m} (t) }$,
that follow the theoretical correlation from Eq.~\eqref{def_Cb}.
These correlated time series are generated using the method
spelled out in~\cite{RomeroSancho1999}.
Once the background noise has been constructed,
the evolution equations for the test particles
are integrated using a fourth-order Runge-Kutta method~\citep{Press2007}.
Finally, in order to avoid late-time biases,
following the Rodrigues rotation formula,
we used a drift operator adapted to the spherical geometry,
which proceeds by rotating the unit vectors
rather than by naively translating them.

\section{Generating virtual dilutions}
\label{app:Virtual}

In Section~\ref{sec:Piecewise},
we developed a piecewise formalism
predicting the average evolution
of the angular separation between two neighbours.
This approach overcame the main difficulty
encountered in Fig.~\ref{fig:PiecewiseSameK},
namely it managed to `bend'
the prediction downwards
to better match the numerical measurements.
Unfortunately, by design,
the piecewise expression from
Eq.~\eqref{AG_linearMomentsRel_avg}
can only predict the expectation
of the angular separation,
namely ${ \langle \cos (\phi) \rangle }$.
It does not predict any higher-order moments,
nor can it generate virtual dilutions,
that is random walks of ${ \cos (\phi) }$
constructed without integrating the evolution equations
(hence, very cheap to produce)
and mimicking the statistics of the
numerical simulations

At first, it could seem that the approach
used in Eq.~\eqref{AG_linearMomentsRel}
could be generalised to predict
higher-order moments.
This would amount to push the perturbative
expansion to higher orders,
obtaining a recurrent system of equations
predicting as many moments of ${ \cos (\phi) }$
as wished.
Unfortunately, the further the expansion is pushed,
the more sensitive it becomes to inaccuracies
in the analytical predictions from Eq.~\eqref{calc_C_neigh}.
In practice, only a first-order expansion,
as in Eq.~\eqref{AG_linearMomentsRel},
was found to be robust enough.

In this section, our goal is to design
a procedure to generate virtual samples
of the dilution process.
Relying still on the Markovian assumption,
we want to construct virtual sequences of angular
separations,
${ \phi_0 \!\rightarrow\! \phi_{1} \!\rightarrow\! ... \!\rightarrow\! \phi_{n} }$. 
For simplicity, in all this Appendix we will restrict ourselves
to the case where both test particles
have identical orbital parameters.
As shown in Fig.~\ref{fig:CorrelationSameK},
accounting for additional differences in orbital parameters
would yield faster dilutions.
As in Eq.~\eqref{choice_Deltat},
we take the transition time ${ \Delta t }$
between $\phi_{i}$ and $\phi_{i+1}$
to be ${ \Tc (\bK) }$,
with $\bK$ the parameters of the test particles.
The key difference with Section~\ref{sec:Piecewise}
is that now $\phi_{i}$ represents an individual sample
of a neighbour separation
rather than the average over many realisations.
Similarly to Eq.~\eqref{AG_conditionalE},
the Markovian assumption implies that the
\PDF\ of $\phi_{i+1}$, ${ \rho_{i+1} (\phi_{i+1}) }$, satisfies
\begin{equation}
\rho_{i+1} (\phi_{i+1}) = \!\!\int\!\! \rd \phi_{i} \, \rho_{i} (\phi_{i}) \, P(\phi_{i+1} | \phi_{i}) ,
\label{Virt_MasterEq}
\end{equation}
where ${ P (\phi_{i+1} | \phi_{i}) }$
is the probability that two test particles
separated by an angle ${ \phi_{i} }$ at time ${ t \!=\! i \Delta t }$
evolve to a separation $\phi_{i+1}$ at time ${ t \!=\! (i \!+\! 1) \Delta t }$.
We call this \PDF\ the transition probability.
We note that this \PDF\ does not depend
on $i$, i.e.\ the underlying physical process
does not explicitly depend on time.
This is again a consequence of our Markovian assumption.

Equation~\eqref{Virt_MasterEq}
is hard to solve analytically,
however once the transition \PDF\@,
${P (\phi_{i+1} | \phi_{i}) }$, is known,
generating samples of random walks
consistent with it is easy.
In the following, our goal will be to obtain
a simple estimate for that \PDF\@,
both accurate and straightforward to sample,
so that one could easily generate virtual dilutions.

\subsection{Finding a suitable ansatz}

Because Section~\ref{sec:Piecewise} only gave us estimates
of the average properties of neighbour separation,
it cannot provide us with a shape for the transition \PDF\@.
We must therefore rely on numerical simulations to characterise it.
Unfortunately, given the numerical cost of full \VRR\ simulations
(see Appendix~\ref{sec:Simulations})
we cannot use them to obtain enough data
to check our estimations of the transition \PDF\@.
As such, we use a restricted toy model
to perform this exploration.
Such a simplified model amounts to assuming
that only the harmonics ${ \ell = 2 }$ of the bath
drives the \VRR\ dynamics.
The main properties of that model
are spelled out in Appendix~\ref{sec:ToyModel}.

Of course, the ${ \ell = 2 }$ toy model
is only an approximation of the full \VRR\ process
of neighbour separation,
for example, reducing the dilution rate
compared to the dynamics driven by the full harmonic range.
Yet, because the underlying mechanisms remain the same,
we expect the functional form of the transition \PDF\
to be similar between the two cases.

\begin{figure}
\centering
\includegraphics[width=0.45 \textwidth]{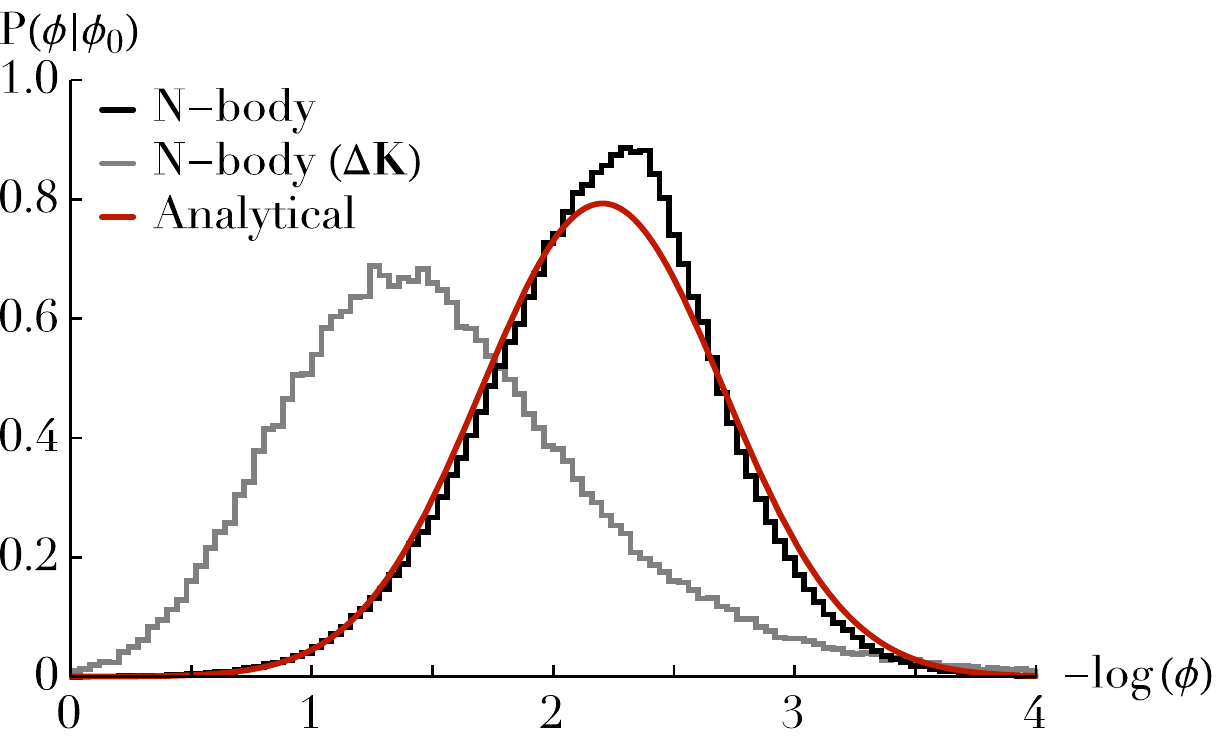}   
\caption{Illustration of the conditional PDF, ${ P(\phi|\phi_0) }$,
for the angle $\phi$, in radians, separating two test particles
after a time ${ \Delta t }$ (see Eq.~\eqref{choice_Deltat})
as measured in the ${ \ell = 2 }$ toy model.
Here, the two test particles have the same orbital parameters,
were initially separated by ${ \phi_0 = 5^{\circ} }$,
and the histogram was obtained over ${ 500\,000 }$ realisations
of the toy model.
The analytical prediction is given by the log-normal ansatz
from Eq.~\eqref{ansatz_lognormal},
with parameters estimated using Eq.~\eqref{calc_C_neigh}.
 In that plot, we also added the same measurement
 for simulations 
 where the two test particles differ by a factor ${1.3}$
 in their coupling factor to the bath,
 which naturally accelerates
 the growth of their angular separation.
}
   \label{fig:HistoPhiSameK}
\end{figure}

In Fig.~\ref{fig:HistoPhiSameK},
we show a numerical measurement in the toy model
of the transition \PDF\@, ${ P (\phi | \phi_{0}) }$.
Rather than considering ${ \cos (\phi) }$ as the variable
of interest, we used ${ - \log (\phi) }$,
whose scale invariance better handles
the regime of very small angular separations.
In the absence of any separation in orbital parameters,
we note that the transition \PDF\ for ${ - \log (\phi) }$
is consistent with a normal distribution.
In that case, we hint therefore from the numerical simulations
that the transition \PDF\@, ${ P (\phi_{i+1} | \phi_{i}) }$,
can be approximated with a log-normal distribution.
As such, we assume
\begin{equation}
P (\phi | \phi_{0}) = \frac{1}{\phi \, \sigma \sqrt{2 \pi}} \exp \bigg[ - \frac{(- \log (\phi) - \mu)^{2}}{2 \sigma^{2}} \bigg] .
\label{ansatz_lognormal}
\end{equation}
This a key assumption to be able to generate
virtual particles' separations.

Even though Fig.~\ref{fig:HistoPhiSameK}
only corresponds to one particular set of orbital parameters
and initial angles,
we investigated these transition probabilities
for a wide range of orbital parameters and initial angles,
and always recovered normal-like distributions.
Of course, the ansatz 
from Eq.~\eqref{ansatz_lognormal}
ultimately breaks down
when the initial angle, ${ \phi_{0} }$, becomes too large,
since the domain of the normal distribution is not bounded
and overflows the range ${ -\log (\pi) \leq - \log (\phi) \leq + \infty }$.
In practice, we found that this ansatz
works well for angles ${ \phi \lesssim 45^{\circ} }$.

\subsection{Estimating the parameters of the transition distribution}

Having identified a plausible ansatz
for the transition \PDF\@,
we must now use the analytical prediction
from Eq.~\eqref{calc_C_neigh}
to estimate the two parameters
of Eq.~\eqref{ansatz_lognormal},
namely $\mu$ and $\sigma$.

The moments of the distribution from Eq.~\eqref{ansatz_lognormal}
generically read
\begin{equation}
\langle \phi^n | \phi_{0} \rangle = \exp \!\big[\! -n \mu + \tfrac{1}{2} n^2 \sigma^2 \big] .
\label{logNormal_Moments}
\end{equation}
From the estimation of
${ \langle \phi^2 | \phi_{0} \rangle }$ and ${ \langle \phi^4 | \phi_{0} \rangle }$
one can then estimate the parameters
of the log-normal distribution from Eq.~\eqref{ansatz_lognormal}.
Indeed, one has
\begin{align}
\mu = \tfrac{1}{4} \log \!\bigg[ \frac{\langle \phi^{4} | \phi_{0} \rangle}{\langle \phi^{2} | \phi_{0} \rangle^{4}} \bigg] ; \;\;\;
\sigma^{2} = \tfrac{1}{4} \log\! \bigg[ \frac{\langle \phi^{4} | \phi_{0} \rangle}{\langle \phi^{2} | \phi_{0} \rangle^{2}} \bigg] .
\label{link_moments}
\end{align}

\begin{figure}
    \centering
   \includegraphics[width=0.45 \textwidth]{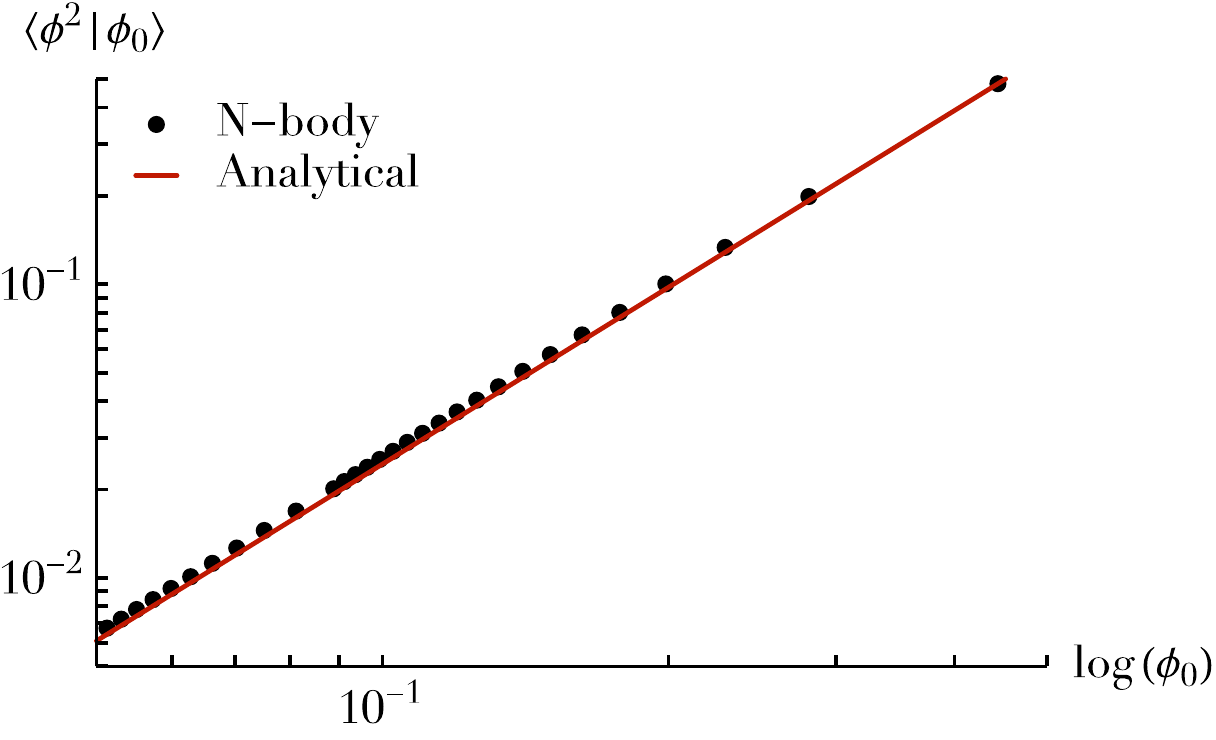}
   \\
   \includegraphics[width=0.45\textwidth]{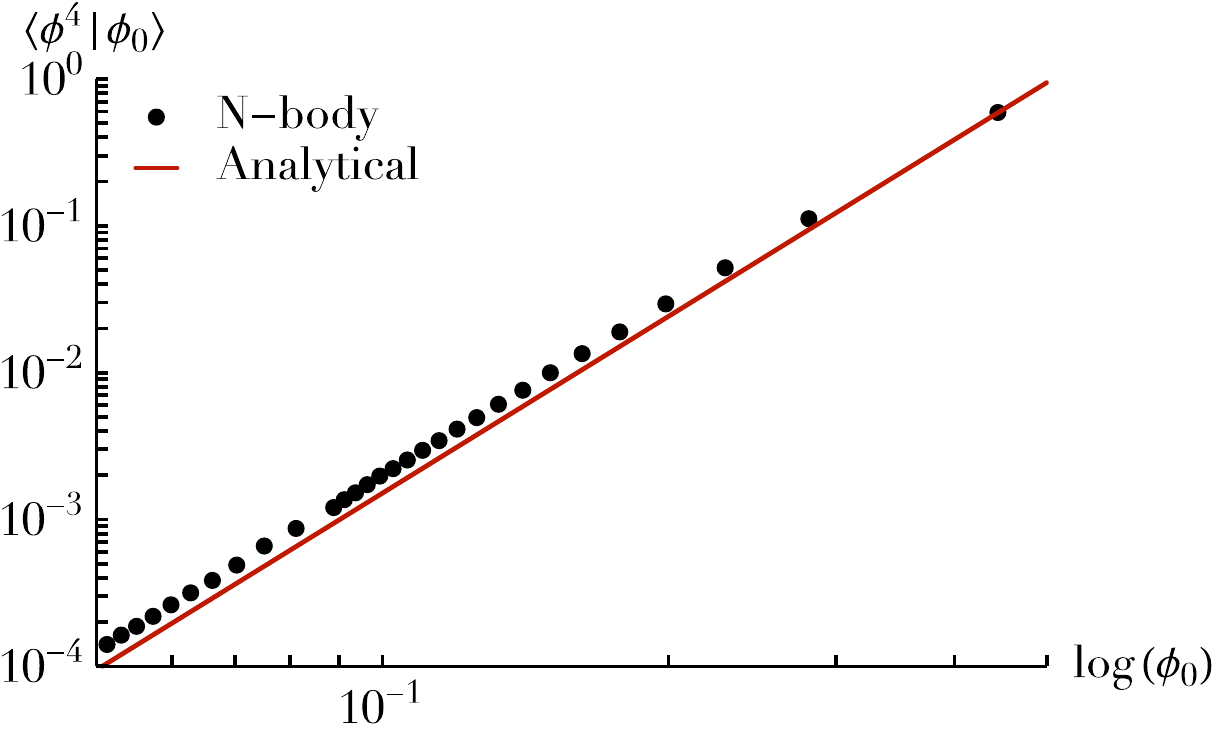}
   \caption{Illustration of the moments
   ${ \langle \phi^{2} | \phi_{0} \rangle }$
   and ${ \langle \phi^{4} | \phi_{0} \rangle }$,
   of the transition \PDF\@, ${ P (\phi | \phi_{0}) }$
   (see Eq.~\eqref{Virt_MasterEq}),
   as a function of the initial separation angle $\phi_0$
   for test particles with the same parameters
   evolving within the ${ \ell = 2 }$ toy model.
   All the angles are expressed in radians.
   Here, the $N$-body measurements were 
   obtained from histograms similar to the one
   of Fig.~\ref{fig:HistoPhiSameK},
   while the analytical prediction was obtained
   following Eq.~\eqref{identification_moments}.
   We observe a good agreement in a wide range
   of initial angular separations.
   }
   \label{fig:FitPhi}
\end{figure}

To estimate these moments,
we rely once again on a perturbative development of Eq.~\eqref{calc_C_neigh}.
For identical particles, we have $C^\bO_\ell = 1$.
Therefore, taking ${ \ell_{\alpha} = 1 }$ in Eq.~\eqref{CbD},
we obtain
\begin{align}
\langle \cos (\phi) & | \phi_{0}  \rangle  \!=\! \cos (\phi_0) \exp \!\bigg[\! \sum_\ell \! \frac{\cos (\phi_0) \!-\! P_\ell (\cos (\phi_0))}{3\cos (\phi_0)} \! \big( E^L_{1\ell\ell} \big)^{2}
\nonumber
\\
& \, \times \!\!\int\!\! \rd \bK \, n(\bK) \mJ_\ell^{2} \big[ \bK_1,\bK \big] \Tc^2 (\bK) \chi [\Delta t / \Tc (\bK)]
 \bigg ].
\label{logNormal_ell1}
\end{align}
Placing ourselves in the limit of small angular separations,
we can expand both sides of Eq.~\eqref{logNormal_ell1}
to obtain
\begin{equation}
1 - \tfrac{1}{2}\langle \phi^2 | \phi_{0} \rangle + \tfrac{1}{24}\langle \phi^4 | \phi_{0} \rangle \simeq 1 - \tfrac{1}{2} \beta_2 \, \phi_0^2 + \tfrac{1}{24} \beta_4 \, \phi_0^4,
\label{logNormal_MomentsExpand}
\end{equation}
where the coefficients
\begin{equation}
\beta_2 = -\frac{\partial^2 \langle \cos (\phi) | \phi_{0}) \rangle}{\partial \phi_0^2} ; \;\;\;
\beta_4 = \frac{\partial^4 \langle \cos (\phi) | \phi_{0} \rangle}{\partial \phi_0^4}
\label{logNormal_defBeta}
\end{equation}
are somewhat too intricate to be written down explicitly here
but can easily be computed numerically.
The next stage of the calculation
is to identify terms order by order in Eq.~\eqref{logNormal_defBeta}.
As such, we write
\begin{equation}
\langle \phi^{2} | \phi_{0} \rangle = \beta_{2} \, \phi_{0}^{2} ; \;\;\;
\langle \phi^{4} | \phi_{0} \rangle = \beta_{4} \, \phi_{0}^{4} . 
\label{identification_moments}
\end{equation}
Of course, this identification is only approximate
as the average term ${ \langle \phi^{2} | \phi_{0} \rangle }$
also contains residual contributions in ${ \phi_{0}^{4} }$,
in an unknown proportion.
Figure~\ref{fig:FitPhi} compares
numerical measurements of
${ \langle \phi^{2} | \phi_{0} \rangle }$
and
${ \langle \phi^{4} | \phi_{0} \rangle }$
with the predictions from Eq.~\eqref{identification_moments},
for a wide range of values of $\phi_{0}$.
While the agreement in Fig.~\ref{fig:FitPhi} is good,
it was only performed for the ${ \ell = 2 }$ toy model.
It therefore remains somewhat unclear
whether or not such relations would still
always be satisfied in the full \VRR\ model.

\subsection{Sampling virtual dilutions}

In Eq.~\eqref{ansatz_lognormal},
we identified a simple ansatz for the
transition \PDF\@, ${ P (\phi | \phi_{0}) }$.
And following Eqs.~\eqref{link_moments} 
and~\eqref{identification_moments},
we can use the analytical prediction from Eq.~\eqref{logNormal_ell1}
to explicitly estimate the two parameters of that \PDF\@.
We note that the coefficients $\beta_{2}$ and $\beta_{4}$
from Eq.~\eqref{identification_moments}
solely depend on the conserved orbital parameters
of the test stars,
i.e.\ for a given pair of particles,
they only have to be computed once.

\begin{figure}
    \centering
   \includegraphics[width=0.45 \textwidth]{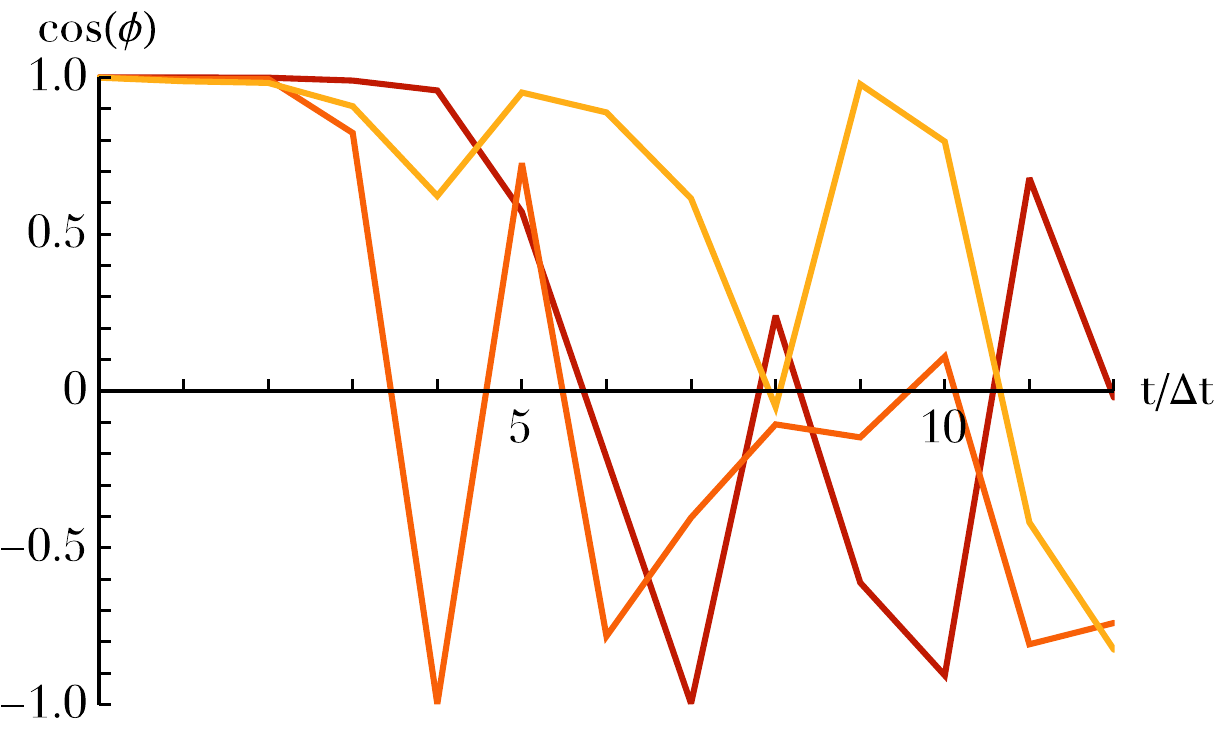}
   \caption{Illustration of three virtual random walks
   in orientation
   generated using the ansatz from Eq.~\eqref{ansatz_lognormal},
   and set to match the numerical simulation
   presented in Fig.~\ref{fig:RandomWalksNBody}.
   Here, we find that test particles
   separate on comparable timescales,
   but, as expected from the Markovian assumption,
   the virtual dilutions cannot mimic
   any of the short time variations
   observed in the simulations.
   }
   \label{fig:RandomWalksFake}
\end{figure}

In Fig.~\ref{fig:RandomWalksFake},
we illustrate examples virtual random walks
generated using that estimated transition \PDF\@.
These virtual random walks should be compared
to the ones from Fig.~\ref{fig:RandomWalksNBody}
directly observed in numerical simulations.
Of course, given that the Markovian approach
only predicts the angular separation every ${ \Delta t }$,
the virtual separations from Fig.~\ref{fig:RandomWalksFake}
cannot capture the short time variations
present in the numerical simulations.
As the motion of the test particles
is essentially ballistic on timescales ${ \lesssim \Delta t }$,
the most important features of the random walks
are none the less correctly captured.
Indeed, we recover
that the timescale for the full separation
of the test particles, a few ${ \Delta t }$,
is similar between the two figures.
We also find that both figures exhibit large angular oscillations
on timescales ${ \gtrsim \Delta t }$.

Once one can create virtual separations
of test particles,
it only remains to average over them
to obtain a Markovian prediction
for the expectation of the neighbour separation,
i.e.\ the quantity ${ \langle \cos (\phi) \rangle }$.
This is illustrated in Fig.~\ref{fig:CorrelationSameK},
where we recover a good agreement
between the numerical simulations 
(for the full \VRR\ model)
and the present approach based
on virtual realisations of the stochastic dynamics
of pair separation.

\section{A stellar cusp around S\lowercase{gr}A*}
\label{sec:PowerLaw}

Let us detail here the calculations presented in Section~\ref{sec:Applications}
to compute the rate of neighbour separation in an infinite power-law stellar distribution
around a supermassive \BH\@.

In order to mimic SgrA*,
we take the mass of the central \BH\ to be
${ \MBH = 4.3\!\times\! 10^{6} \Msun }$~\citep{Gillessen2017}.
The stellar population is assumed to be single mass,
with ${ \mstar = 1 \, \Msun }$,
so that the mass of the background star
is given by the \PDF\
${ f_{m} (m) = \deltaD (m - \mstar) }$.
The stellar distribution of eccentricities
is supposed to be thermal~\citep{Merritt2013},
i.e.\ ${ f_{e} (e) = 2 e }$.
We also assume that the number of stars per unit $a$
follows a power-law distribution of the form
\begin{equation}
n_{a} (a) = \frac{N_{0}}{a_{0}} \, \bigg( \frac{a}{a_{0}} \bigg)^{2 - \gamma} ,
\label{def_na}
\end{equation}
with $a_{0}$ a given scale semi-major axis.
In that expression, we also introduced ${ N_{0} = g (\gamma) N ( \!<\! a_{0}) }$
with the normalisation function
\begin{equation}
g (\gamma) = 2^{-\gamma} \, (3 - \gamma) \, \sqrt{\pi} \, \frac{\Gamma [1 \!+\! \gamma]}{\Gamma [\gamma - \tfrac{1}{2}]} .
\label{def_g_gamma}
\end{equation}
Here, ${ N ( \!<\! a_{0}) }$ stands for the number of stars
physically within a sphere of radius $a_{0}$ from the centre. 
In practice, for the numerical applications,
we used ${ a_{0} \!=\! \rh \!=\! 2 \, \mathrm{pc} }$
the sphere of influence of SgrA*,
and as such used ${ N( \!<\! a_{0}) = 4.3 \!\times\! 10^{6} }$.
Following the normalisation of ${ n(\bK) }$ from Eq.~\eqref{res_Cb},
we write the background bath's \DF\ as
${ n (m , a , e) \!=\! f_{m} (m) \, f_{e} (e) \, n_{a} (a) / (4 \pi) }$.

When adding \IMBHs\ to the background distribution,
we assume that they are all of the same individual mass
${ m_{\bullet} = 100 \Msun }$.
For simplicity, we also assume that they follow
a power law distribution in semi-major axes,
as well as a thermal distribution of eccentricities.
Finally, we systematically normalise the \PDFs\
so that the total enclosed stellar mass
within SgrA*'s sphere of influence remains the same,
i.e.\ we have
\begin{equation}
M_{\star} ( \!<\! a_{0}) + M_{\mathrm{IMBH}} ( \!<\! a_{0}) = \MBH ,
\label{enclosedmass_IMBH}
\end{equation}
with ${ M_{\star} ( \!<\! a_{0}) = m_{\star} N( \!<\! a_{0}) }$,
and similarly for the \IMBHs\@.
Equation~\eqref{enclosedmass_IMBH}
is used in Fig.~\ref{fig:AstroIMBH}
to define the mass fraction of \IMBHs\@.

\end{document}